%% file: 41200corr.tex
\documentclass[longauth]{aa} 

\usepackage{graphicx}
\usepackage{txfonts}
\usepackage{natbib}
\usepackage[colorlinks=true, linkcolor={blue}, citecolor={blue}, urlcolor={blue}]{hyperref}
\usepackage{subcaption}
\usepackage{xspace}

\newcommand{\madcows}{MaDCoWS\xspace}
\newcommand{\micron}{\mbox{{$\mu$}m}\xspace}
\newcommand{\yc}{$\Tilde{\vary}_0$\xspace}

\begin{document}

   \title{Atacama Cosmology Telescope measurements of a large sample of candidates from the Massive and Distant Clusters of WISE Survey:}
   \subtitle{Sunyaev-Zeldovich effect confirmation of MaDCoWS  candidates using ACT}
    \titlerunning{SZ effect confirmation of MaDCoWS candidates using ACT}   

\author{John Orlowski-Scherer\inst{1}\and 
    Luca Di Mascolo\inst{2,3,4,5}\and
    Tanay Bhandarkar\inst{1}\and
    Alex Manduca\inst{1}\and
    Tony Mroczkowski\inst{6}\and 
    Stefania Amodeo \inst{7}\and
    Nick Battaglia\inst{7} \and
    Mark Brodwin\inst{8}\and
    Steve K. Choi\inst{7,9}
    Mark Devlin\inst{1}\and
    Simon Dicker\inst{1}\and
    Jo Dunkley\inst{10,11}\and
    Anthony H. Gonzalez\inst{12}\and
    Dongwon Han\inst{13}\and
    Matt Hilton\inst{14,15}\and
    Kevin Huffenberger\inst{16}
    John P. Hughes \inst{17}\and
    Amanda MacInnis\inst{13}\and
    Kenda Knowles \inst{18} \and
    Brian J. Koopman\inst{19}\and
    Ian Lowe\inst{1}\and
    Kavilan Moodley\inst{14,15}\and
    Federico Nati \inst{20}\and
    Michael D. Niemack \inst{7,9,21}
    Lyman A. Page\inst{11}
    Bruce Partridge\inst{22}\and
    Charles Romero\inst{1}\and
    Maria Salatino\inst{23,24}
    Alessandro Schillaci\inst{25}\and
    Neelima Sehgal\inst{14}\and
    Crist\'obal Sif\'on\inst{26}
    Suzanne Staggs\inst{11}\and
    S. A.\ Stanford\inst{27}\and
    Robert Thornton\inst{1,28}\and
    Eve M. Vavagiakis \inst{7} \and
    Edward J. Wollack \inst{29}\and
    Zhilei Xu\inst{1,30}\and 
    Ningfeng Zhu\inst{1}
}
\authorrunning{Orlowski-Scherer}

\institute{
    {Department of Physics and Astronomy, University of Pennsylvania, 209 South 33rd Street, Philadelphia, PA, 19104, USA}
    \and
    {Max-Planck-Institut f\"{u}r Astrophysik (MPA), Karl-Schwarzschild-Strasse 1, Garching 85741, Germany}
    \and {Astronomy Unit, Department of Physics, University of Trieste, via Tiepolo 11, Trieste 34131, Italy}
    \and {INAF – Osservatorio Astronomico di Trieste, via Tiepolo 11, Trieste 34131, Italy}
    \and {IFPU - Institute for Fundamental Physics of the Universe, Via Beirut 2, 34014 Trieste, Italy}
    \and {European Southern Observatory (ESO), Karl-Schwarzschild-Strasse 2, Garching 85748, Germany}
    \and
    {Department of Astronomy, Cornell University, Ithaca, NY 14853, USA}
    \and
    {Department of Physics and Astronomy, University of Missouri, Kansas City, MO 64110} 
    \and
    {Department of Physics, Cornell University, Ithaca, NY 14853, USA}
    \and
    {Department of Astrophysical Sciences, Peyton Hall,
    Princeton University, Princeton, New Jersey 08544, USA}  
    \and
    {Joseph Henry Laboratories of Physics, Jadwin Hall, Princeton University, Princeton, NJ 08544, USA}
    \and
    {Department of Astronomy, University of Florida, Gainesville, FL, USA 32611}
    \and
    {Physics and Astronomy Department, Stony Brook University, Stony Brook, NY 11794}
    \and
    {Astrophysics Research Centre, University of KwaZulu-Natal, Westville Campus, Durban 4041, South Africa}
    \and
    {School of Mathematics, Statistics \& Computer Science, University of KwaZulu-Natal, Westville Campus, Durban 4041, South Africa}
    \and
    {Department of Physics, Florida State University, Tallahassee FL, USA 32306}
    \and
    {Department of Physics and Astronomy, Rutgers, the State University of New Jersey, Piscataway, NJ 08854, USA}
    \and
    {Centre for Radio Astronomy Techniques and Technologies, Department of Physics and Electronics, Rhodes University, Drosty Rd, Grahamstown 6139, South Africa}
    \and
    {Department of Physics, Yale University, New Haven, CT 06520, USA}
    \and
    {Department of Physics, University of Milano-Bicocca, Piazza della Scienza 3, 20126 Milano (MI), Italy}
    \and
    {Kavli Institute at Cornell for Nanoscale Science, Cornell University, Ithaca, NY 14853, USA}
    \and
    {Department of Physics and Astronomy, Haverford College, 370 Lancaster Ave, Haverford, PA 19041, USA}
    \and
    {Department of Physics, Stanford University, Stanford 94305 CA USA}
    \and
    {Kavli Institute for Particle Astrophysics and Cosmology, Stanford CA, 94305 US}
    \and
    {Department of Physics, California Institute of Technology, Pasadena, CA 91125, USA}
    \and
    {Instituto de F\'isica, Pontificia Universidad Cat\'olica de Valpara\'iso, Casilla 4059, Valpara\'iso, Chile
    }
    \and
    {Department of Physics and Astronomy, University of California, Davis, CA, 95616, USA}
    \and
    {Department of Physics, West Chester University of Pennsylvania, West Chester, PA 19383, USA}
    \and
    {NASA Goddard Space Flight Center, 8800 Greenbelt Road, Greenbelt, MD 20771 USA}
    \and
    {MIT Kavli Institute for Astrophysics and Space Research, Massachusetts Institute of Technology, 77 Massachusetts Avenue, Cambridge, MA 02139 USA}
}

   \date{Received April 28, 2021; accepted June 21, 2021}

 
  \abstract
   {Galaxy clusters are an important tool for cosmology, and their detection and characterization are key goals for current and future surveys.
   Using data from the Wide-field Infrared Survey Explorer (WISE), the Massive and Distant Clusters of WISE Survey (MaDCoWS) located 2,839 significant galaxy overdensities at redshifts $0.7\lesssim z\lesssim 1.5$, which included extensive follow-up imaging from the Spitzer Space Telescope to determine cluster richnesses. Concurrently, the Atacama Cosmology Telescope (ACT) has produced large area millimeter-wave maps in three frequency bands along with a large catalog of Sunyaev-Zeldovich (SZ)-selected clusters as part of its Data Release 5 (DR5).}
   {We aim to verify and characterize MaDCoWS clusters using measurements of, or limits on, their thermal SZ (tSZ) effect signatures.  We also use these detections to establish the scaling relation between SZ mass and the MaDCoWS-defined richness.}
   {Using the maps and cluster catalog from DR5, we explore the scaling between SZ mass and cluster richness.  We do this by comparing cataloged detections and extracting individual and stacked SZ signals from the MaDCoWS cluster locations. We use complementary radio survey data from the Very Large Array, submillimeter data from {\it Herschel}, and ACT 224~GHz data to assess the impact of contaminating sources on the SZ signals from both ACT and MaDCoWS clusters. We use a hierarchical Bayesian model to fit the mass-richness scaling relation, allowing for clusters to be drawn from two populations: one, a Gaussian centered on the mass-richness relation, and the other, a Gaussian centered on zero SZ signal.}
   {We find that MaDCoWS clusters have submillimeter contamination that is consistent with a gray-body spectrum, while the ACT clusters are consistent with no submillimeter emission on average. Additionally, the intrinsic radio intensities of ACT clusters are lower than those of MaDCoWS clusters, even when the ACT clusters are restricted to the same redshift range as the MaDCoWS clusters. We find the best-fit ACT SZ mass versus MaDCoWS richness scaling relation has a slope of $p_1 = 1.84^{+0.15}_{-0.14}$, where the slope is defined as $M\propto \lambda_{15}^{p_1}$ and $\lambda_{15}$ is the richness.  We also find that the ACT SZ signals for a significant fraction ($\sim57\%$) of the MaDCoWS sample can statistically be described as being drawn from a noise-like distribution, indicating that the candidates are possibly dominated by low-mass and unvirialized systems that are below the mass limit of the ACT sample. Further, we note that a large portion of the optically confirmed ACT clusters located in the same volume of the sky as MaDCoWS are not selected by MaDCoWS, indicating that the MaDCoWS sample is not complete with respect to SZ selection. Finally, we find that the radio loud fraction (RLF) of MaDCoWS clusters increases with richness, while we find no evidence that the submillimeter emission of the MaDCoWS clusters evolves with richness.}
    {We conclude that the original MaDCoWS selection function is not well defined and, as such,  reiterate the MaDCoWS collaboration’s recommendation that the sample is suited for probing cluster and galaxy evolution, but not cosmological analyses. We find a best-fit mass-richness relation slope that agrees with the published MaDCoWS preliminary results. Additionally, we find that while the approximate level of infill of the ACT and MaDCoWS cluster SZ signals ($1-2\%$) is subdominant to other sources of uncertainty for current generation experiments, characterizing and removing this bias will be critical for next-generation experiments hoping to constrain cluster masses at the sub-percent level. 
    }
   \keywords{galaxies: clusters --- galaxies: clusters: intracluster medium --- cosmic background radiation\vspace{-21pt}}

   \maketitle
%


\section{Introduction}\label{sec:intro}

Astronomers have long sought an efficient and effective way to identify galaxy clusters as well as a convenient observational proxy for their mass  \citep{Abell1958,Rykoff2012,Andreon2015,Saro2015,Geach2017,Simet2017,Rettura2018,Gonzalez2019,Chiu2020}, particularly at high redshift, where their formation and distribution are sensitive probes of cosmology.

Recently, the Sunyaev-Zeldovich (SZ) effect (see \citealt{Sunyaev1970, Sunyaev1972} as well as \citealt{Birkinshaw1999,Carlstrom2002, Mroczkowski2019} for reviews) has been used to uncover large populations of distant clusters. In particular, the thermal SZ (tSZ) effect allows redshift-independent detections of clusters due to inverse-Compton scattering of photons from the Cosmic Microwave Background (CMB) as they pass through hot gas in the clusters.\footnote{Throughout the paper we use ``SZ effect'' to refer exclusively to the tSZ effect, as opposed to the kinetic SZ effect.} Cluster masses can then be estimated from the amplitude of the SZ signals under the assumption of a universal pressure profile \citep[e.g.,][]{Arnaud2010}; we refer to such estimates as "SZ masses." The {\it Planck} satellite, which provides the only all-sky SZ survey to date, has been limited by both sensitivity and angular resolution ($10\arcmin$ at $100$~GHz) and has identified clusters with $z < 1$ \citep{Planck2016XXVII}. Ground-based surveys, such as the Atacama Cosmology Telescope (ACT; \citealt{Fowler2007,Swetz2011, Thornton2016}) and the South Pole Telescope (SPT; \citealt{Carlstrom2011, Benson2014}), have achieved $1-2\arcmin$ resolution and are sensitive to high-redshift clusters but until recently only surveyed a small fraction of the sky. These surveys are also now more sensitive than {\it Planck} over large portions of the sky \citep[see, e.g.,][]{Naess2020}.

Meanwhile, surveys from optical through infrared (IR) wavelengths as well as analysis methods have progressed, offering new data, new selection techniques (e.g., weak lensing shear), and more advanced richness selection criteria.  One such survey, the Massive and Distant Clusters of WISE Survey \citep[MaDCoWS;][]{Gonzalez2019}, relies on data from the Wide-field Infrared Survey Explorer (WISE). MaDCoWS provides an IR-selected sample of candidate clusters at redshifts $0.7<z<1.5$. The MaDCoWS sample aims to extend richness selection to a higher average redshift than previous surveys.  

ACT observed roughly $40\%$ of the sky as of the fifth data release (hereafter referred to as DR5)\footnote{\url{https://lambda.gsfc.nasa.gov/product/act/actpol_prod_table.cfm}}, which includes cluster data taken through the 2018 observing season \citep{Hilton2021}.  This data release enables large, statistical comparisons between cluster richness, as measured by optical or IR surveys and their SZ mass. DR5 provides SZ measurements for a large fraction of the MaDCoWS candidates, well beyond the handful of systems targeted for individual SZ follow-up in \cite{Gonzalez2019}, \cite{Dicker2020}, \cite{DiMascolo2020}, and \cite{Ruppin2020}, for example.

In this work we use data from ACT DR5 to establish how SZ mass scales with the MaDCoWS definition of richness for a large sample of MaDCoWS cluster candidates.
The work presented here complements the recent work by \cite{Madhavacheril2020}, who report the mean mass, determined through stacked CMB lensing, of the MaDCoWS candidates located within the ACT survey region and above a richness of 20 (Sect.~\ref{sec:data:madcows}). Additionally, this work probes the mass-richness scaling relation, and hence cluster abundance, at a higher redshift than previous studies \citep[e.g.,][]{Sehgal2013}.

The paper is structured as follows. An overview of the ACT DR5 and MaDCoWS samples, as well as the ancillary data we use to test for radio and submillimeter source infill, is provided in Sect.~\ref{sec:data}. 
In Sect.~\ref{sec:crossmatches} we detail the process used to identify clusters detected in both  ACT and MaDCoWS samples and from this estimate the completeness of the MaDCoWS sample.
In Sect.~\ref{sec:photometry} we describe the forced photometry process used to obtain masses for the MaDCoWS clusters from the ACT maps.
In Sect.~\ref{sec:baises} we review the corrections performed on the forced photometry mass estimates, including contamination by active galactic nuclei (AGN) and dusty submillimeter galaxies via ancillary radio and submillimeter data.
In Sect.~\ref{sec:scalingrels} we describe the process used to infer the mass-richness scaling relation. 
In Sect.~\ref{sec:discuss} we discuss the inferred scaling relations and completeness of the MaDCoWS catalog, as well as the impact of contamination by IR and submillimeter sources on the SZ mass.
In Sect.~\ref{sec:conclusions} we provide conclusions and give an outlook for further extensions to this work.\\

Throughout this work, we assume a flat $\Lambda\ $cold dark matter cosmology with $\Omega_{\mathrm{M}}=0.307$, $\Omega_{\Lambda}=0.693$, and $H_0=67.7~\mathrm{km\,s^{-1}\,Mpc^{-1}}$ from \citet{PlanckXIII}.

\section{Data}\label{sec:data}

In this work we primarily use the MaDCoWS cluster catalog \citep{Gonzalez2019catalog} and the catalog and maps from ACT \citep{Hilton2021}.
Additionally, we use data from the {\it Herschel} Space Observatory \citep{Valiante_2016, Smith_2017} and the Very Large Array (VLA) \citep{Condon1998, Lacy2020} to constrain dust and radio infill of the SZ signal, respectively. We also use the ACT $224$~GHz maps to assess and constrain the impact of dust in-fill.  

\subsection{MaDCoWS}\label{sec:data:madcows}

The MaDCoWS galaxy cluster catalog comprises 2839 cluster candidates spanning redshifts $0.7 \lesssim z \lesssim 1.5$, selected using WISE \citep{Wright2010} all-sky survey data \citep{Gonzalez2019}.\footnote{The MaDCoWS catalog is available as a supplement on Vizier \citep{Gonzalez2019catalog}.}  To reduce contamination by lower-redshift galaxies, MaDCoWS uses optical data from the Panoramic Survey Telescope and Rapid Response System \citep[Pan-STARRS;][]{Chambers2016} at declination $\delta > -30^\circ$, and SuperCOSMOS \citep{Hambly2001c, Hambly2001b,Hambly2001a} at $\delta < -30^\circ$ to reject low-redshift interlopers. MaDCoWS also uses data from the Sloan Digital Sky Survey \citep[SDSS]{York2000} for the same purpose over sections of the SDSS footprint. In total, $2433$ cluster candidates were identified by the WISE--Pan-STARRS search, and 250 by the WISE--SuperCOSMOS search. The MaDCoWS catalog includes photometric redshifts for $1869$ of its candidates, derived from \textit{Spitzer} imaging. Spectroscopic measurements of a limited subsample of the MaDCoWS cluster candidates indicate that the photometric uncertainty is $\sigma_{z} / (1+z) \approx 0.036$. In addition to photometric redshifts, the \textit{Spitzer} follow-up also enabled an estimate of cluster richness \citep[{$\lambda_{15}$}; see Sect. $6.3$ of][]{Gonzalez2019}. Briefly, the MaDCoWS richness parameter $\lambda_{15}$ is the number of galaxies within a comoving 1~Mpc radius aperture for the candidate's redshift having a flux density $> 15\,\mu$Jy after applying the color selection criteria described in \cite{Wylezalek2013} and subtracting the expected number of field galaxies. The color selection was designed to select only high-redshift clusters.
Within the MaDCoWS catalog,  1869 of the 2839 cluster candidates have both richness and redshift estimates.  For the purposes of determining a mass-richness scaling relationship, this subset was further restricted to those MaDCoWS cluster candidates lying in the ACT footprint, which totaled 1035. Additionally, 70 clusters lie in regions that are masked due to point source contamination: this leaves {\bf 965} MaDCoWS cluster candidates that were used in the analysis. To determine the radio and submillimeter properties of both the ACT and MaDCoWS clusters, we consider the full ACT and MaDCoWS cluster catalogs, necessarily restricted to those clusters for which we have radio and submillimeter data. Finally, in Sect.~\ref{sec:photometry}, we stacked on the MaDCoWS cluster locations to verify that the MaDCoWS cluster candidates did, on average, produce an SZ signal. In order to ensure that known ACT clusters did not dominate this stacked signal, we excluded MaDCoWS clusters that were also detected in ACT from the stacking analysis. We did however include clusters without a redshift measurement, resulting in the stacking analysis using a slightly different number of clusters ($948$) from the mass-richness fit.

\subsection{ACT}\label{sec:data:act}

ACT is a $6$-meter, off-axis Gregorian telescope located in the Atacama Desert in Chile that has been operating since 2007 \citep{Fowler2007}. The Advanced ACTPol (AdvACT) receiver, which was deployed in 2016, is its latest camera  \citep{Henderson2016, Thornton2016}. It performs polarization sensitive observations centered at $98$, $150$, and $224$~GHz, corresponding to a diffraction-limited resolution of $2.2\arcmin$, $1.4\arcmin$, and $1.0\arcmin$, respectively. Throughout, we use f090, f150, and f220 to refer to the maps made at those frequencies and $98$, $150$, and $224$~GHz when referring specifically to the frequencies. ACT has undertaken a number of large area, unbiased cluster surveys using the SZ effect \citep{Menanteau2010, Marriage2011, Sehgal2011, Hasselfield2013, Menanteau2013, Hilton2018, Hilton2021}. In this work, we use the DR5 cluster catalog \citep{Hilton2021}, which we refer to as the ACT cluster catalog, and whose members we call ACT clusters, as well as maps of the central Comptonization parameter \citep[\yc, often referred to as "SZ maps" in this work; see Sect. 2.3 of][]{Hilton2021}. To construct these SZ maps, we  use the ACT maps filtered at the reference 2.4\arcmin\ scale to perform forced photometry at the locations of clusters reported in the MaDCoWS catalog. This matched filtering essentially reduces the SZ detection of given cluster candidate to a single quantity, \yc. The SZ map is constructed such that each pixel records the $\Tilde{\vary}_0$ value that a cluster would have if it were detected at a given location in the map. Therefore, we simply extract $\Tilde{\vary}_0$ and S/N$_{2.4}$ (the signal-to-noise measured in this 2.4$\arcmin$ scale map) values at the coordinates of each MaDCoWS cluster to produce a forced photometry catalog. Sub-pixel interpolation is performed using a bivariate third-order spline method. We warn that following Sect.~$4.2$ of \citet{Hilton2021}, the SZ masses estimated here from these maps may be underestimated by $\sim5-10\%$, and as such caution should be exercised when comparing the reported masses here to other cluster catalogs. Additionally, it is known that SZ measured masses are biased low by about $30\%$ as compared to weak-lensing calibration \citep{Miyatake2019}. There are currently efforts underway to measure ACT cluster masses via weak-lensing: When available they will represent the most accurate, least biased cluster masses available.

The DR5 catalog contains $4195$ SZ-selected, optically confirmed clusters with signal-to-noise $> 4$ and with redshifts in the range $0.04 < z < 1.91$ over 13,211\,deg$^2$ of the sky. The catalog has a 90\% completeness mass limit of $3.8 \times 10^{14} \text{M}_{\odot}$ at $\text{z} = 0.5$.

While the ACT cluster search was conducted using matched filters with a number of different scales, a fixed reference scale with $\theta_{\rm 500c} = R_{\rm 500c} / D_A = 2.4\arcmin{}$ was used for characterizing the SZ signal and its relation to mass. This scale is equivalent to a cluster with $M_{500c} = 2 \times 10^{14}$\,$M_{\sun}$ at $z = 0.4$, assuming the \citet{Arnaud2010} pressure profile and associated scaling relation. In this work, we use the map of the central Comptonization parameter \yc\ at this reference scale  and the associated signal-to-noise map to estimate the masses of MaDCoWS clusters using forced photometry (see Sect.~\ref{sec:photometry}).

In addition to the \yc maps, we also used the individual frequency maps, f090 and f150, as well as \yc maps made with each frequency ($98$ and $150$~GHz) individually. We also used the f220 maps constructed from  observations at $224$~GHz. While the f220 maps are noisier ($50-60~\mu$K arcmin) than the f090 and f150 data  \citep[ $\lesssim 30~\mu$K arcmin typical; see][]{Naess2020}, the band is centered near the null in the SZ effect, providing a clean band for quantifying the dust emission in the ACT and MaDCoWS clusters, as is discussed in Sect.~\ref{sec:submm_emission}. For all these maps, the pixel size is $0.5\arcmin$.

\subsection{{\it Herschel}}\label{sec:data:herschel}
We used the {\it Herschel} Astrophysical Terahertz Large Area Survey (H-ATLAS) DR1 \citep{Valiante_2016} and DR2 \citep{Smith_2017} to measure the thermal emission from dust in the ACT and MaDCoWS clusters within the H-ATLAS footprint. H-ATLAS covers 660 deg$^2$ at $100$, $160$, $250$, $350$, and $500$~$\mu$m using the PACS and SPIRE cameras. We used only the $250$, $350$, and $500$~$\mu$m bands, all from the SPIRE camera. The SPIRE resolution is $18.2\arcsec$, $25.2\arcsec$, and $36.3\arcsec$ at $250$, $350$, and $500$~$\mu$m, respectively, with pixel size equal to the resolution. Due to the relatively small size of the H-ATLAS field, only $34$ ACT and $66$ MaDCoWS clusters have {\it Herschel} coverage. 

\subsection{Very Large Array}\label{sec:data:vla}

In order to determine if radio source in-fill impacts the SZ signals from MaDCoWS candidates, we examine data from the National Radio Astronomy Observatory (NRAO) VLA Sky Survey \citep[NVSS;][]{Condon1998} and the Very Large Array Sky Survey \citep[VLASS;][]{Lacy2020}. 

NVSS is a $1.4$~GHz survey with $45\arcsec$ FWHM angular resolution ($15\arcsec$ pixels) that covers approximately 82\% of the sky at declinations $\delta \geq -40 ^\circ$. The NVSS catalog includes a set of 2326 continuum images made with a large restoring beam to provide the sensitivity needed for completeness. 

VLASS is an on-going $3$~GHz radio survey producing Stokes I, Q, and U maps with an angular resolution $\approx2.5\arcsec$ ($1\arcsec$ pixel size). Like NVSS before it, the survey covers the entire sky visible to the VLA, a $\sim$34,000\,deg$^2$ ($\delta > -40^\circ$) area. The survey's first observations began in September 2017. VLASS is expected to detect, by the project's completion in 2024, $\sim$5,000,000\ sources and record data with a continuum image RMS of 70 $\mu$Jy/beam combined and 120 $\mu$Jy/beam per-epoch. The first epoch survey of the entire VLASS footprint has been completed, and data products are available.\footnote{The VLASS data are available on the Canadian Astronomy Data Centre site, \href{http://www.cadc-ccda.hia-iha.nrc-cnrc.gc.ca/en/search/?collection=VLASS&noexec=true\#resultTableTab}{http://www.cadc-ccda.hia-iha.nrc-cnrc.gc.ca/en/search/?collection=VLASS\&noexec=true\#resultTableTab}.} ``Quicklook'' 2D Stokes I images covering the entire survey were used to conduct our investigations into MaDCoWS radio source in-fill.

\section{Co-detections in the MaDCoWS and ACT catalogs}\label{sec:crossmatches}

\begin{figure}
    \centering
    \includegraphics[clip,trim=1mm 0mm 15mm 0mm,width=\columnwidth]{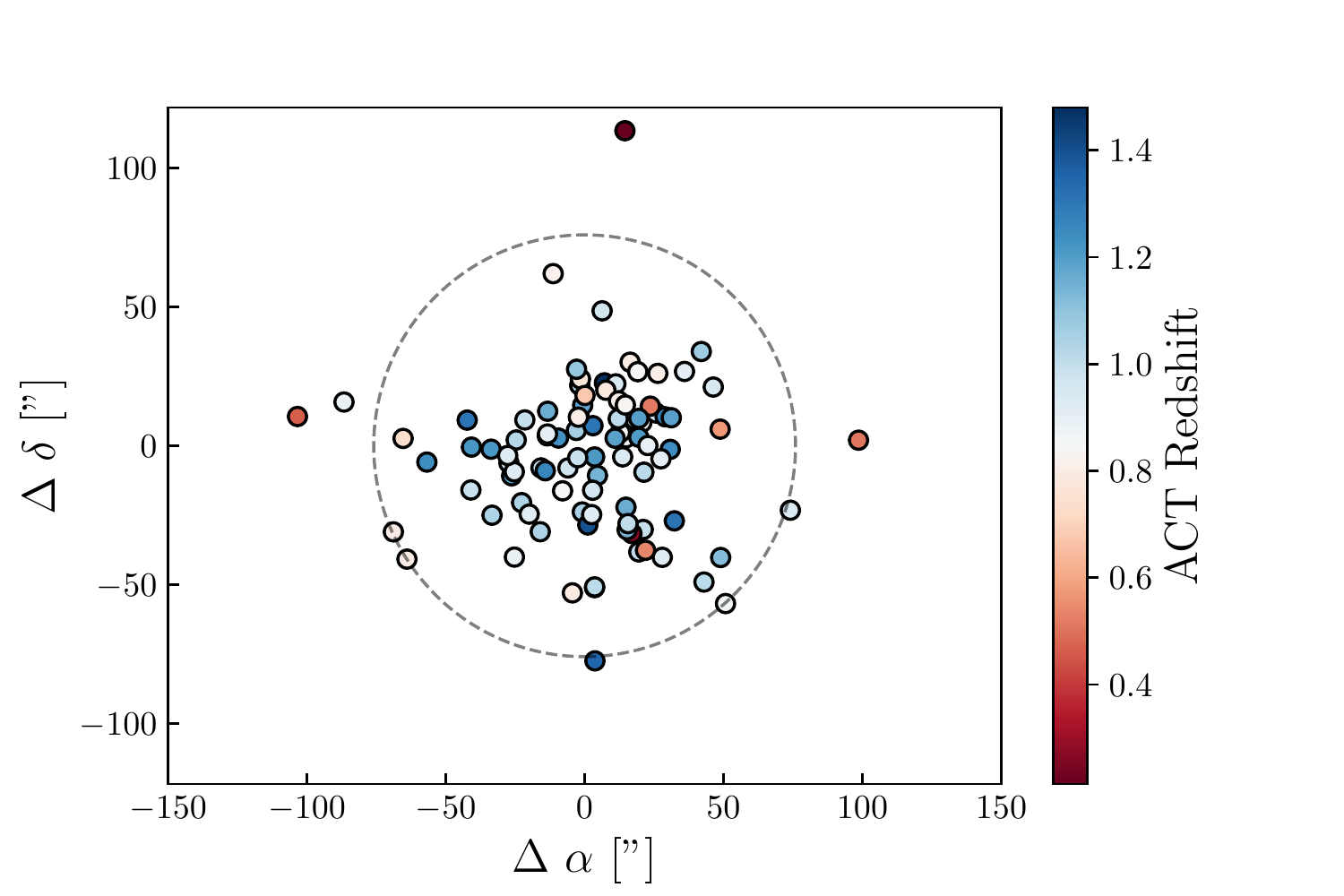}
    \caption{Offsets in the right ascension and declination of ACT clusters \citep{Hilton2021} and their co-detected MaDCoWS counterparts \citep{Gonzalez2019catalog}. The color bar indicates the redshift of the co-detection as recorded by ACT in the DR5 cluster catalog. The black-dotted circle is the radius ($1.2\arcmin$) that includes 89 (95\%) of the co-detections. }
    \label{fig:Offsets}
\end{figure}

In order to understand the completeness of MaDCoWS, we identify ACT-selected clusters that we consider to be matches with MaDCoWS candidates, which we refer to as co-detections. We consider a MaDCoWS candidate and an ACT cluster to be a co-detection (i.e., the same cluster) if the positional difference between the ACT entry and MaDCoWS entry was less than or equal to $2.5\arcmin$. The criterion of a $2.5\arcmin$ matching scale was chosen as it is approximately the resolution limit given the ACT catalog filter reference scale ($2.4\arcmin$). It should be noted that \citet{Hilton2021} find that $99.7\%$ of the ACT cluster centers are within $1.9\arcmin$ of the optical centers. Using this criterion, we identified 96 co-detections. We report these 96 cluster co-detections in Appendix~\ref{Catalog:ACTCOWS}. Restricting our search to only include matches to MaDCoWS clusters in the Pan-STARRS footprint, that number is reduced to 80.

We explored co-detections with larger positional difference values as well, increasing the accepted positional difference from 2.5\arcmin\ to 5\arcmin. However, doing so only resulted in an additional 7 co-detection candidates, which we deemed to be only chance superpositions.  In Fig. \ref{fig:Offsets}, we show the typical offsets between the co-detections, indicating the match is generally within 0.7\arcmin.

As discussed in Sect.~\ref{sec:data:madcows}, after restricting the MaDCoWS cluster catalog to the ACT footprint, masking point sources, and removing clusters without measured richnesses or redshifts, a total of $965$ MaDCoWS clusters remain, which are used for the mass richness scaling relation. We note that discrepancies in the redshift determinations existed for several of the co-detections.  Wherever these occurred, we used the redshift reported in \cite{Hilton2021} since these generally included newer and more complete data, and correspondingly smaller uncertainties.

To estimate what the background rate of line-of-sight coincidences between ACT and \madcows clusters is, we simulated 100,000 surveys with the same angular density of ACT and MaDCoWS clusters as our paper (0.32 and 0.074 clusters per square degree, respectively), spread randomly over 13,211\,deg$^2$. We then simply counted the number of ACT and MaDCoWS clusters lying within $2.5\arcmin$ of each other. We found that there was a chance coincidence of at least one cluster $87\pm 11\%$ of the time, and that on average there were $2\pm 2$ chance coincidences per survey. For each of our actual co-detections, we computed the difference in the measured ACT and MaDCoWS $z$ divided by the quadrature sum of the $z$ uncertainties:

\begin{equation}
    \sigma_{z} = \frac{|z_{\textsc{ACT}}-z_{\textsc{\madcows}}|}{\sqrt{ \sigma_{\textsc{ACT}}^2 + \sigma_{\textsc{\madcows}}^2}}
.\end{equation}

We found that two co-detections (ACT-CL J0002.3+0131 and ACT-CL J0009.8-0205) are significantly discrepant ($\sigma_{\textsc{z}} = 7.0$ and $7.1$, respectively), and could be line-of-sight coincidences. Additionally, ACT-CL J0009.1-4147 is marginal at $\sigma_{\textsc{z}} = 3.1$, given the sample size. The rest were $ \sigma_{\textsc{z}} \lesssim 2$, with most around 1. In principle spectroscopic follow-up would be able to disentangle line-of-sight coincidences.

\begin{figure*}
    \centering
    \includegraphics[scale=0.5]{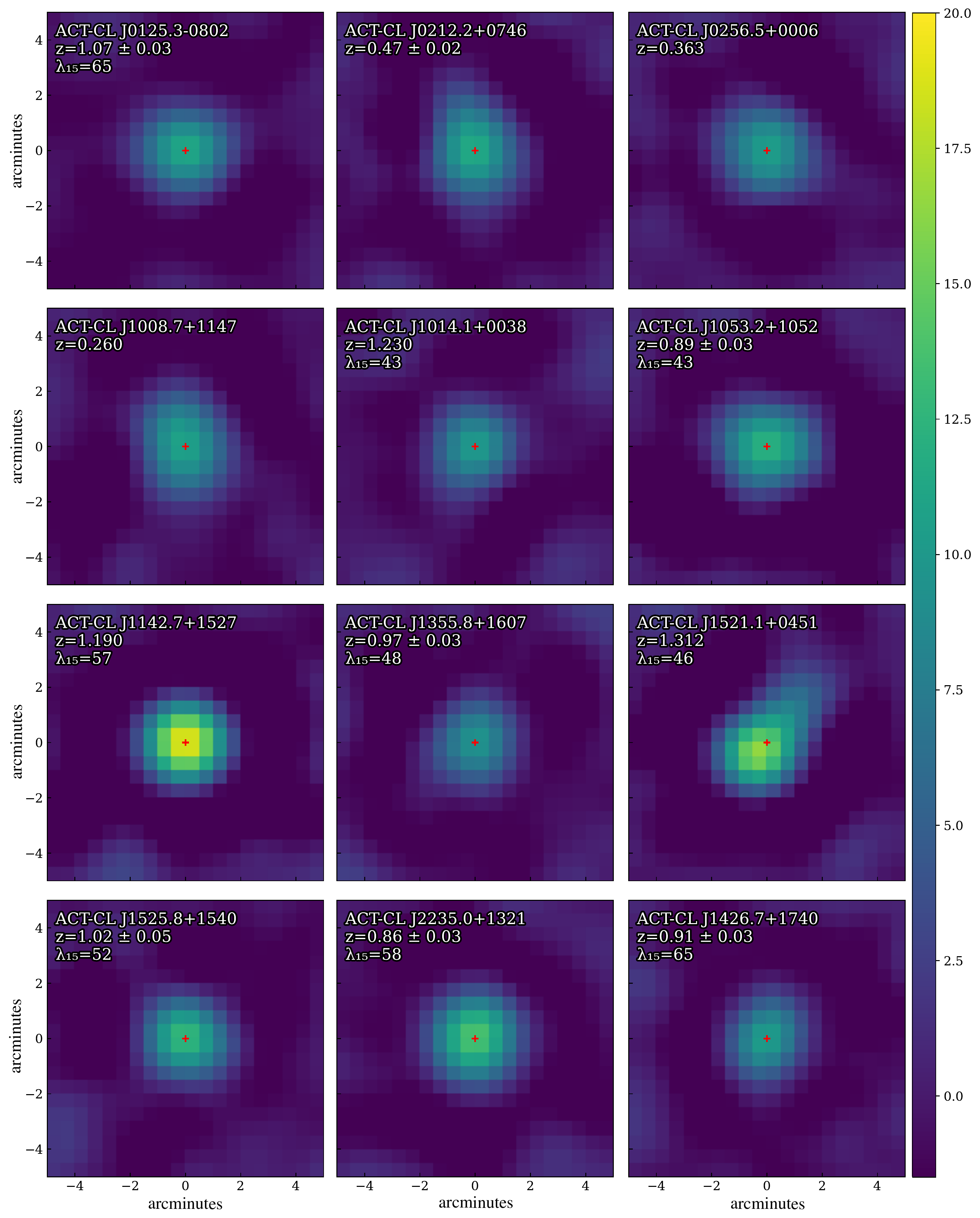}
    \caption{ACT S/N maps of the dozen highest-significance co-detections, where the S/N is with respect to \yc. Each panel notes the ACT cluster name and redshift from \citet{Hilton2021}. Some of these clusters do not have a measured MaDCoWS richness, and as such none is reported. The three clusters lacking richness measures were found to be lower-redshift clusters. Red crosses denote the ACT-identified cluster center. The color bar scale is in units of \yc.}
    \label{fig:stamps}
\end{figure*}

Given the ACT cluster catalog, the number of co-detections sets an upper limit on the completeness with respect to the ACT catalog of the MaDCoWS cluster catalog and informs our understanding as to what extent these two surveys probe the same population of clusters. We first consider whether MaDCoWS detected all ACT clusters. Restricting the ACT catalog to match the MaDCoWS catalog in redshift ($0.7<z<1.5$) and footprint ($\delta > -30^\circ$, corresponding to the Pan-STARRS follow-up region) yields 712 ACT clusters compared to 80 co-detections restricted to the Pan-STARRS region. Relaxing the footprint constraint to include areas of SuperCOSMOS follow-up yields 1,102 ACT clusters, compared to 96 co-detections. 
While this does not put a hard constraint on the completeness of MaDCoWS, it does suggest that it is $\lesssim 10\%$. A primary reason for this low completeness is expected to be the large non-Gaussian scatter between the mass and detection significance in the MaDCoWS search.MaDCoWS clusters were detected as galaxy excesses traced by the bright tip of the luminosity function, with the strength of the signal significantly affected by both Poisson statistics and blending of galaxies at the resolution of WISE. In other words, the selection function is not dominated by the richness of the clusters, but rather by other factors. This in turn means that the selection function is only weakly dependent on mass, so that it is not necessarily the case that all high-mass (i.e., ACT) clusters will be detected by MaDCoWS.

Qualitatively, we measure an intrinsic scatter ($\sigma_{\ln{\lambda}|\text{S/N}}=0.26\pm 0.01$) of the same order as the intrinsic scatter on the mass-richness scaling relation ($\sigma_{\ln(M) | \lambda} = 0.22 \pm 0.10$; see Sect.~\ref{sec:scaling_results}). As such, the relationship between mass and S/N is quite scattered. Furthermore, given that S/N is a detection limited quantity (i.e., we only consider clusters with S/N > 5 when fitting for the richness-S/N relation), then the measured intrinsic scatter of this richness-S/N relation is going to be biased low, as we have excluded clusters with low S/N for their richness. All together, the effect is that the MaDCoWS selection function does not track mass particularly closely, and as such the number of ACT clusters co-detected by MaDCoWS is lower than one would expect.

\begin{figure}[]
    \centering
    \includegraphics[clip,trim=0mm 0mm 9mm 0mm,width=\columnwidth]{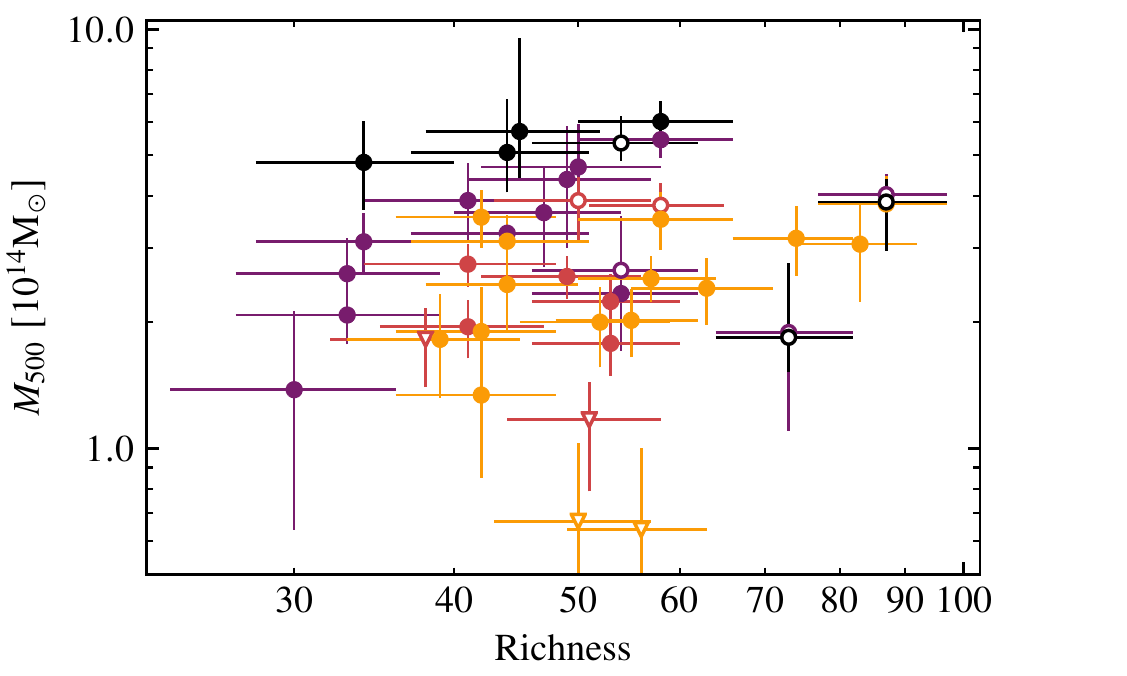}
    \caption{Mass vs. richness relation for a selection of MaDCoWS clusters with SZ mass estimates. The purple circles correspond to the CARMA MaDCoWS cluster sample from \citet{Gonzalez2019}. The VACA LoCA points from \citet{DiMascolo2020} are shown in red, and the MUSTANG2 measurements from \citet{Dicker2020} are shown in orange. The black data points are {\it Chandra} observations of \madcows clusters. Points that are open are known active mergers, and points that are represented with triangles are consistent with no signal. 
    We note that Fig. \ref{fig:scaling} provides a similar comparison for the complete sample of MaDCoWS candidates in the ACT survey footprint.
    } 
    \label{fig:massrich}
\end{figure}

We highlight in Fig. \ref{fig:stamps} a few prominent co-detections. Additionally, we compare the SZ masses for a number of ACT clusters to those from the literature in Appendix~\ref{app:szmasscomp}.
Overall, the masses inferred from these targeted observations agree within $1\sigma$ with the ACT-inferred mass estimates. One noteworthy exception is that of MOO~J1142+1527, where the mass estimates using Combined Array for Research in Millimeter-wave Astronomy (CARMA) \citep{Gonzalez2019}, New IRAM Kids Arrays (NIKA2)+CARMA \citep{Ruppin2020}, ACT, and MUSTANG2 \citep{Dicker2020} differ at approximately the $2\sigma$ level. We note that \cite{Moravec2020} report this cluster as an ongoing merger and that it may require multiple SZ components to describe. 
In Fig.~\ref{fig:massrich}, we show the data used to infer scaling relations using CARMA and MUSTANG2 \citep[respectively]{Gonzalez2019, Dicker2020}.  Additionally, there is some evidence in this figure that, for the ACT co-detections as well as the clusters from the previous SZ follow-up campaigns, there appears to be a bimodal split in the SZ mass of the high $\lambda_{15}$ systems, which is more evident when plotted in log-space (see Fig.~\ref{fig:scaling} below, left panel). The effect may be in part due to merging and pre-merger systems, which can have low SZ signals for their richness \citep{Dicker2020}. The suggestion of bimodal behavior in Fig.~\ref{fig:massrich} becomes more evident in Fig.~\ref{fig:forced_photo}, in which the SZ signals of the entire MaDCoWS catalog as measured with ACT are plotted. The masses of the high-richness ($\lambda_{15} \gtrsim 55$) systems cluster into two branches, one higher \yc and higher slope, and one lower \yc and lower slope. On the other hand, this bimodality may simply be scatter in the relatively low number of candidates at high richness. High resolution follow-up observations of these clusters could provide insight into whether the apparent bimodality is in fact due to merger history.

\section{Forced photometry at MaDCoWS cluster candidate locations}\label{sec:photometry}

\begin{figure}
\centering 
\includegraphics[clip,trim=2mm 2mm 12mm 9mm,width=\columnwidth]{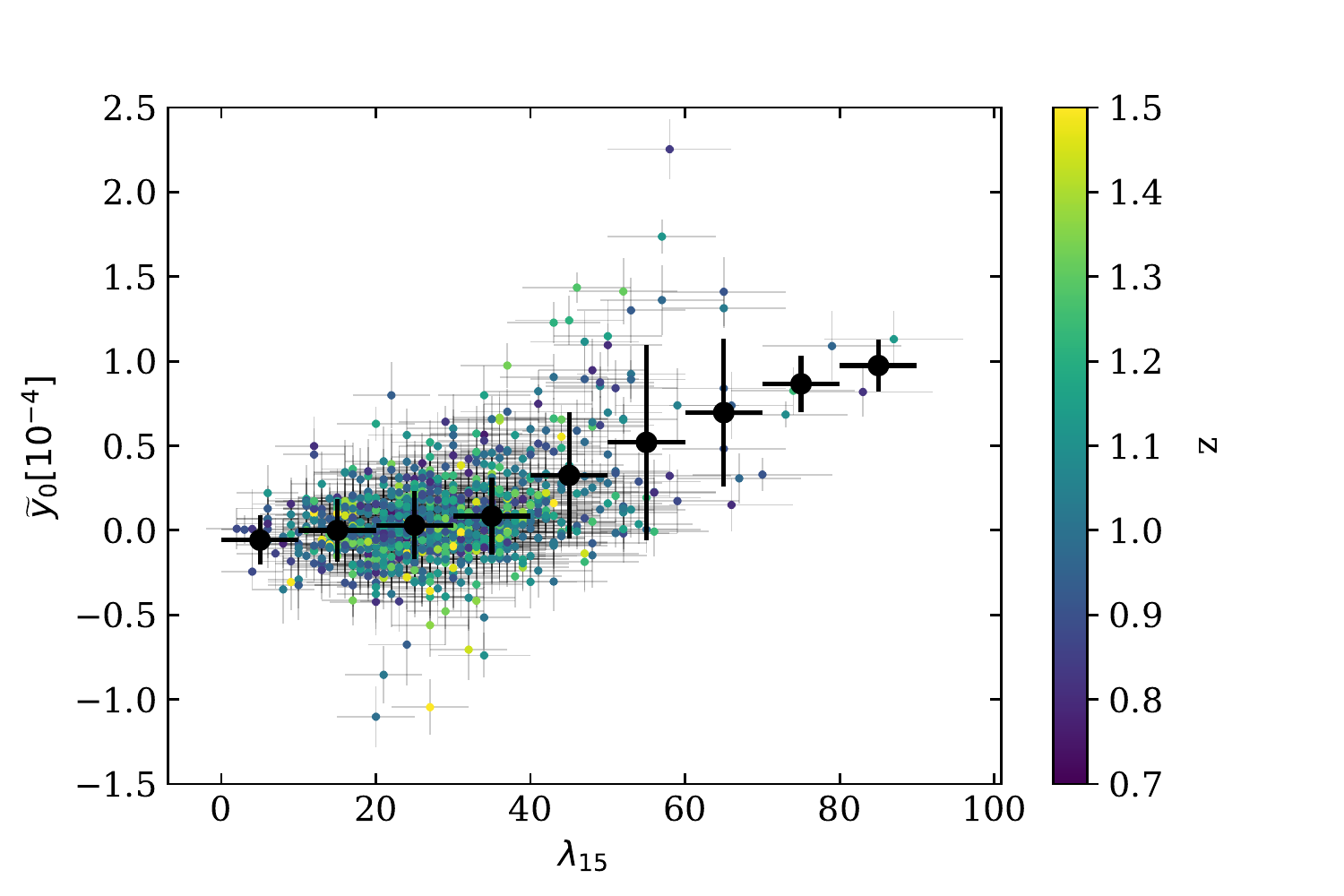}
\caption{ $\Tilde{\vary}_0$--$\lambda_{15}$ relation for the forced photometry ACT$\times$MaDCoWS catalog. 
The color bar on the right indicates the redshift of each candidate. The black points indicate the average \yc in bins of richness. We note that these are not quite the same as those computed from stacking in Fig.~\ref{fig:stacks} as those do not include the co-detections. See Sect.~\ref{fig:forced_photo} for a discussion of why they were not included. The binned \yc uncertainties were computed via bootstrapping, while the $\lambda_{15}$ error bars simply show the bin width. The redshift given is the ACT catalog redshift for those clusters detected in ACT; otherwise, it is the \madcows-reported redshift.}\label{fig:forced_photo}
\end{figure}

\begin{figure*}
\centering
\includegraphics[clip,trim=2.0cm 1.0cm 2.0cm 1cm,width = \linewidth]{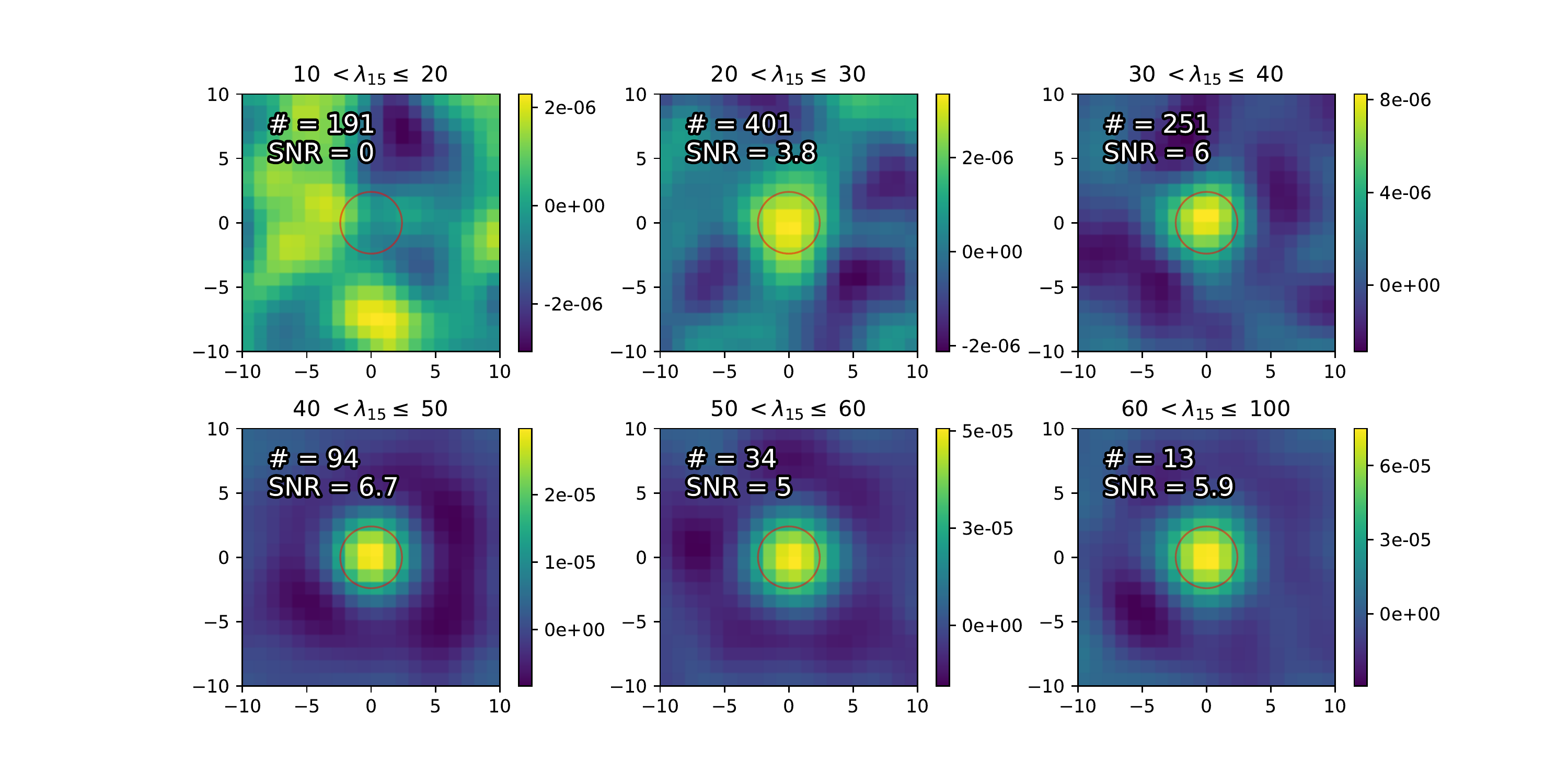}
\caption{Stacks on MaDCoWS cluster positions in bins of richness on the ACT \yc\ maps. The color bar scale is in units of \yc. Due to the order of magnitude difference in scale maximum between bins, the scaling is not consistent between plots. While there is no detection in the $10 < \lambda_{15} \leq 20$ bin, there is a clear signal in all the other bins. The red circle indicates the central $2.4\arcmin$ in diameter. The x axis shows the offset in RA in arcminutes, while the y axis shows the offset in declination in arcminutes. The text in the top left is the number of \madcows cluster candidates in the stack and the S/N.} \label{fig:stacks}
\end{figure*}

As discussed in Sect.~\ref{sec:data:act}, to form the forced photometry catalog we simply record the \yc value in the SZ map at a \madcows candidate location. The resulting distribution of \yc versus $\lambda_{15}$, for the 965 candidates that have reported richness values, is shown in Fig. \ref{fig:forced_photo}. 
The forced photometry catalog has no S/N limit, that is, it includes all 965 MaDCoWS cluster candidates that fall within the ACT survey footprint, excluding those that fall within masked regions (e.g., due to the dust mask or point sources; see \citealt{Hilton2021}). This means that the catalog is free from SZ-selection bias, and 
by fitting the \yc\ values at the MaDCoWS cluster locations through a Bayesian approach, we are able to infer the mass-richness relation (see Sect. \ref{sec:scalingrels}) while addressing sample impurity and sources of contamination. Due to noise in the map, as well as radio or dusty sources in or near clusters, the \madcows clusters may have negative \yc values; we account for this in Sects.~\ref{sec:submm_emission}, \ref{sec:radio}, and \ref{sec:regression}.

Conversion from \yc to mass was done following \citet{Hilton2021}. We note that there is a small calibration difference between our masses and those of \citet{Gonzalez2019}; they used the \citet{Andersson2011} scaling relation, while our work uses the \citet{Arnaud2010} scaling relation. The difference between these two should be $\lesssim 5\%$ \citep{Andersson2011}, which is subdominant to other sources of uncertainty in our main results (Sect.~\ref{sec:scaling_results}).

We verified the presence of SZ signal, on average, by stacking the ACT SZ maps on the MaDCoWS candidate locations.  For this, we used the {\tt Pixell} software suite.\footnote{\url{https://github.com/simonsobs/pixell}} We removed candidates outside the ACT footprint or lacking a richness estimate, as well as co-detections to ensure that the signal would not be dominated by known ACT clusters, after which $984$ clusters remained for stacking. We divided these clusters into richness bins as shown in Fig.~\ref{fig:stacks}, starting at a richness of $\lambda_{15} > 10$. There are only $28$ clusters in the remaining $984$ with $\lambda_{15} \leq 10$, and the signal in this stack is consistent with $0$ to within $1\sigma$. In each bin, we then stacked $20\arcmin \times 20\arcmin$ cutout maps, centered on the MaDCoWS cluster positions. The results are shown in Fig. \ref{fig:stacks}. We computed the average \yc over the central $1.2\arcmin$ radius in the stacks in each richness bin; uncertainties on this figure were evaluated via bootstrapping (see Appendix \ref{app:boots}). The lowest-richness bin ($10 < \lambda \leq 20$) contains $191$ cluster candidates; the stack on this bin is consistent with zero signal. In each of the remaining bins there is a clear detection at $\geq 3\sigma$ (see Table~\ref{tab:stacks} for the exact \yc signals).

\begin{table}[]
    \centering    
    \caption{Stacks of \yc\ values for the \madcows cluster candidates.}
    \begin{tabular}{cccc}
    \hline\hline\noalign{\smallskip}
        $\lambda_{15,\text{low}}$  & $\lambda_{15,\text{high}}$  & \# in bin & \yc [$10^{-4}$]\\
        \hline
        10 & 20 & 191 & $(0 \pm 1) \times 10^{-2}$ \\
        20 & 30 & 401 & $(30 \pm 8)\times 10^{-3}$\\
        30 & 40 & 251 & $(54 \pm 9)\times 10^{-3}$\\
        40 & 50 & 94 & $(20 \pm 3)\times 10^{-2}$\\
        50 & 60 & 34 & $(40 \pm 8)\times 10^{-2}$\\
        60 & 100 & 13 & $(53 \pm 9)\times 10^{-2}$\\
    \noalign{\smallskip}
    \hline
    \end{tabular}
    \label{tab:stacks}
    \caption*{The above stacked values of \yc\ exclude the co-detections, as discussed in Sect.~\ref{sec:photometry} and shown in Fig.~\ref{fig:forced_photo}. The value $\lambda_{15,\text{low}}$ denotes the lower edge (exclusive) used for binning, and $\lambda_{15,\text{high}}$ denotes the upper edge (inclusive) used for binning. The uncertainty on \yc is calculated via bootstrapping.}
\end{table}

\section{Mass estimate biases and corrections}\label{sec:baises}

The targeted SZ follow-up in  \cite{Gonzalez2019}, \cite{DiMascolo2020} and \cite{Dicker2020} mainly probes the high-richness tail in the distribution, and hence may present both Malmquist and Eddington biases \citep[see e.g.,][]{Malmquist1922,Kelly2007} in the richness selection. Moreover, the CARMA sample exhibits a Malmquist bias in its SZ flux measurements; as noted in \cite{Gonzalez2019}, the CARMA sample was constructed by performing shallower observations of the higher-richness objects, based on the expectation that the integration times should be shorter, and only reported the results for robust detections. Correcting for these biases was one of the prime motivations of this paper. Since the MaDCoWS clusters were not SZ selected, it is unnecessary to  de-boost our \yc or mass estimates. However, three primary effects still need to be corrected in the forced photometry mass estimates:

Firstly, the cluster locations that MaDCoWS reports are the peaks of a smoothed galaxy density map; the identified cluster location can be offset from the center of the cluster mass, and hence the center of the SZ signal \citep{George2012, Sehgal2013, Viola2015}, which leads to a suppression of the SZ signal. This scatter can be due to measurement uncertainty of the cluster's MaDCoWS centroid ($\sim 15 \arcsec$ in each of RA and Dec., \citealt{Gonzalez2019}, Sect. 5.3) or SZ peak ($1.5\arcmin$ total \citealt{Hilton2021}), or it can be due to astrophysical reasons; in other words, the hot, virialized gas that is responsible for the SZ signal may not have the same spatial distribution as the galaxy number density used to determine the MaDCoWS centroid \citep{George2012, Sehgal2013, Viola2015}. Either way, the result is that for a set of clusters, stacking on the optical centroids produces a signal that is suppressed as compared to stacking on their SZ peak locations, which \citet{Ge2019} have found to be around the $\approx 10\%$ level. This agrees with the typical suppression that we find in Sect.~\ref{sec:miscentering}.

Secondly, the matched filter used in the forced photometry method as described in Sect.~\ref{sec:photometry} cannot account for compact sources at cluster locations.
If emission from compact sources, such as radio sources or dust, reduces the $98$ and/or the $150$~GHz emission, the effect will be to bias the \yc estimate low; the emission "infills" the SZ decrement causing the mass estimate from forced photometry to be biased low. The significantly negative \yc clusters (see Fig.~\ref{fig:forced_photo}) suggest that this might be occurring. Further, stacking the f220 maps on the \madcows candidate locations produces a significant positive signal that is not present when stacking on the ACT cluster locations. We ascribe this excess to dusty submillimeter emission spatially correlated with the \madcows cluster locations. Stacks on ACT and \madcows cluster locations at radio frequencies show that the ACT clusters have higher observed radio flux density on average (Sect.~\ref{sec:radio}). As such, the source of the excess \madcows $224$~GHz emission is likely not radio. From the f220 stacks and similar stacks on the H-ATLAS maps, we find that the MaDCoWS candidates have more significant infill at submillimeter wavelengths than the general population of ACT clusters. Further, we find that the spectral form of the submillimeter emission in the \madcows cluster stacks is well described by a gray body (Eq.~\ref{eq:4}). Such infill is typically due to dusty submillimeter galaxies \citep[e.g.,][]{Casey2014}.  We combined the stacks on the f220 and H-ATLAS data set to estimate this contamination and remove it from the mass-richness scaling relation (see Sect.~\ref{sec:submm_emission}).

Thirdly and finally, in Sect. \ref{sec:radio} we consider the effect of contamination at radio wavelengths on the MaDCoWS cluster candidates.  Bright radio contamination, while declining at millimeter wavelengths, could potentially remain relatively significant at $98$ and $150$~GHz,  once again in-filling the measured \yc toward lower values. We describe our correction for this effect in Sect.~\ref{sec:radio}.

\subsection{Centroid offset}\label{sec:miscentering}
\begin{figure}
    \centering
    \includegraphics[width=\columnwidth]{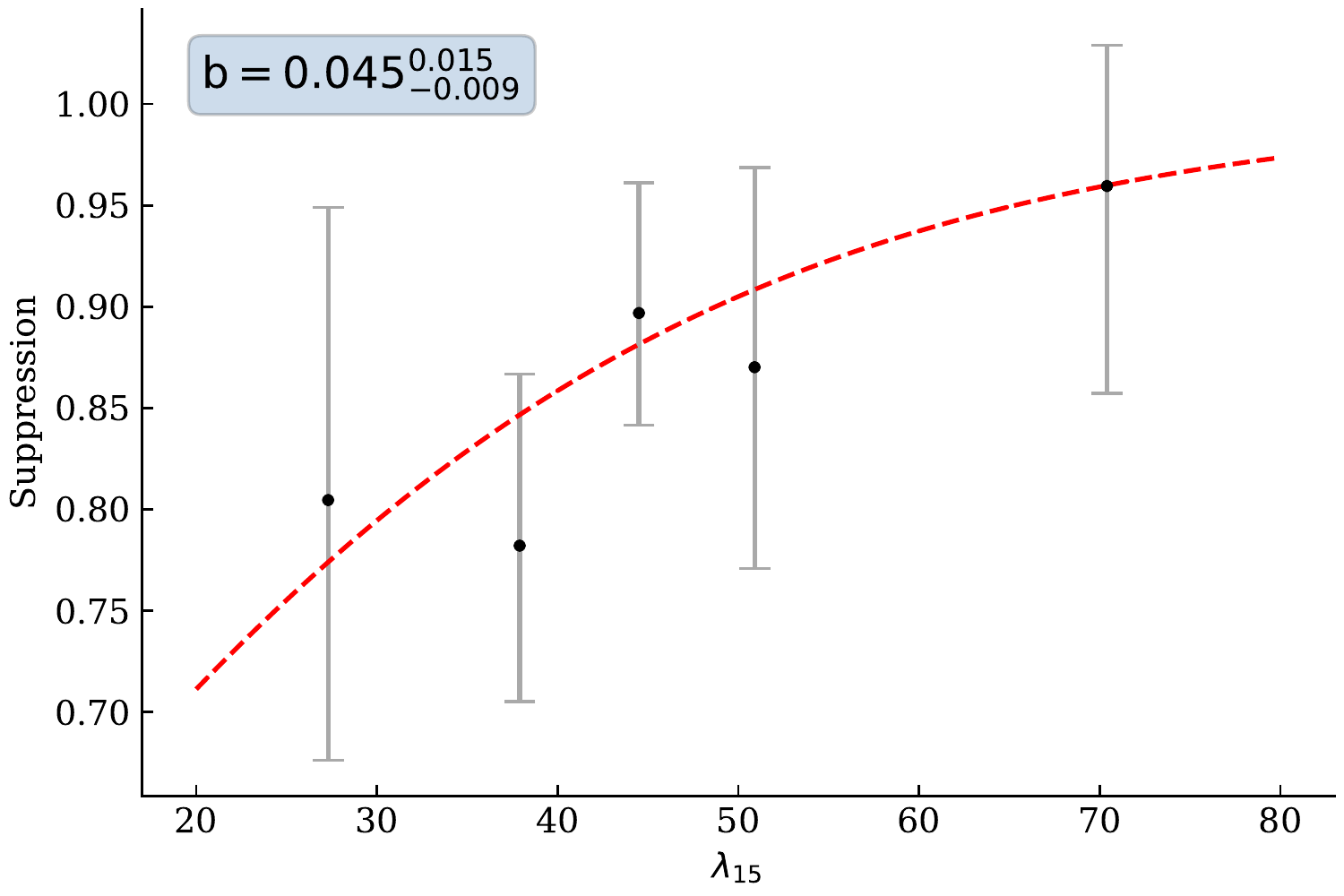}
    \caption{The MaDCoWS to ACT $\Tilde{\vary}_0$ suppression ratio due to miscentering based on the $96$ co-detections. The position of data points on the $x$ axis shows the center of the richness bin. The $y$ axis is the ratio of the aperture $\delta T_\textsc{cmb}$  in the central $1.2\arcmin$ radius of the MaDCoWS centered stacks to that of the ACT centered stacks, where $\delta T_\textsc{cmb}$ is the fluctuation in CMB temperature from the mean. The data have been fit to a one-parameter sigmoid model of the form $f(x) = \frac{1}{1+e^{(-bx)}} $. The dashed red line shows this fit, and the legend reports $b$.} 
    
    \label{fig:scatter_supression}
\end{figure}

 To correct for suppression of the SZ signals due to positional offsets in MaDCoWS-determined cluster centroids as compared to the ACT-determined centroids, we stacked the \yc maps on the $96$ ACT/MaDCoWS co-detections twice, once on the ACT identified cluster locations and once on the MaDCoWS identified location. At each cluster location, we created a $20\arcmin$ square sub-map of the $\Tilde{\vary}_0$ map centered on the cluster location. We normalized each sub-map to have unity amplitude, so that the suppression was not dominated by the particular scatter of the brightest clusters. We performed this stacking analysis on clusters with measured richness greater than $20$, divided into $5$ bins of even richness range. This left only about $15$ clusters in each bin, leading to relatively large uncertainties as shown in Fig.~\ref{fig:scatter_supression}.


To compute the suppression, for each of the stacks above we computed the aperture $\delta T_\textsc{cmb}$ within a $1.2\arcmin$ radius of the stack center. For a given richness bin, the ratio of this statistic for the MaDCoWS centered stacks to the ACT centered ones sets the suppression. We computed the variance via bootstrapping. We then fit the richness-suppression relation to a sigmoid model of the form $f(x) = \frac{1}{1+e^{(-bx)}} $ by maximizing the likelihood function:

\begin{equation}
 -\frac{1}{2}\sum_n\frac{(y_{n} - f(x_{n}))^2}{\sigma_{n}^2} + \log(\sigma_{n}^2)    
,\end{equation}
where $y_n$ is the ratio of the ACT to \madcows signal in the $n^{\textrm{th}}$ richness bin and $\sigma_{n}$ is the uncertainty in the $n^{\textrm{th}}$ that data point. The results are shown in Fig.~\ref{fig:scatter_supression}. In our mass-richness fitting routine, we include a parameter to account for this suppression effect. We do not directly adjust the measured fluxes using this model; rather the model developed above enters into the fit as a richness-dependent prior on that suppression parameter (Sect.~\ref{sec:priors}). 

\subsection{Submillimeter emission} \label{sec:submm_emission}
Since the SZ effect at the frequencies of interest (98 and 150~GHz) manifests as a decrement of the CMB temperature, there is the potential that dusty, submillimeter sources could fill in or partially suppress the SZ signal.  It is also known that dusty submillimeter galaxies \citep[e.g.,][]{Erler2018}, as well as radio AGN (considered in the next section), frequently reside within the centers of clusters, where feedback processes are strongest \citep[see e.g.,][]{Coble2007, Sayers2013, Gralla2014, Zakamska2019, Gralla2020}.

To determine if the SZ signals of the MaDCoWS candidates are suppressed by a dusty contribution, we stacked the f220 maps on the MaDCoWS centers. The f220 map was matched filtered in the same way as the \yc map to remove point sources, maintaining consistency. We then stacked the f220 maps on both the ACT (all $4195$) and MaDCoWS (all $1572$ in the ACT footprint) cluster positions and computed the aperture flux density in units of $\delta T_\textsc{cmb}$ within the $1.2\arcmin$ radius of the stack center, corresponding to the ACT cluster finder reference filter scale of $2.4\arcmin$.


\begin{figure}
    \centering
    \includegraphics[clip,width=\columnwidth]{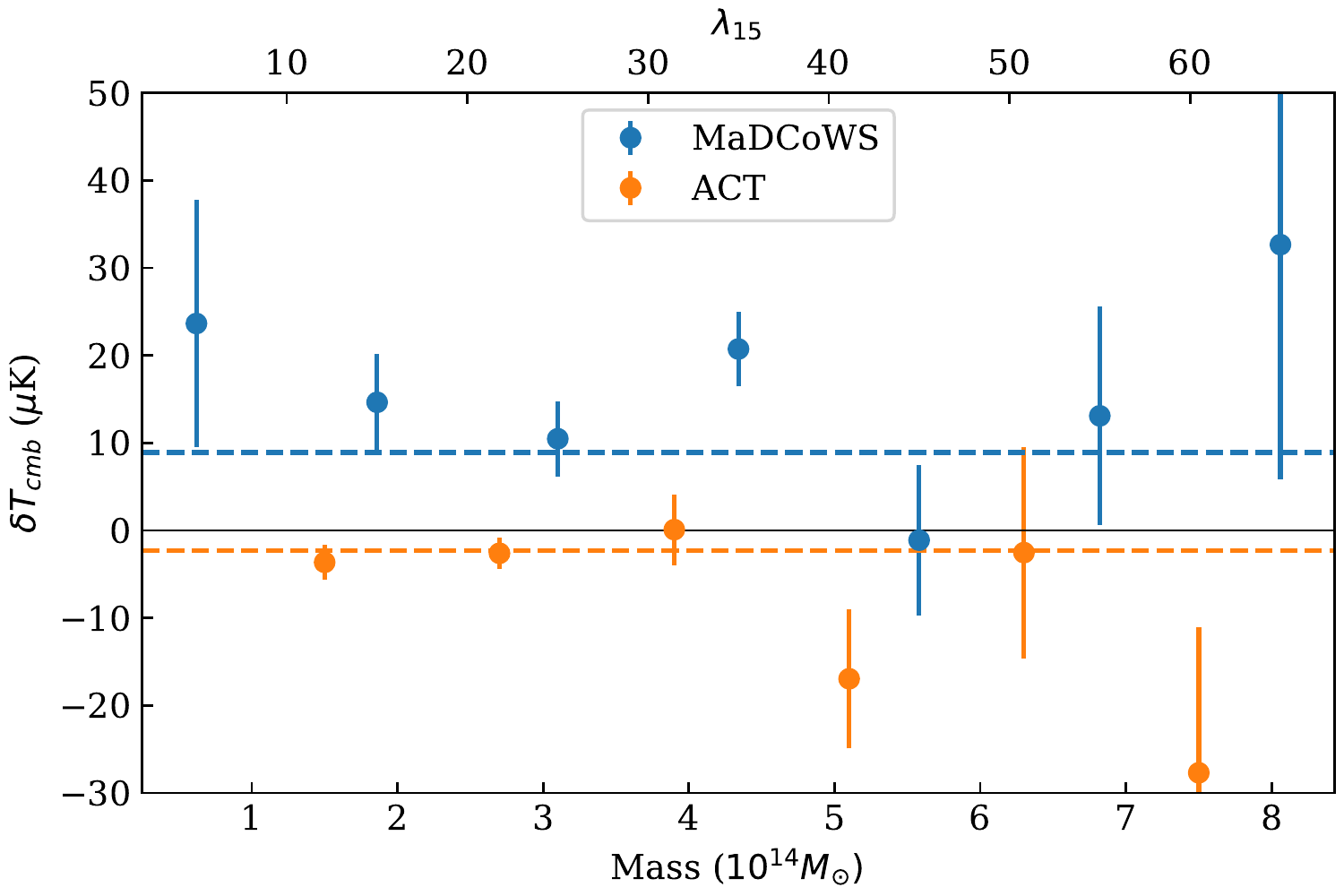}
    \caption{Average $224$~GHz emission for \madcows (blue) and ACT (orange) clusters. \madcows clusters were binned in richness, while the ACT clusters were binned in mass. Richness is plotted on the lower x axis while mass is plotted on the upper; the two scales are not equivalent, and are simply co-plotted for convenience. Error bars were estimated via bootstrapping. The \madcows clusters show a statistically significant excess emission at $224$~GHz on the whole, while the ACT clusters show a small decrement. In neither the ACT nor the \madcows clusters is there a trend with mass or richness. The dashed lines indicate the average $224$~GHz emission across all \madcows (blue) and ACT (orange) clusters. We attribute the signal from the \madcows cluster candidates at $224$~GHz to IR emission and follow up with {\it Herschel} data (Sect.~\ref{sec:submm_emission}).}
    
    \label{fig:220_scaling}
\end{figure}


\begin{figure*}
    \centering
    \begin{subfigure}[b]{0.3\textwidth}
         \centering
         \includegraphics[clip,trim=5mm 5mm 8mm 10mm,width=\textwidth]{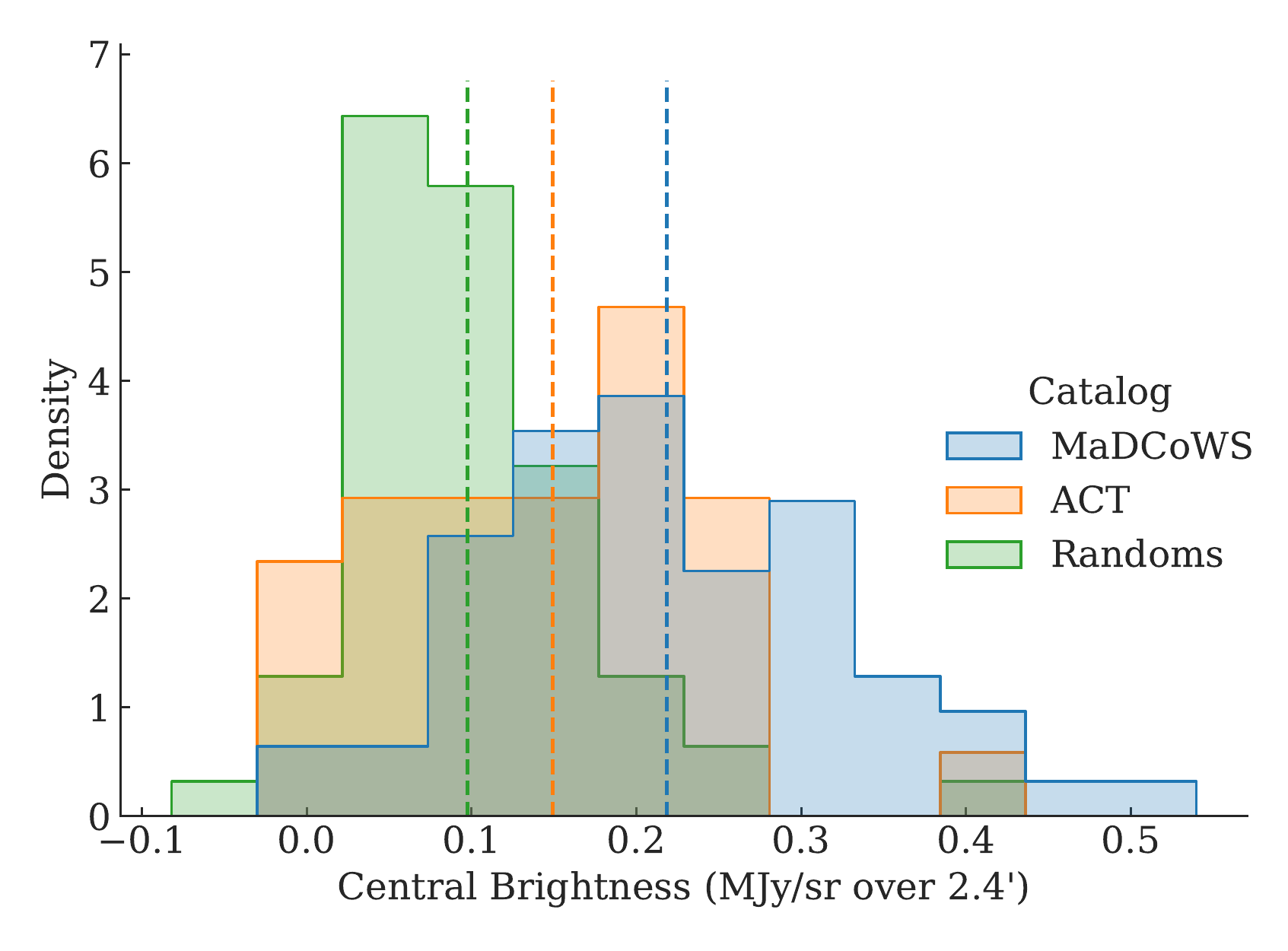}
         \caption{$250~\mu$m}
         \label{fig:kde_250}
     \end{subfigure}
     \hfill
     \begin{subfigure}[b]{0.3\textwidth}
         \centering
         \includegraphics[clip,trim=11mm 5mm 8mm 10mm,width=\textwidth]{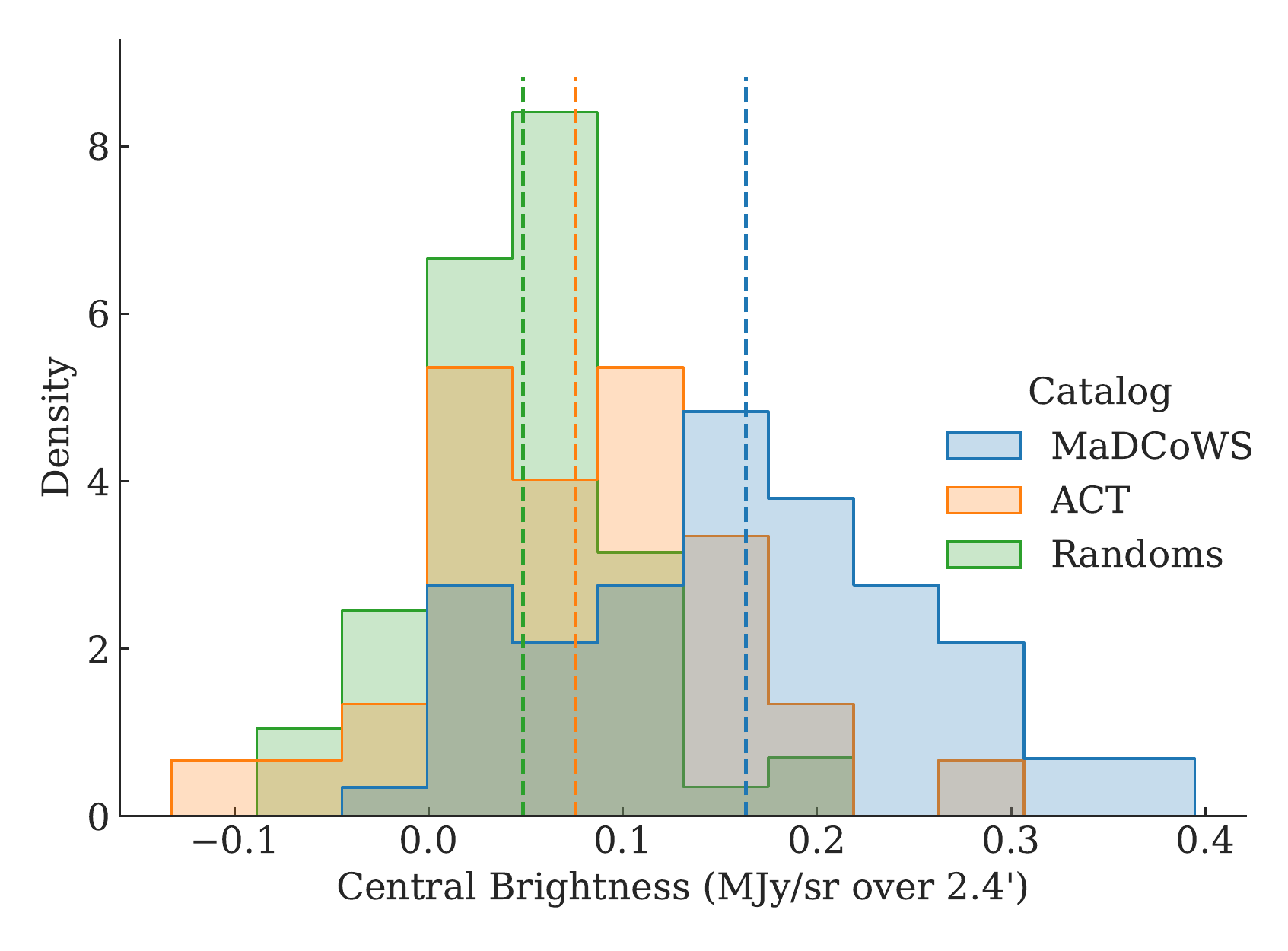}
         \caption{$350~\mu$m}
         \label{fig:kde_350}
    \end{subfigure}
\hfill
     \begin{subfigure}[b]{0.3\textwidth}
         \centering
         \includegraphics[clip,trim=11mm 5mm 8mm 10mm,width=\textwidth]{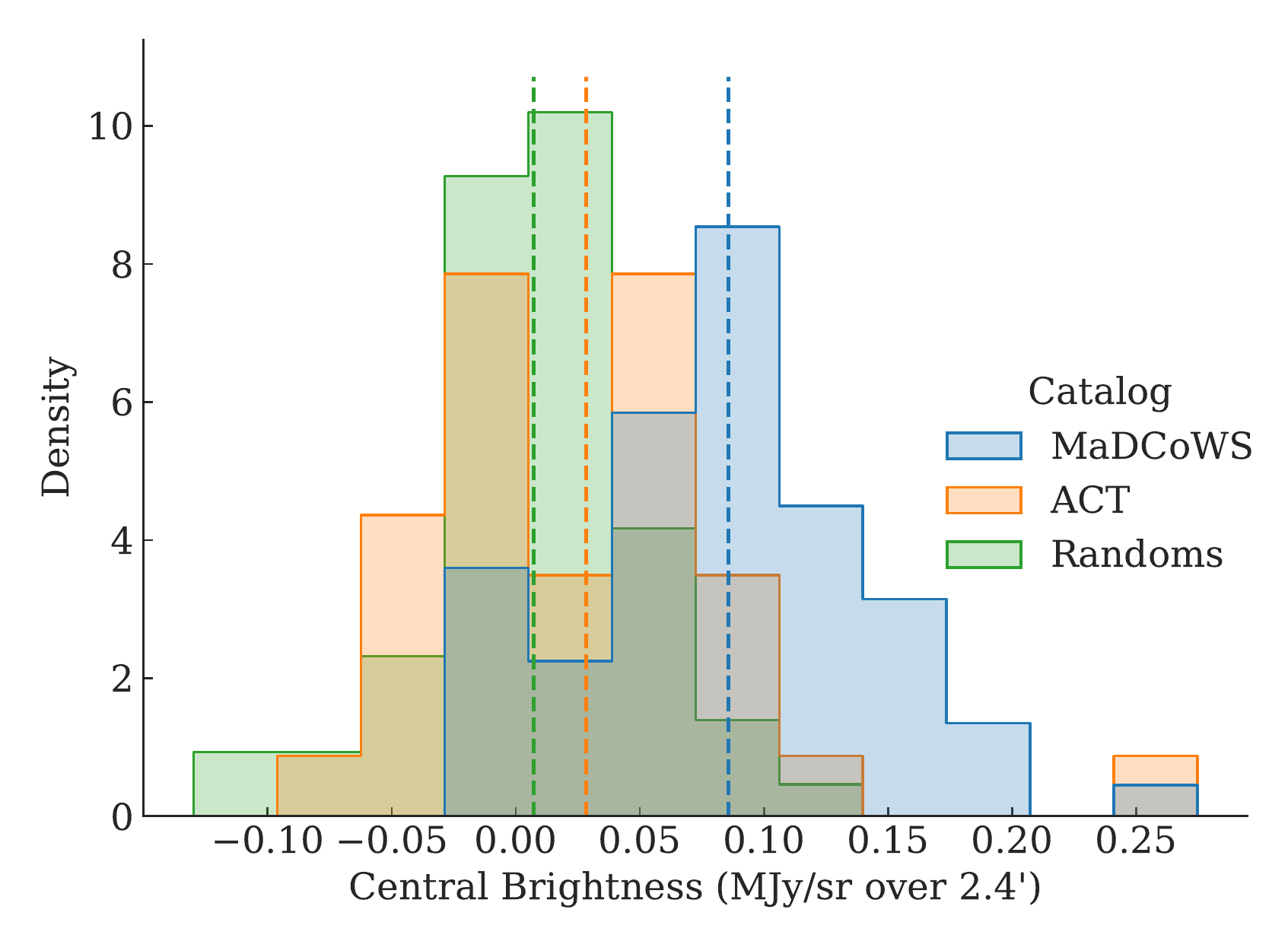}
         \caption{$500~\mu$m}
         \label{fig:kde_500}
     \end{subfigure}
        \caption{Histograms of average surface brightness in {\it Herschel} submillimeter observations corresponding to a $1.2\arcmin$ radius aperture flux, centered on ACT, MaDCoWS, and random cluster locations. The dashed lines show the average brightness for a given frequency and catalog. At $250~\mu$m, the average surface brightness is $0.10\pm 0.08$, $0.15\pm 0.10$, and $0.22\pm 0.11$~MJy/sr for the random, ACT, and \madcows samples, respectively. At $350~\mu$m, those surface brightnesses are respectively $0.05\pm 0.05$, $0.08\pm 0.08$, and $0.16\pm 0.09$~MJy/sr for the random, ACT, and \madcows samples. Finally, at $500~\mu$m they are $0.01\pm 0.04$, $0.03\pm 0.07$, and $0.09\pm 0.06$~MJy/sr for the random, ACT, and \madcows. The statistically higher \madcows emission at each {\it Herschel} frequency, along with the higher emission in the ACT 224~GHz channel (Fig.~\ref{fig:220_scaling}), indicates that the \madcows clusters may be contaminated by dusty sources.}
        \label{fig:brighthist}
\end{figure*}

We used bootstrapping to estimate the uncertainty in this measure. We found that for the \madcows clusters, the signal was $8.9 \pm 1.2~\mu\text{K}$, while for the stack on ACT clusters, the signal was $-2.3 \pm 0.6~ \mu\text{K}$, where the units are $\delta T_\textsc{cmb}$. Additionally, we binned both the \madcows and ACT data into 6 bins. For the \madcows clusters, we binned them in richness, while for the ACT, we binned them in cluster mass. We then repeated the analysis using these bins. The results are shown in Fig.~\ref{fig:220_scaling}. For both the \madcows and ACT clusters, there is no obvious trend in 224~GHz flux density with richness or mass, respectively. The excess emission in the MaDCoWS clusters we attribute to dust emission. The decrement in 224~GHz emission from the ACT clusters may be due to a small bias in the cluster finder algorithm; since the CMB serves as a source of noise when searching for the SZ signal, the algorithm preferentially finds clusters in regions of lower primary CMB signal (i.e., "cold spots"). Therefore, when stacking on the f220 maps, the result is a preferential stack on regions of low 224~GHz emission, leading to a decrement.

\begin{figure}
    \centering
    \begin{subfigure}[b]{0.45\textwidth}
         \centering
         \includegraphics[clip,trim=0.0cm 0.0cm 0cm 1.0cm,width=\textwidth]{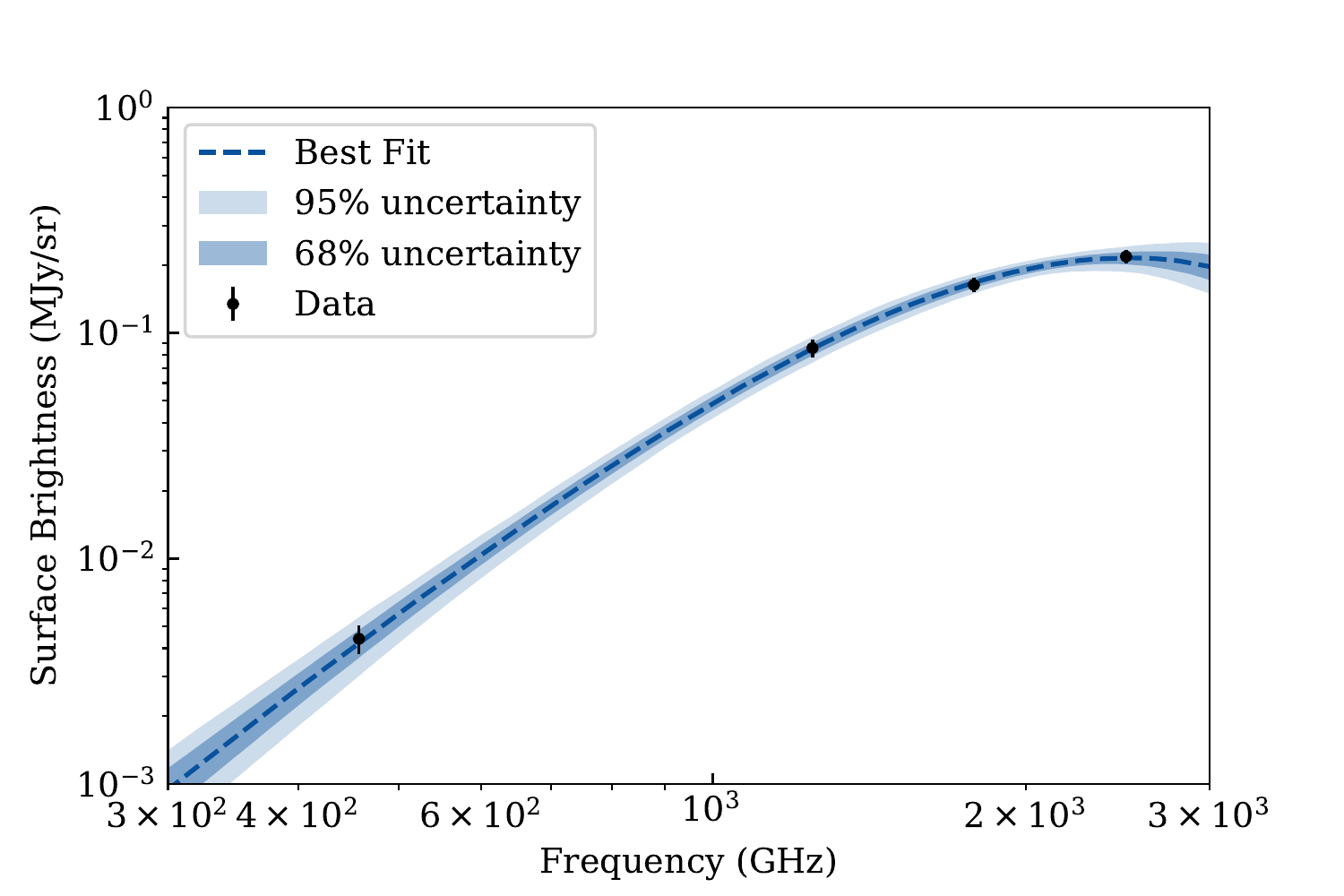}
         \caption{Maximum likelihood best fit of the form Eq. \ref{eq:4} to the mean emission from stacks on the MaDCoWS clusters at $224$ (ACT), $600$, $857$, and $1200$~GHz ({\it Herschel}). The total number of clusters in the stack is the $66$ clusters in the H-ATLAS footprint for the three {\it Herschel} bands and $1572$ for the ACT $224$~GHz band. The dashed blue line is the best fit, and the light and dark blue bands represent the $68$ and $95\%$ confidence limit, respectively. The y axis is the surface brightness averaged over a $1.2\arcmin$ radius aperture at the center of the stack. Error bars were estimated via bootstrapping.
         }
         \label{fig:greybody_bestfit}
     \end{subfigure}
     \hfill
     \begin{subfigure}[b]{0.45\textwidth}
         \centering
         \includegraphics[clip,trim=0.0cm 0.0cm 0.0cm 0.0cm,width=\textwidth]{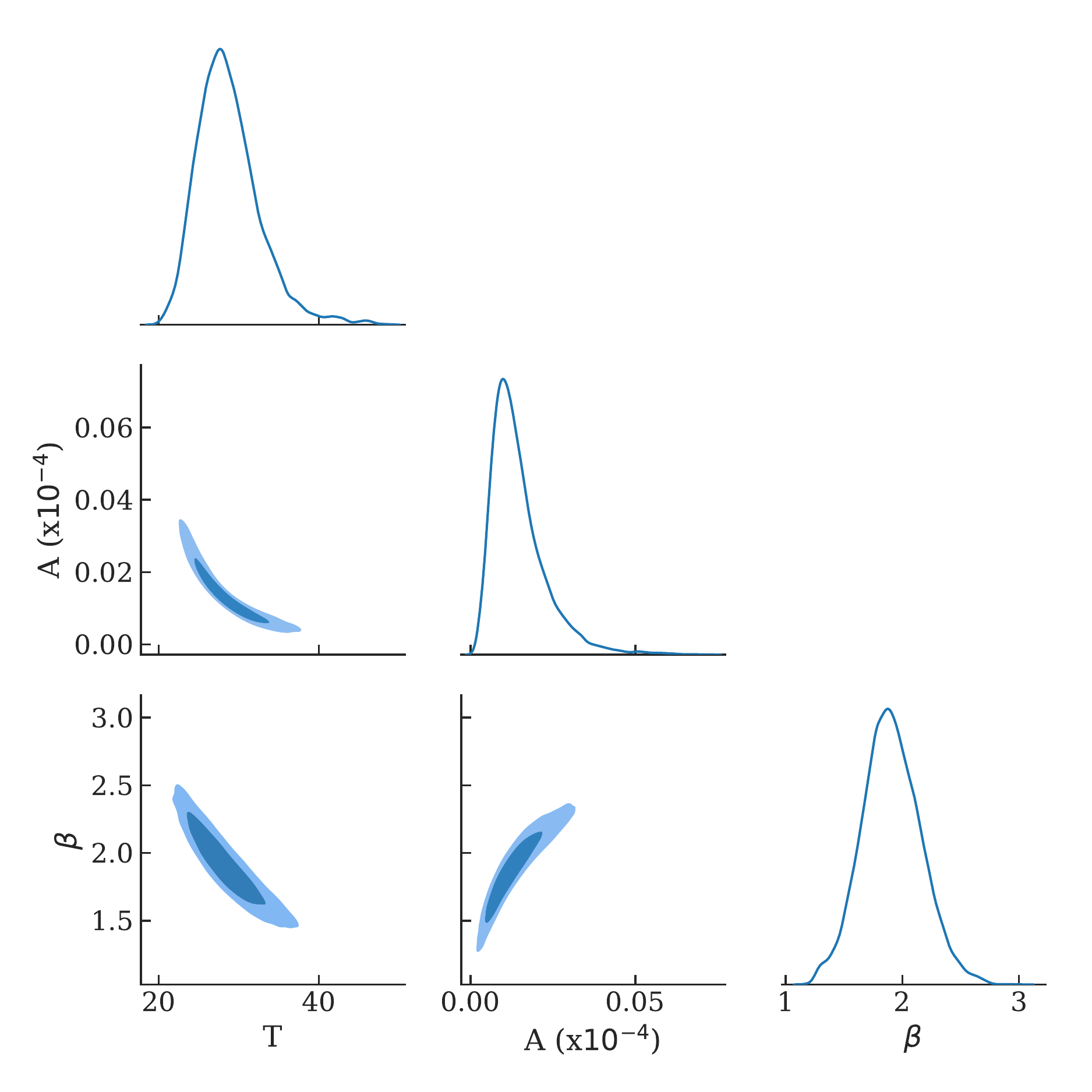}
         \caption{Constraints on the dust temperature, $T = 28^{+4}_{-3}$~K, the dust spectral index, $\beta = 1.9^{+0.3}_{-0.2}$, and the normalization constant, $A = 1.3^{+0.1}_{-0.1}\times 10^{-6}$. The light and dark blue contours show the $68$ and $95\%$ confidence intervals, respectively, for the 2D projections of the posterior probability. The 1D posterior probabilities are shown on the diagonal.}
         \label{fig:greybody_corner}
     \end{subfigure}
        \caption{Gray-body fit constraints.}
        \label{fig:greybody}
\end{figure}

To quantify the dust emission, we stacked on the MaDCoWS cluster locations in the H-ATLAS maps \citep{Valiante_2016,Smith_2017}. For comparison, we also stacked the maps on ACT cluster locations and a sample of random locations generated by offsetting each MaDCoWS cluster location by $5\arcmin$ in a random direction. Due to the small size of the H-ATLAS survey, we were only able to perform this analysis for $34$ ACT clusters and 66 \madcows cluster candidates, limiting our ability to determine if the in-fill has a richness or mass dependence. In order to compare with the emission at 224~GHz, we first converted each of the 250, 350, and 500~$\mu$m maps to MJy/pixel and then smoothed to 1.0$\arcmin$ (i.e., the ACT $220$~GHz resolution) using a Gaussian kernel with the integral normalized to unity. We then stacked these maps on the MaDCoWS and ACT cluster locations using inverse-noise weighting and calculated the aperture flux density within 1.2$\arcmin$ of the stack center (i.e., in the central 2.4$\arcmin$ diameter corresponding to the reference filter scale used in \cite{Hilton2021}). We computed the uncertainty in each stack via bootstrapping. For the ACT and randoms stack, the signal was consistent with zero; for the \madcows there was a statistically significant excess (see Fig.~\ref{fig:brighthist}). We converted the average fluctuation of $\delta T_\textsc{cmb}$ in the $224$~GHz stack described above to emission (in MJy sr$^{-1}$) using the derivative of the blackbody function (see, e.g., \citealt{Mroczkowski2019}):
\begin{equation}
    \frac{\Delta I}{\Delta T_{\textsc{cmb}}} = \frac{I_0}{T_{\textsc{cmb}}} \frac{x^4 e^x}{(e^x-1)^2},
\end{equation}
where $T_\textsc{cmb} = 2.7255$~K is the monopole temperature of the primary CMB, $\Delta I$ is the change in intensity above background, $\Delta T_{\textsc{cmb}}$ is the fluctuation in temperature about the CMB monopole, and $x = (h\nu)/(k_{\text{B}} T_{\textsc{cmb}}) \approx \nu/(56.8~\rm GHz)$ is the dimensionless frequency. The normalization factor of the primary CMB spectrum is
\begin{equation}
    I_0 = \frac{2(k_{\text{B}} T_{\textsc{cmb}})^3}{(hc)^2} \approx 270.33 \left[ \frac{T_{\textsc{cmb}}}{2.7255~\text{K}}\right]^3 \text{MJy/sr}.
\end{equation}
The $\Delta I$ can then be converted to surface brightness $\Delta S_{\nu} \left[ \text{Jy/bm}\right]$ via
\begin{equation}
    \Delta S_{\nu} = \int \Delta I_{\nu} d\Omega = \langle \Delta I_{\nu} \rangle \Omega_{\text{beam}}.
\end{equation}
For all three stacks, we estimated uncertainties via bootstrapping. Given the excess submillimeter emission detected in the \madcows clusters, we then performed a maximum likelihood fit to the resulting surface brightnesses of the \madcows stacks to a gray-body model of the form
\begin{equation}
    \label{eq:4}    
    \Delta S(\nu) = A \, B_{\nu}(T) \, \left[ 1 - e^{-(\nu / \nu_0)^{\beta}} \right],
\end{equation}
where $A$ is an amplitude normalization, $\nu_0 = 3000$~GHz is a reference frequency \citep{Draine2006}, $\beta$ is the dust emissivity spectral index, $T$ is the dust temperature, and $B_{\nu}(T)$ is the Planck blackbody function. To account for the effect of redshift, we fit using the rest frame frequencies,  we converted from the observed frequencies by multiplying by $(1+\langle z \rangle)=2.08$, with $\langle z \rangle$ the average $z$ for the subsample of MaDCoWS clusters in the H-ATLAS region for which there are photometric redshifts. We assumed a Gaussian function for the likelihood and estimated the uncertainties in our fit parameters using a Markov chain Monte Carlo (MCMC) method, implemented in the \texttt{emcee} \citep{MacKey2013} package. We placed flat, uninformative priors on $T$, $A$, and $\beta$, enforcing $0~\text{K} \leq T < 200~\text{K}$,  $-1 < A < 1$, and $-5 < \beta < 10$. Results of the fit are presented in Fig.~\ref{fig:greybody}. 
Sub-figure~\ref{fig:greybody_bestfit} shows the data with error bars in black, the best fit in blue, and the $68$ and $95\%$ confidence limits. In Subfigure~\ref{fig:greybody_corner} we show constraints on the fit parameters $T_{\rm rest}$,  $\beta$, and $A$.

We find that for the \madcows clusters in the H-ATLAS footprint, the best-fit temperature is $T = 28^{+4}_{-3}$~K. This is in good agreement with other measurements of the dust properties of IR and optically selected clusters \citep[see, e.g.,][]{Smith2013, Erler2018, Amodeo2021, Fuzia2021,Vavagiakis2021}. Our constraint on the dust emission spectral index, $\beta = 1.9^{+0.3}_{-0.2}$ is somewhat higher than other measurements, but does agree within uncertainties \citep[see, e.g.,][]{Magnelli2012,Smith2013,Sayers2019}.

When performing the mass-richness scaling relation fit, we repeated the analysis above by including the dust model in the full joint probability distribution. This allowed for accounting of degeneracies between the dust model parameters and other parameters (see Sect.~\ref{sec:regression}).\\

To get an estimate of the level of bias due to dusty emission in-filling the SZ signal in \madcows cluster candidates, we converted the modeled emission at 98~GHz ($77\pm 3~\mu$Jy) and 150~GHz ($370\pm 50~\mu$Jy) to Compton $y$ via 
\begin{equation}
    y \approx \Delta I_{\nu}\times
    \left[I_0\left(\frac{x}{\tanh{(0.5x)}}-4\right)\left(\frac{x^4 e^x}{(e^x-1)^2}\right) \right]^{-1}.
\end{equation}
We then weighted the computed \yc at $98$ and $150$~GHz by their average relative contributions to \yc ($\approx 66$ and $33$\%, respectively) and compared that weighted average to the \yc signal for a $2\times 10^{14}M_{\odot}$ cluster at redshift of 1, following \citet{Hilton2021}. We find that the in-fill is $1.5\pm 0.5\%$ for the average \madcows cluster candidate. The above computation ignores the effects of the matched filter, which will in general suppress the in-fill for a source not centered on the cluster, and should be considered an approximation only.


\subsection{Radio emission}\label{sec:radio}

\begin{figure}
    \centering
    \begin{subfigure}[b]{7.9cm}
         \centering
         \includegraphics[clip,trim=0.5cm 0.3cm 1.2cm 0.5cm, width=\textwidth]{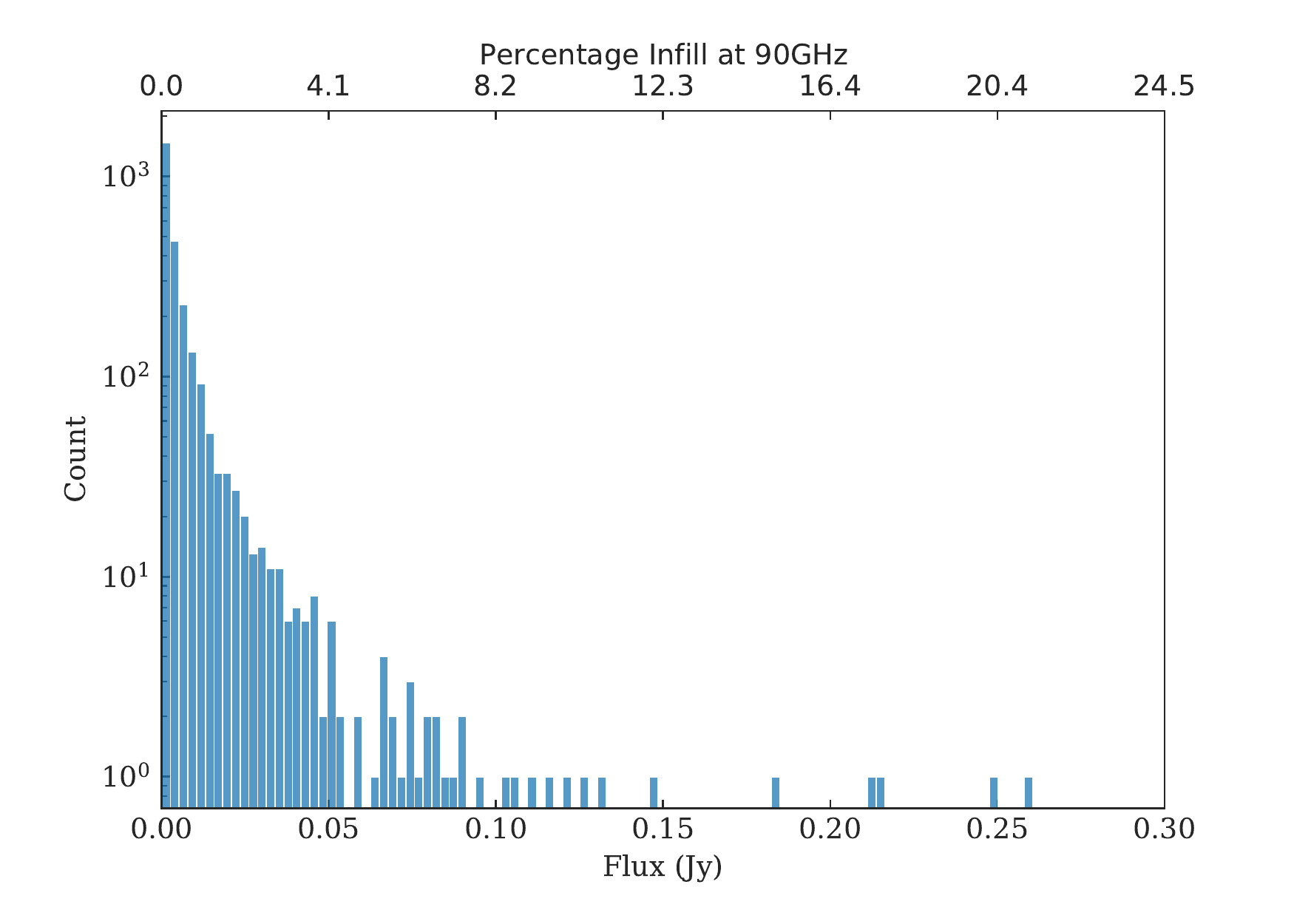}
         \caption{ACT $1.4$~GHz fluxes}
         \label{fig:ACT_hist}
     \end{subfigure}
     \hfill
     \begin{subfigure}[b]{7.9cm}
         \centering
         \includegraphics[clip,trim=0.5cm 0.3cm 1.2cm 0.5cm, width=\textwidth]{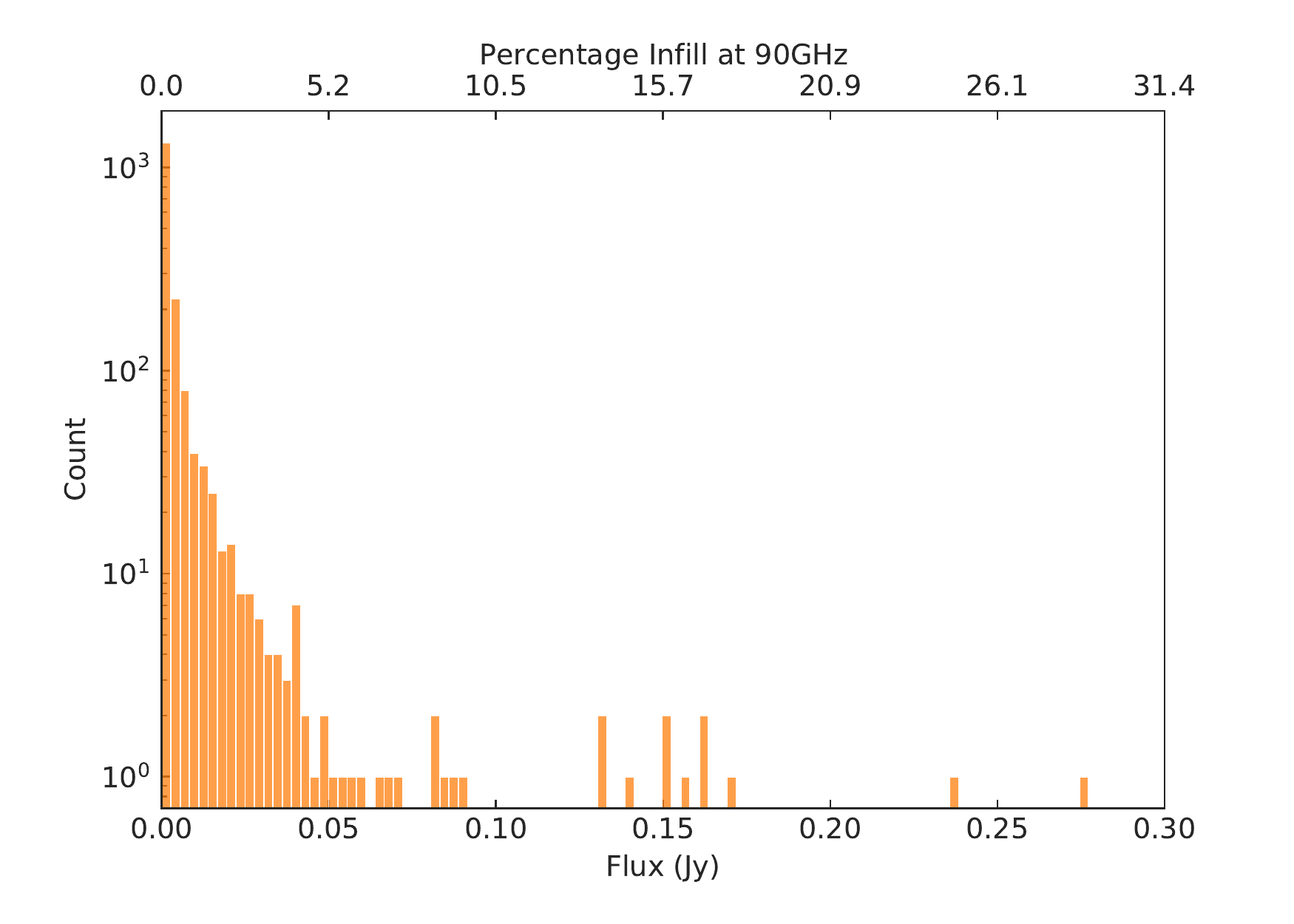}
         \caption{MaDCoWS $1.4$~GHz fluxes}
         \label{fig:mdcw_hist}
    \end{subfigure}
\hfill
     \begin{subfigure}[b]{7.9cm}
         \centering
         \includegraphics[clip,trim=0.0cm 0.0cm 1.2cm 0.0cm, width=\textwidth]{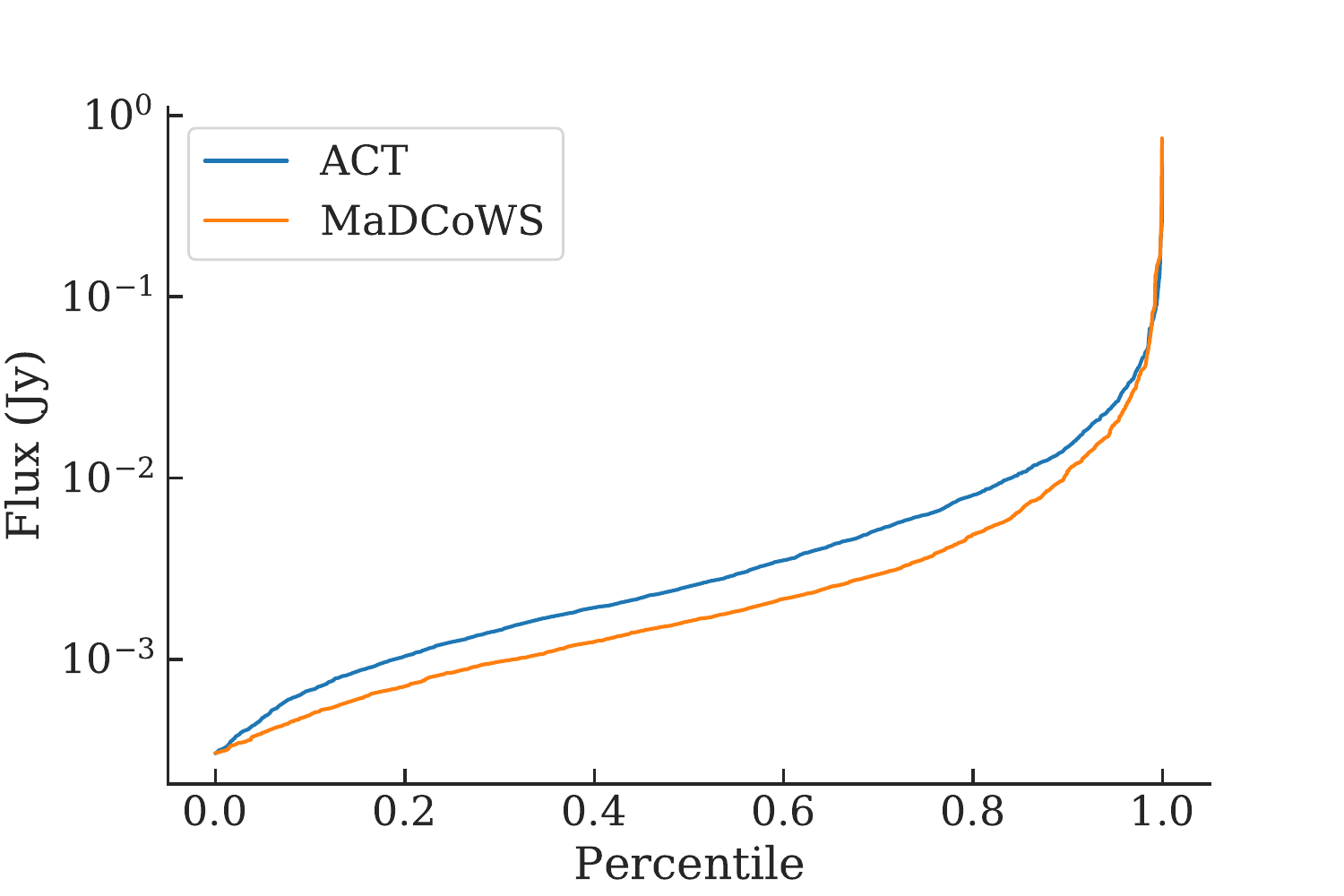}
         \caption{$1.4$~GHz cumulative histogram}
         \label{fig:cumul_hist}
     \end{subfigure}

        \caption{Histograms of the $1.4$~GHz aperture flux density of the ACT (Fig.~\ref{fig:ACT_hist}) and MaDCoWS (Fig.~\ref{fig:mdcw_hist}) cluster catalogs. Figure~\ref{fig:cumul_hist} shows the normalized cumulative histogram of the same data, cut at the aperture noise of $0.3$~mJy, to better illustrate the difference in infill between the two samples. The fluxes shown are all for a $1.2\arcmin$ radius aperture, while the percentage infill is the approximate infill of the SZ signal as described in Sect.~\ref{sec:radio_disc}. We note that due to differences in average redshift and spectral index of the ACT and \madcows clusters, the same measured flux density at $1.4$~GHz does not correspond to the same percentage infill. The ACT clusters show on average significantly more radio infill at $1.4$~GHz than their MaDCoWS analogs.
        }
        \label{fig:NVSS_histograms}
\end{figure}

In order to assess and quantify low frequency radio source in-fill strong enough to impact our $98$ and $150$~GHz measurements of the SZ signal, we require constraints on the radio flux density and spectral indices of the source population. Using both NVSS ($1.4$~GHz) and VLASS (nominally $3$~GHz) allows us to estimate both. 
From these surveys (Sect.~\ref{sec:data:vla}), we evaluate the distribution of radio in-fill associated with members of the ACT and MaDCoWS samples. For each cluster there are two statistics with which we are concerned: 1) the intrinsic luminosity of sources within the cluster in its frame of reference, and 2) the observed flux density in our frame of reference. We are concerned with the intrinsic radio luminosities as they tell us whether the ACT and \madcows clusters were drawn from populations with the same intrinsic radio properties. We must also consider the observed fluxes, as they set the level of radio in-fill in the cluster. For clarity, whenever we refer to fluxes or surface brightness or use units of flux density or surface brightness ([Jy] or [Jy/sr]), we are referring to the observed flux density or surface brightness in our frame of reference. Whenever we refer to radio luminosities or use units of luminosity ([W/Hz)]), we are referring to the intrinsic radio luminosity. Our process is to first compute the observed flux density for each cluster, and then convert that into intrinsic radio luminosity.

To compute the radio flux, for each cluster we produced a cutout image (postage stamp) of the NVSS map centered at the cluster position and smoothed it to $1\arcmin$ to match the ACT $224$~GHz beam scale, in order to account for the effect of smoothing by the ACT beam on the f220 maps. We then calculate the aperture flux density in the central $1.2\arcmin$ radius of the smoothed stamps. Additionally, for each stamp we estimate the local background flux density by computing the $1.2\arcmin$ radius aperture flux density at 20 random locations in the stamp lying outside the central aperture and taking the median. We then subtracted the background flux density from the central flux density to form an estimate of the radio flux density for that cluster (see Fig.~\ref{fig:NVSS_histograms} for histograms of the ACT and MaDCoWS radio fluxes in NVSS). For ACT, we found a median flux density at $1.4$~GHz of $6.1\pm 0.4$~mJy for the $3341$ clusters in the NVSS footprint. For MaDCoWS, we found a median flux density of $3.9\pm 0.4$~mJy for the $1780$ clusters in the NVSS footprint, where the uncertainties were estimated via bootstrapping the stack. 

We used an identical method to measure the background subtracted fluxes for the clusters in the VLASS data. For each cluster with background subtracted flux density greater than the NVSS confusion limit of $2.5$~mJy in both data sets, we proceeded to compute a spectral index $\alpha$.  However, in order to compute $\alpha$, one requires more precise knowledge of the flux-weighted band center $\nu_{\textsc{vlass}}$ for the wide bandwidth (2-4~GHz) VLASS data. For example, if the measured flux density in VLASS is lower than the flux density one would find by assuming a fiducial spectral index, the resulting $\alpha$ is steeper, and $\nu_{\textsc{vlass}}$ will shift lower than that initially assumed; if the flux density is higher, $\alpha$ is flatter, and $\nu_{\textsc{vlass}}$  shifts higher.  This leads us to rely on a recursive or iterative approach when estimating $\nu_{\textsc{vlass}}$ and the resulting $\alpha$.
Assuming a flat instrument passband from 2-4~GHz and emission of the form $S_{\nu} = C_0\nu^{\alpha}$ for $C_0$ the amplitude and $\alpha$ the spectral index, we compute the flux-weighted band average as:
\begin{equation} \label{eq5}
\begin{split}
\nu_{\textsc{vlass}}  = \frac{\int_{2~\mathrm{GHz}}^{4~\mathrm{GHz}} C_0\nu^{\alpha+1}  d\nu}{\int_{2~\mathrm{GHz}}^{4~\mathrm{GHz}} C_0\nu^{\alpha}d\nu}
 = \left( \frac{\alpha+1}{\alpha+2} \right) \left( \frac{4^{\alpha+2}-2^{\alpha+2}}{4^{\alpha+1}-2^{\alpha+1}} \right)
\end{split}
.\end{equation}

We note that we assume the passband to be flat for simplicity. For a given cluster with a measured radio flux density in both the NVSS and VLASS surveys, we first computed the spectral index using the nominal VLASS band center of 3~GHz.\footnote{Here we define spectral index such that negative values for $\alpha$ show the typical behavior of declining at higher frequencies (i.e., the flux density $S_{\nu} \propto (\nu/\nu_0)^{\alpha}$).} Using that spectral index and Eq.~\ref{eq5}, we computed a new VLASS band center. We then calculated a new spectral index using this band center, and repeated the process until the difference between consecutive spectral indices was $>0.1\%$. The uncertainty on the spectral index was computed from the fluxes $(S)$ and uncertainties $(\sigma)$ following, for example, \cite{Zajacek2019}, as
\begin{equation}
    \sigma_{\alpha}=\frac{1}{|\log{(\nu_{\textsc{nvss}}/ \nu_{\textsc{vlass}})}|}\sqrt{(\sigma_{\textsc{nvss}}/S_{\textsc{nvss}})^2+(\sigma_{\textsc{vlass}}/S_{\textsc{vlass}})^2}.
\end{equation}
The population of spectral indices and fluxes that was generated in this manner was then used to correct the mass-richness scaling relation as described in Sect.~\ref{sec:priors}. For both the ACT and \madcows clusters, we find a fairly broad distribution of spectral indices computed between $1.4$ and $3$~GHz, with mean for the \madcows candidates of $\langle \alpha_\text{\madcows} \rangle = -0.9$ and standard deviation $\sigma_{\alpha} = 0.7$, and mean for the ACT clusters of $\langle \alpha_\text{ACT} \rangle = -1.2$ and standard deviation $\sigma_{\alpha} = 0.8$.
These mean spectral indices are broadly consistent with values typically found for radio AGN \citep{Coble2007,Sayers2013}, though we note our spectral index is also consistent with that found for star-forming regions \citep[typically $\alpha<-0.6$; see][]{CalistroRivera2017}, so we cannot conclude whether AGN or star formation dominates the observed radio spectra.\\

Similarly to Sect.~\ref{sec:submm_emission}, we computed the bias in the measured SZ signal due to radio in-fill of the ACT and \madcows clusters. We computed the in-fill percentage twice, once assuming our measured average spectral index of $\alpha_{\textsc{meas}} = -0.91$ and the other assuming a typical radio spectra of $\alpha_{\textsc{typ}} = -0.7$. We refer to the in-fill percentages assuming their respective spectral indices as $\textsc{s}_{\textsc{meas}}$ and $\textsc{s}_{\textsc{typ}}$, respectively. The high uncertainty on this number and strong dependence of the $98$ and $150$~GHz flux density on $\alpha$ mean that the bias for a specific cluster can vary quite a bit depending on its spectral index. For example, a 100~mJy source at 1.4~GHz with a spectral index of $-0.91$ produces a decrement of about 10\%, while one with a spectral index of $-1.2$ produces only about 3\%, and one with a spectral index of $-0.7$ produces a 26\% in-fill, assuming $M = 2\times 10^{14} M_{\odot}$, $z = 1$. For the \madcows clusters, we continue to use the reference cluster with $M = 2\times 10^{14} M_{\odot}, z = 1$ to compute the in-fill. For the ACT clusters, we use their individual measured masses and redshifts. 
Extrapolating flux density using spectral indices derived at $1.4$ and $3.0$~GHz out to $98$ and $150$~GHz can be risky not only due to the uncertainty in the spectral index but also because it is not assured that the index is consistent between those two frequency ranges. \citet{Sayers2013} found that spectral indices for radio galaxies in large clusters were generally consistent when computed between 1.4 and 30~GHz and 30 and 140~GHz, so that as an estimate it is justified to use the measured spectral indices to extrapolate our flux densities from $1.4$ to $98$ and $150$~GHz. For precise determination of the radio infill, multifrequency observations near $98$ and $150$~GHz will be required to determine the effective spectral index of sources near the SZ frequencies.

The average in-fill for an ACT cluster is $\textsc{s}_{\textsc{meas}} = 0.8\pm 0.03\%$ and $\textsc{s}_{\textsc{typ}}= 2.0\pm0.1\%$, while for the \madcows cluster candidates the average in-fill is $\textsc{s}_{\textsc{meas}} =0.45 \pm 0.05 \%$ and $\textsc{s}_{\textsc{typ}}= 0.83\pm0.05\%$. However, while the average bias is quite low, for an appreciable number of the clusters the bias is non-negligible. For the 95th percentile of ACT clusters, $\textsc{s}_{\textsc{meas}} = 7.0\pm 0.5\%$ and $\textsc{s}_{\textsc{typ}}= 18\pm1\%$. We note that of the nine \madcows clusters with \yc $< -0.5$, 5 had significant radio in-fill (flux density $\gtrsim 20$~mJy at 1.4~GHz), including five of the top six with most negative \yc, suggesting that the significantly negative \yc measurements could be due to radio in-fill. For example, one such cluster with very negative \yc, MOO J2247+0507, has a flux density at 1.4~GHz of $160\pm 1$~mJy and $\alpha = -0.61\pm 0.8$ (its measured $\alpha$) and $z = 1.02$; assuming it has a mass of $2\times 10^{14}~\rm M_{\odot}$, the bias at 98~GHz is $s_{\textsc{J2247}} = 70\pm 17\%$, where the uncertainties have been propagated from the flux density uncertainty only. 

We emphasize that we have not included the effect of the matched filter, so the numbers presented above represent only an approximation of the effect of the infill. Further, since our radio fluxes were computed using aperture photometry, the measured fluxes and spectral indices are averages for all compact sources in the cluster. High resolution, multifrequency follow-up of cluster candidates would be required to precisely remove this infill.
\\

\begin{figure}
    \centering
    \includegraphics[clip,trim=0.0cm 0.0cm 1.2cm 0.0cm,width=\columnwidth]{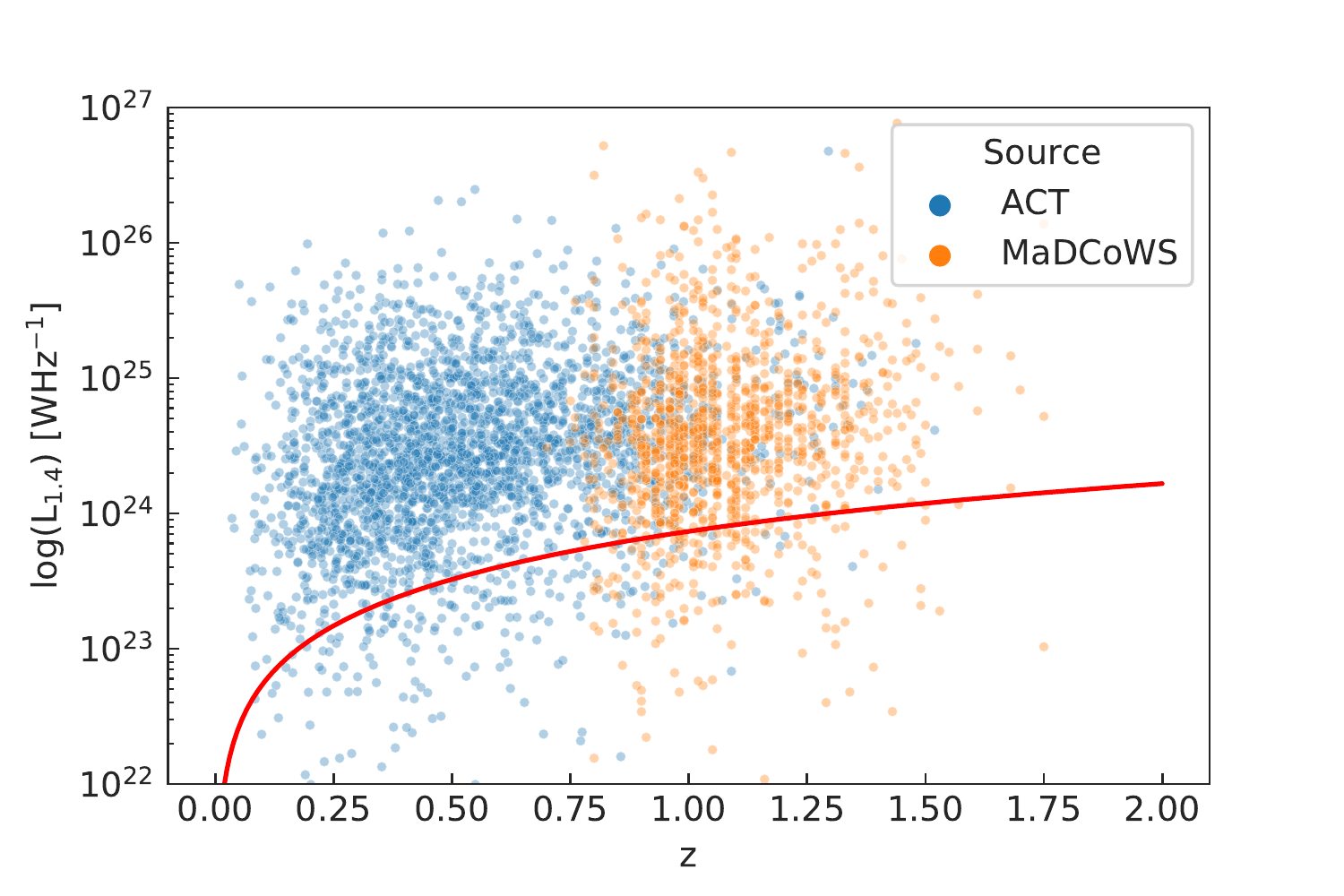}
    \caption{Intrinsic radio luminosities at 1.4~GHz for ACT (blue, $3341$ clusters) and \madcows (orange, $1780$ clusters). The red line shows a 0.3~mJy flux density at 1.4~GHz in the observed frame of reference  converted to 1.4~GHz in the emitted frame of reference for the given $z$ according to Eq.~\ref{eq:8}, and assuming a spectral index of $-0.91$. The choice of 0.3~mJy corresponds to the variance of the background aperture flux density in the NVSS maps for a $1.2\arcmin$ radius aperture, and thus roughly corresponds to the noise floor for our background subtracted fluxes. Therefore, the red line should guide the eye as to which radio luminosities are above the noise. We note that many clusters are not shown as they have very low or negative values for their radio fluxes due to subtraction of the background. }
    \label{fig:radio_lum}
\end{figure}

Consistent with the above, we computed the intrinsic radio luminosities. To do so, we assumed a simple power law of the form $S_{\nu} = C_0\nu^{\alpha}$. This then leads to the usual $K$-correction; $K(z)= (1+z)^{-(1+\alpha)}$ for redshift $z$. The intrinsic luminosity at frequency $\nu_1$ can then be computed from the observed flux density $S_{\nu_2}$ at frequency $\nu_2$ using the luminosity distance $D_L(z)$:
\begin{equation} \label{eq:8}
    L_{\nu_1} = \frac{4\pi D_L(z)}{(1+z)^{-(1+\alpha)}}\left( \frac{\nu_1}{\nu_2}\right)^{\alpha} S_{\nu_2}
.\end{equation}
For convenience we select $\nu_1$ to be $1.4$~GHz. For each cluster then we calculated its intrinsic luminosity using the NVSS measured flux, its measured redshift, and the spectral index for that cluster as computed above. All the ACT clusters have measured redshifts; for the \madcows candidates that do not have redshifts, we used the mean of the sample, $\langle z \rangle=1.01$. For clusters that did not have measured spectral indices, we used the average spectral index of that cluster's catalog, either ACT ($\langle \alpha_\text{ACT} \rangle = -1.21$) or \madcows ($\langle \alpha_\textsc{\madcows} \rangle = -0.91$). The results are shown in Fig.~\ref{fig:radio_lum}. The average luminosity for ACT clusters is $5.4\pm 0.3 \times 10^{24}~\textsc{W Hz}^{-1}$, while for the \madcows clusters it is $9.1\pm 1.0 \times 10^{24}~\textsc{W Hz}^{-1}$, where the statistical uncertainties have been computed via bootstrapping. Restricting the ACT clusters to the same redshift range as the \madcows ($0.7\leq z \leq 1.5)$ raises the average luminosity of ACT clusters to $6.5 \pm 0.7 \times 10^{24}~\textsc{W Hz}^{-1}$, suggesting that, even accounting for redshift, the \madcows clusters are on average more radio loud than their ACT counterparts, although the average redshift of the ACT clusters after this restriction ($\langle z \rangle =0.89$) is still lower than that of the \madcows sample ($\langle z\rangle = 1.01$). 

\section{The SZ mass-richness scaling}\label{sec:scalingrels}
Given the results above, we attempt to investigate the \madcows mass-richness scaling relation. In doing so, we hope to understand if richness provides a good proxy for mass in the \madcows cluster catalog. Additionally, we address whether the preliminary mass-richness scaling relation found in \citet{Gonzalez2019} is consistent with the scaling relation of the entire \madcows catalog, and whether the full scaling relation follows self-similarity. 
\subsection{Regression technique}
\label{sec:regression}
We adopt a hierarchical Bayesian approach that builds upon the work of \citet{Kelly2007} and \citet{sereno2016}, to which we refer for a thorough discussion of the fitting technique. Here we provide a summary of key details central to our analysis.

At each step of the modeling process and for each of the clusters within the considered sample, we consider an independent variable $\xi$ drawn from a mixture of Gaussian probability distributions \citep{Kelly2007}. This corresponds to the true value of the logarithm of the cluster richness, $\ln{\lambda}$. We therefore fit richness as $\ln{\lambda}$, which is related to the observed richness $\lambda_{\mathrm{obs}}$ through a Poisson probability distribution, $P(\lambda_{\mathrm{obs}}|\xi) = \mathcal{P}(e^{\xi})$.

In a similar manner to richness, we fit the mass as $\ln{M}$.  We define the dependent quantity $\eta$ as the true value of the logarithm of the cluster mass, $\ln{M}$, which we assume to be connected to the independent variable $\xi$ through a normal probability distribution $P(\eta|\xi)$ with mean
\begin{equation}
    \left<\eta|\xi\right> = p_0+p_1\xi
\end{equation}
and variance corresponding to the intrinsic scatter of the true quantities about the mean scaling relation. Given the definition of the variables $\xi$ and $\eta$, our choice of probability distribution is equivalent to using a log-normal model for describing the relation between the true mass and richness for a given cluster. Following, for example, \citet{Evrard2014} and \citet{Simet2017}, it is hence possible to express the variance due to intrinsic scatter as
\begin{equation}
    \mathrm{Var}(\ln{M}|\lambda) = \sigma^2_{\mathrm{int}} =  \frac{p_1^2}{\lambda}+\sigma^2_{\ln{M}|\lambda}.
    \label{eq:scatter}
\end{equation}
Here, the first term on the right-hand side accounts for the contribution to the total scatter due to Poisson noise on richness, while the second term describes the scatter inherent to the independent variable $\eta=\ln{M}$. This is introduced to account for any additional deviation from the reconstructed scaling that is not accounted for in the observational uncertainties or intrinsic variable properties (e.g., due to unknown biases in the considered observables).

\begin{figure*}
    \centering
    \includegraphics[clip,trim=0cm 2mm 5mm 0cm,width=\textwidth]{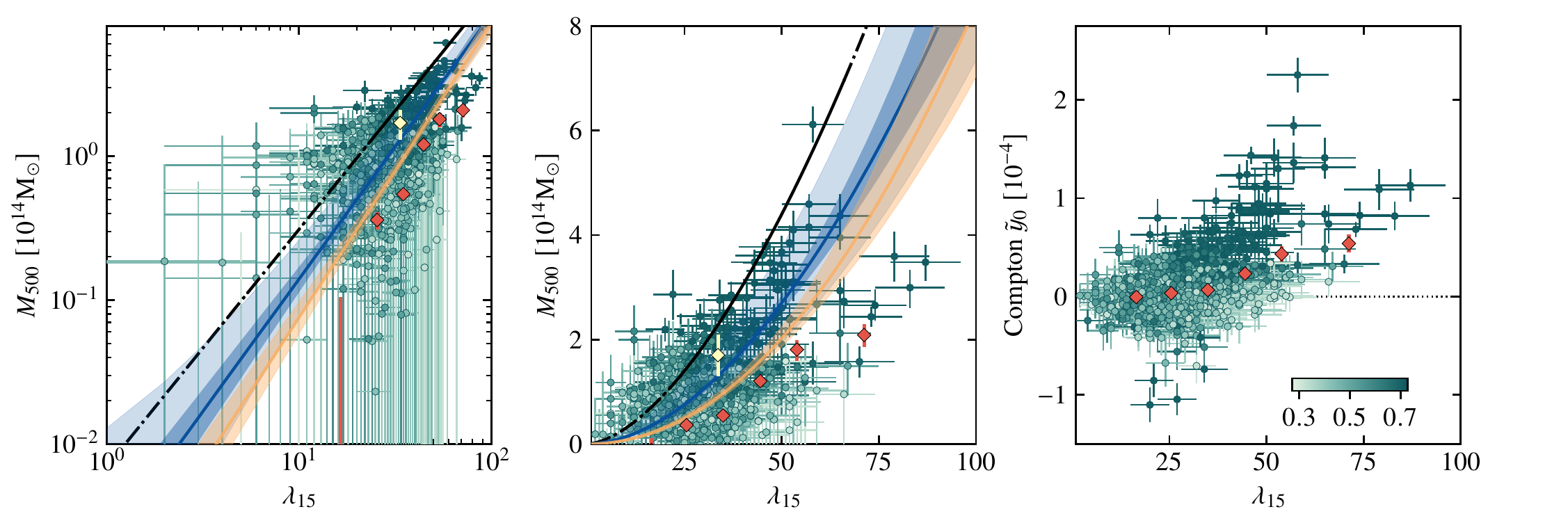}
    \caption{Mass-richness diagram displayed three ways. The left and center panels are respectively log-log and linear plots of \yc converted to mass versus richness, while the right panel is a linear plot of unconverted \yc versus richness. On left and in the center, the best-fit scaling relation is shown as computed when including (blue) and excluding (yellow-orange) the weight parameter of Eq.~\ref{eq:16} corresponding to noise-like data points. For each band, the solid line corresponds to the best-fit scaling, while the bands correspond to the 68\% credible interval. 
    The lighter blue band denotes the confidence interval  $\sigma_{\ln{M}|\lambda}$ due to the intrinsic scatter around the mean scaling relation obtained when considering the mixed model of Eq.~\ref{eq:16}. The dot-dashed line denotes the scaling reported by \citet{Gonzalez2019}, with the solid section marking the range of richness employed to derive the relation. The green points correspond to the masses or \yc values without any correction for the steepness of the halo mass function \citep{Hilton2021}. These are color coded according to their weight $w$  (color bar in the bottom right corner of the right panel). In red are the masses or \yc values computed per richness bin. For comparison, we include, as a yellow diamond, the average mass estimate computed by \citet{Madhavacheril2020} from CMB lensing, shown at the mean value of $\lambda_{15}$ used in that work. We note that the left and center plots do not show the negative \yc points, but those points are included in all fits shown.}
    \label{fig:scaling}
\end{figure*}

In principle, it should be possible to define a relation between the measured masses and the dependent variable $\eta$ through a normal distribution centered on $e^{\eta}$ and with variance $\sigma^2_{M}$ equal to the square of the observational uncertainties on the measured mass $M_{\mathrm{obs}}$,
\begin{equation}
    P(M_{\mathrm{obs}}|\eta) = \mathcal{N}(e^{\eta},\sigma^2_{M}).
\end{equation} 
However, Fig.~\ref{fig:scaling} shows that multiple data points from the compiled forced photometry catalog manifest negative values for the central Compton \yc, corresponding to an unphysical negative cluster mass. In the case of low-mass clusters, with SZ signal below the sensitivity threshold of the ACT maps, noise fluctuations can cause this negative \yc. However, as discussed in previous sections, radio sources as well as dust are found to contaminate the SZ signal of the clusters, leading to clusters with significantly negative \yc. Further, in Sect.~\ref{sec:miscentering} we show that miscentering effects may provide a significant suppression of the measured central Compton $\Tilde{\vary}_0$ with respect to the true value. In order to properly account for such contributions, we compare the true scattered quantities $\eta$ to the central Compton parameter $\Tilde{\vary}_0$ computed separately from the f090 and f150 maps. In particular, we assume the joint probability distribution to be described by a bivariate normal distribution,
\begin{equation}
     P(\vary_{\mathrm{f090}},\vary_{\mathrm{f150}}|\xi,\eta,\theta_{\mathrm{radio}},\theta_{\mathrm{dust}}) = \mathcal{N}^{2D}(\{f_{\mathrm{f090}},f_{\mathrm{f150}}\},\Sigma),
    \label{eq:like1}
\end{equation}
where $\vary_{\mathrm{f090}}$ and $\vary_{\mathrm{f150}}$ are \yc derived exclusively from the f090 and f150 maps, respectively. Here, the covariance matrix $\Sigma$ is expressed as
\begin{equation}
    \Sigma = \begin{pmatrix} \sigma^2_{\mathrm{f090}} & \rho\sigma_{\mathrm{f090}}\sigma_{\mathrm{f150}} \\
                             \rho\sigma_{\mathrm{f090}}\sigma_{\mathrm{f150}} & \sigma^2_{\mathrm{f150}}\end{pmatrix}
   \label{eq:post1}
,\end{equation}
with $\sigma_{\mathrm{f090}}$ and $\sigma_{\mathrm{f150}}$ equal to the uncertainties on the $\tilde{\vary}_0$ measured from the f090 and f150 maps, while $\rho=0.21$ is the Pearson product-moment correlation coefficient for the two flattened frequency maps after filtering. The mean term is given by
\begin{align}
    f_{\nu} &= f_{\nu}(\xi,\eta,\theta_{\mathrm{radio}},\theta_{\mathrm{dust}}) \nonumber\\
            &= c(\xi)\cdot m_{\nu}(\eta)+g_{\nu}\left[d_{\nu}(\theta_{\mathrm{dust}})+r_{\nu}(\theta_{\mathrm{radio}})\right].
    \label{eq:like2}
\end{align}

\noindent The first term on the right hand side is the product of the miscentering suppression factor $c(\xi)$ (Sect.~\ref{sec:miscentering}) and the mass-to-Compton $\vary$ conversion $m_{\nu}(\eta)$ in Eq.~5 of \citet{Hilton2021}. All the factors entering Eq.~\ref{eq:like2} (i.e., the relativistic SZ correction and the filter mismatch factor) are computed from the specific sets of cluster masses at each step of the modeling process. The terms in parentheses, $d_{\nu}(\theta_{\mathrm{dust}}=\{A,\beta,T_{\mathrm{rest}}\})$ and $r_{\nu}(\theta_{\mathrm{radio}}=\{C_0,\alpha\})$, provide the estimates of the level of dust and radio infill at the considered frequency, based on the respective spectral properties discussed in Sect.~\ref{sec:submm_emission} and Sect.~\ref{sec:radio}. The function $g_{\nu}$ is the nonrelativistic spectral dependence of the SZ effect \citep{Sunyaev1970, Mroczkowski2019}, which we adopt to convert the radio and dust surface brightness values to units of Compton $\vary$. We note that, in order to compute the infill components as in Eq.~\eqref{eq:like2} shown above, we are assuming the radio and dust sources to be described by point-like signals centered on the cluster centroids. 

Finally, we account for possible elements of the \madcows cluster sample that are not well described by a mass-richness scaling relation (i.e., for which richness is not a good proxy for mass) by integrating the probability term described in Eq.~\eqref{eq:like1} into a mixture model aimed at evaluating how likely well each point is to be drawn from a Gaussian centered on the mass-richness scaling relation versus a Gaussian centered on zero. For each given point, the total posterior probability distribution can then be written as
\begin{equation}\label{eq:16}
    P_{\nu,\mathrm{obs}} = w\cdot P_{\mathrm{true}}+(1-w)\cdot P_{\mathrm{noise}},
\end{equation}
where the weight, $w$, is the probability that a data point is drawn by the mass-richness scaling relation versus the noise-like population, and $P_{\mathrm{true}}=P(\vary_{\mathrm{f090}},\vary_{\mathrm{f150}}|\xi,\eta,\theta_{\mathrm{radio}},\theta_{\mathrm{dust}})$ is the joint probability distribution introduced in Eq.~\ref{eq:post1} for a data point that follows the mass-richness scaling relation. We instead assume the noise-like measurements to be drawn from $P_{\mathrm{noise}}$, for which we assume a normal distribution centered around zero and with standard deviation equal to the observational uncertainty associated with the $\tilde{y}_c$ value of the considered measurement. 

\subsection{Parameter priors}\label{sec:priors}
Overall, our model comprises 965+12+3+3 free parameters, specifically corresponding to:  the weight for each of the 965 data points entering the mixed probability distribution of Eq.~\eqref{eq:16}; 12 parameters associated with three Gaussian kernels used for building the probability mixture discussed at the beginning of Sect.~\ref{sec:regression} (see \citealt{Kelly2007} and \citealt{sereno2016} for details);
the slope, intercept, and intrinsic scatter of the mass-richness scaling relation; and the  parameters of the gray-body spectrum discussed in Sect.~\ref{sec:submm_emission}.

In addition, for every data point we marginalize over a set of four additional parameters, governing the effects due to miscentering (see Sect.~\ref{sec:miscentering}) and radio and dust contamination.
The regression is performed by means of the Hamiltonian Monte Carlo algorithm provided in the \texttt{NumPyro} \citep{bingham2018,phan2019} Python package.
The implementation of the Gaussian mixture model follows the same prescriptions adopted in \citet{Kelly2007}. We use uninformative uniform priors on all the parameters of the scaling relation except for the slope $p_1$, for which we consider a Student's $t$-distribution with one degree of freedom following \citet{Andreon2010} and \citet{sereno2016}, as it does not bias the slope to high values. We then assume the probability weight $w$ to be distributed uniformly in the range $[0,1]$.

The suppression factor due to miscentering for each of the considered clusters is modeled as a truncated normal distribution, bound between 0 and 1, with mode equal to 1 and standard deviation $\sigma=\sqrt{\frac{\pi}{2}}(1+e^{-b\lambda})^{-1}$. The coefficient $b$ corresponds to the best-fit parameter derived in Sect.~\ref{sec:miscentering}, while the pre-factor $\sqrt{\frac{\pi}{2}}$ is introduced so that the mean of the prior distribution equals the average suppression $(1+e^{-b\lambda})^{-1}$ for a given true richness $\lambda=e^\xi$.

For the clusters with a clear identification of radio sources in the NVSS and VLASS fields, we employ normal priors on the estimated normalization $C_0$ and spectral index $\alpha$ parameters, with mean and standard deviation set to the values derived for the specific cluster. Otherwise, we draw for each cluster a realization of $C_0$ from an exponential distribution with mean equal to the average flux density measured from all the sources in the NVSS catalog. An analogous approach is considered for $\alpha$, but used a normal prior with mean and standard deviations measured from the distribution of spectral indices estimated in Sect.~\ref{sec:radio}.

Instead of introducing priors on the single parameters of the dust spectrum (Sect.~\ref{sec:submm_emission}), which would have neglected information on their degeneracy, we reanalyze the ACT 224~GHz and {\it Herschel} stacked measurements jointly with the mass and richness data. For full consistency, we consider exactly the same priors employed in Sect.~\ref{sec:submm_emission}.

\subsection{Scaling relation results}\label{sec:scaling_results}

In Fig.~\ref{fig:scaling} we show the mass-richness scaling relation reconstructed from the ACT+MaDCoWS sample. As the selection of the clusters is unbiased with respect to the Eddington bias, we consider here the cluster masses without any correction for the steepness of the halo mass function. We note that, in this case, the reported masses represent an upper limit for the actual distribution, as the correction would de-boost the high-mass end of the sample.

For the model including the weight parameter from Eq.~\ref{eq:16}, we find that the best-fit slope to be $p_1 = 1.84^{+0.15}_{-0.14}$, where $M\propto \lambda_{15}^{p_1}$. This slope is consistent with \citet{Gonzalez2019}, although with an overall offset toward lower masses. Setting all the weights to unity (that is,  assuming all clusters are drawn from the mass-richness scaling relation) leads to a slightly higher slope estimate ($1.95^{+0.17}_{-0.16}$), as well as a greater overall offset as compared to \citet{Gonzalez2019}.

Regarding the intrinsic scatter of the forced photometry data points (see Eq.~\eqref{eq:scatter} and related description), we estimate $\sigma_{\ln{M}|\lambda}=0.21\substack{+0.08\\-0.11}$ when including the weight parameter of Eq.~\ref{eq:16} to account for sample contamination.
Although already evident from the distribution of the ACT data points in the mass-richness distribution in relation to the scaling relation by \citet{Gonzalez2019}, such a large scatter provides a quantitative view of the limited capabilities of the MaDCoWS richness to provide a robust proxy for cluster masses. A similar scatter of $0.31\substack{+0.03\\-0.03}$ 
is found even when removing the negative Compton $\Tilde{\vary}_0$ measurements from the fit.

\subsection{Cluster weights} \label{sec:weights}

\begin{figure}
    \centering
    \includegraphics[clip,trim=3mm 6mm 3mm 5mm,width=\columnwidth]{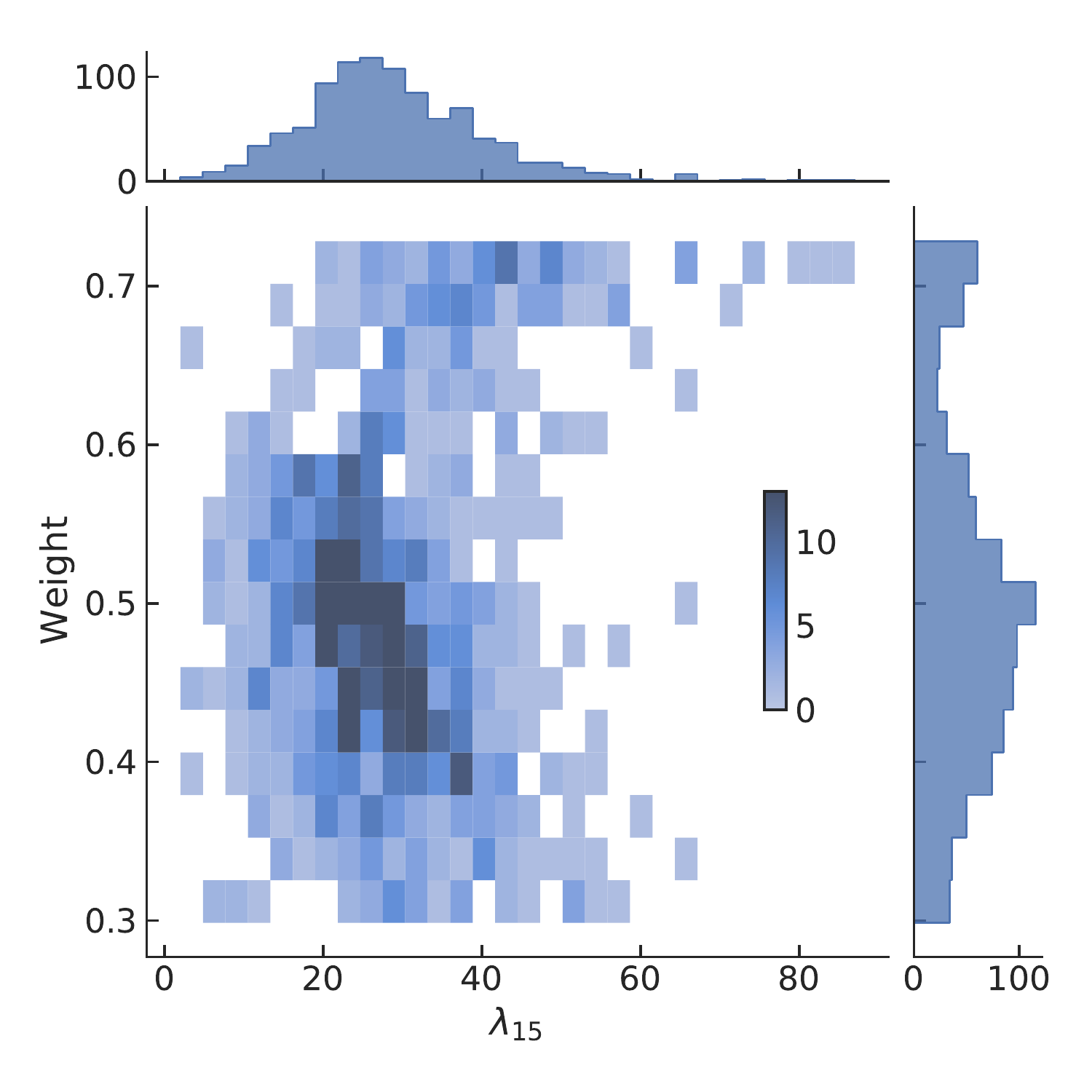}
    \caption{2D histogram of weight and richness ($\lambda_{15}$). The weight is the probability that a \madcows cluster data point is well described by a mass-richness scaling relation, as opposed to being drawn from the noise. Higher probabilities mean they are more likely to be real. The bimodal distribution is clear, with a large population of low probability clusters centered around 50\% and a population of high probability clusters at $\approx 70\%$.}
    \label{fig:prob_hist}
\end{figure}

From the results of the mass-richness fit including the probability term, it is clear that an appreciable fraction of the \madcows candidates are not well described by the mass-richness scaling relation (Fig.~\ref{fig:prob_hist}); in other words, for a large subset of the \madcows catalog, richness is not a good proxy for mass.  Of the 965 candidates used in the fitting, only 419 had a weight greater than 50\%; only 131 have a weight greater than 65\%. The mean weight is $50.2\%$. Since we do not have an estimate of the uncertainties on individual weights, we do not place an uncertainty on these statistics. Moreover, while the mean does not fully encapsulate the bimodal distribution of the cluster probabilities, it does indicate that the \madcows sample is likely composed largely of  cluster candidates that are well below the SZ detection limit of the ACT survey. 
Examining the distribution of the probabilities, we see a clearly bimodal distribution (Fig.~\ref{fig:prob_hist}). One population, containing the majority of the clusters, is roughly normally distributed, centered on a probability of $\approx 50\%$. The other is centered at a higher probability of $\approx 70\%$ with a much narrower distribution. In general, clusters with high-significance \yc[] and richness are given higher probabilities, while those with lower \yc, and specifically high richness and low \yc, are given lower probabilities. Further examining Fig.~\ref{fig:prob_hist}, it is evident that the \madcows sample is well described by a mass-richness scaling relation above $\lambda_{15} \gtrsim 55$, with $50$\% of such clusters having a weight $w > 0.7$, and an average weight for clusters above that richness of $0.63\%$. Additionally,  all the high weight ($w >70\%$) clusters have $\lambda_{15} > 27$. Interestingly, examining the probabilities of individual cluster candidates reveals that the fitter identifies a number of the very negative \yc clusters, seen in the right hand panel of Fig.~\ref{fig:scaling}, as having high weight: these clusters have significant ($\gtrapprox25$~mJy) flux density at $1.4$ or $3.0$~GHz. 

\section{Discussion}\label{sec:discuss}

\subsection{Population differences}\label{sec:counts}

The MaDCoWS sample comprises 2839 candidate high-$z$ clusters, while the ACT DR5 sample comprises a nearly mass-limited, optically confirmed sample of 4195 clusters from across all redshifts. 
As noted in Sect. \ref{sec:crossmatches}, our cross-matching criteria leads to a catalog of 96 co-detected clusters at the intersection of the ACT and MaDCoWS samples, reported in Table \ref{Catalog:ACTCOWS}.

\begin{figure}
    \centering
    \includegraphics[clip,trim=1mm 2mm 15mm 3mm,width=\columnwidth]{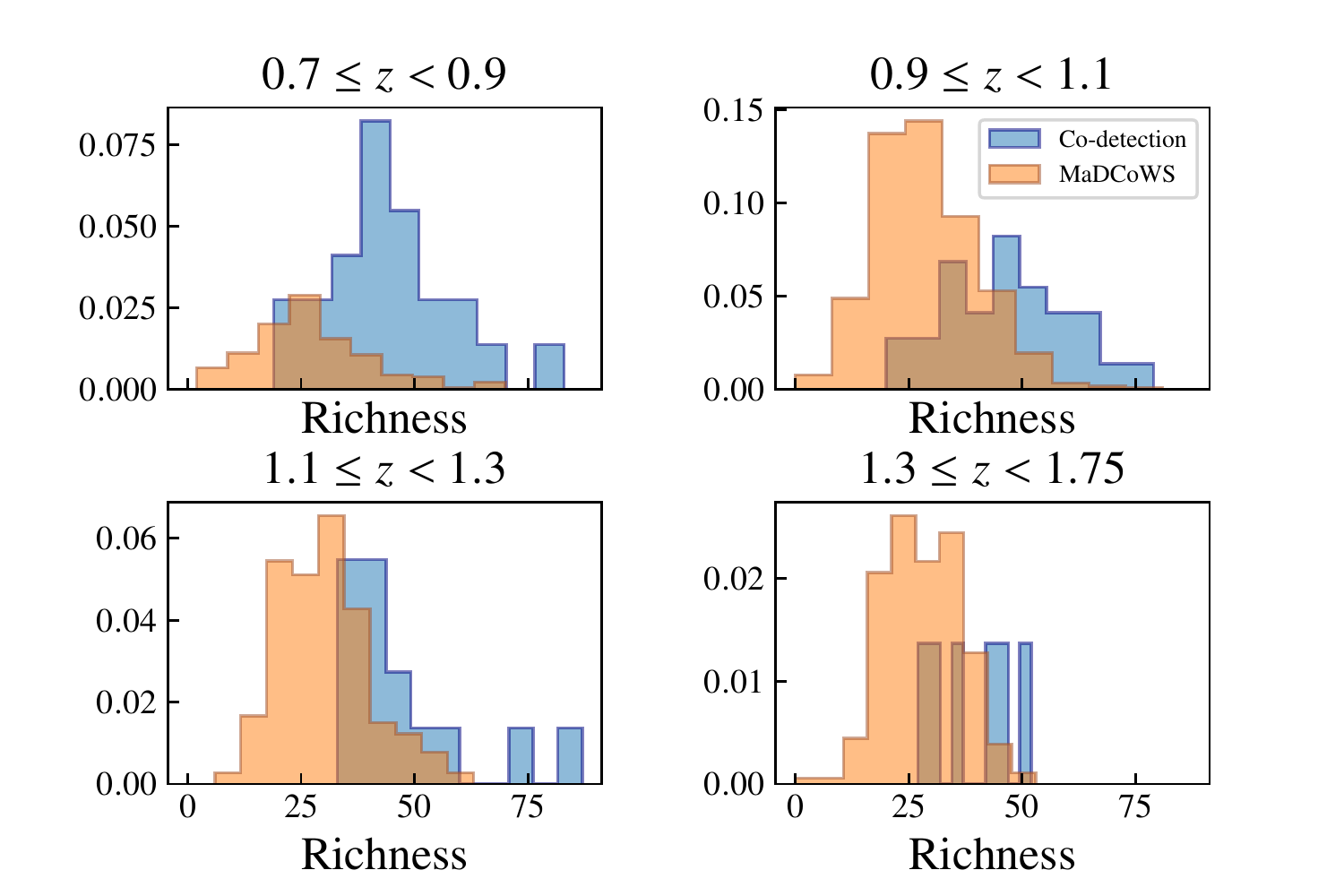}
    \caption{Comparisons of richness populations for the co-detected \madcows and the remaining \madcows in four redshift bins. At higher richness ($\lambda_{15}>60$),  there is no preferred redshift for the co-detections. This indicates that the disparity between the number of \madcows in the ACT region and the number of co-detections is not a product of survey biases.}
    \label{fig:richness_histograms}
\end{figure}

\begin{figure}
    \centering
    \includegraphics[clip,trim=1mm 3mm 3mm 3mm,width=\columnwidth]{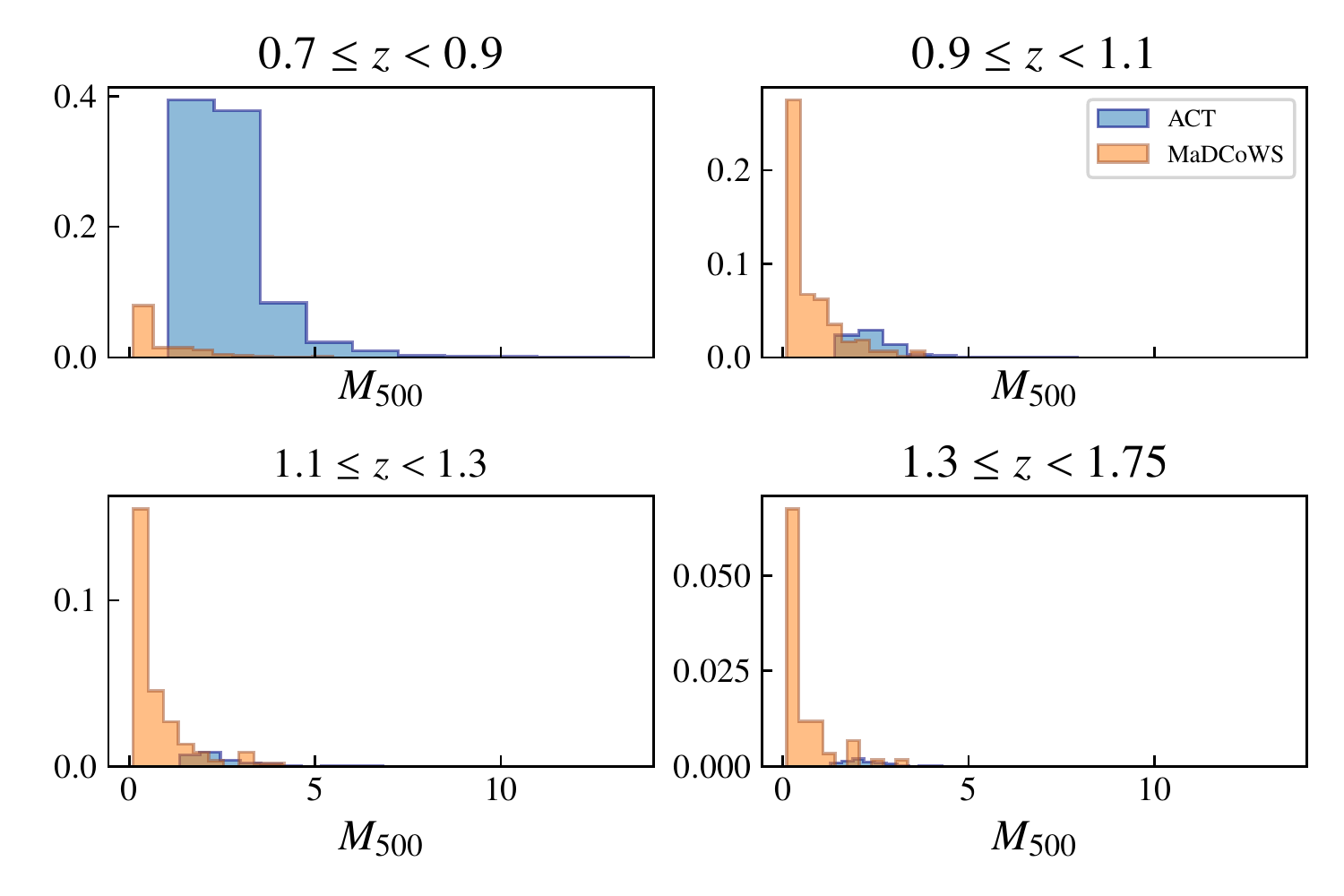}
    \caption{$M_{500}$ of ACT clusters vs. MaDCoWS in the same redshift ranges. These distributions suggest that ACT clusters are skewed toward lower redshifts, while the \madcows distribution is even across higher redshifts. In addition, \madcows clusters tend to be less massive than ACT clusters at a given redshift.}
    \label{fig:m500_histograms}
\end{figure}

We compared both richness and $M_{500}$ measurements at different redshift ranges between the ACT and MaDCoWS populations. In Fig. \ref{fig:richness_histograms}, we plot the distribution of co-detections and MaDCoWS across four redshift ranges. We can see that there is no preferred redshift for the co-detected clusters with high richness. Figure \ref{fig:m500_histograms} shows that ACT clusters are mostly found at $z<0.9$, regardless of mass. We also see that MaDCoWS clusters as a whole tend to be lower in richness (and hence mass) than co-detected clusters, but are found more frequently at higher redshifts than ACT clusters. This is not unexpected, as the lower average mass of clusters at higher redshifts means that they are detected relatively less frequently. The \madcows masses are from forced photometry, and we have cut off clusters with \yc$ < 0$. 

The low rate by ACT of co-detections of \madcows cluster candidates is due to the low SZ signal of many of those candidates, which is reflected in their correspondingly low weight. We discuss possible sources of this low weight in Sect.~\ref{sec:weights_disc}. As for the low rate by \madcows of co-detections of ACT clusters, it is likely that the \madcows catalog is missing them simply due to the \madcows selection function. By construction, the \madcows selection function does not precisely trace the {\it Spitzer} measured richness; the \madcows cluster candidates are first selected from smoothed galaxy density maps created using WISE data, and then their \madcows\ richnesses are measured using {\it Spitzer} follow-up. As such, the catalog is not richness limited, meaning that even under the assumption of some relation between mass and richness, the \madcows cluster catalog cannot strictly be mass-limited. In any case, the \madcows cluster catalog, while astrophysically interesting, should be approached with caution for computing cosmological parameters.


\begin{figure*}
\centering
\includegraphics[clip,trim=2.5cm 0.5cm 3.5cm 1.0cm,height=0.27\textwidth]{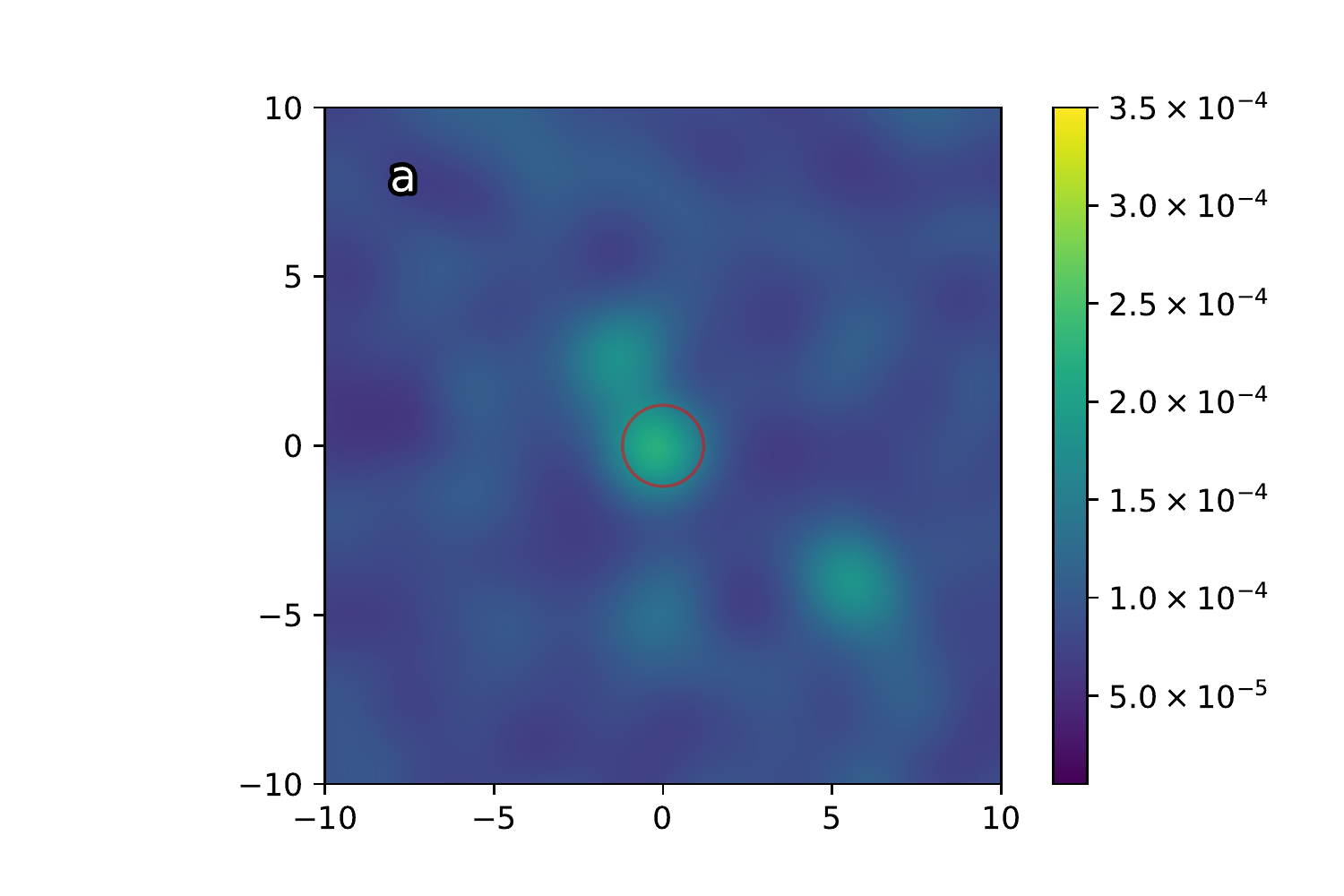}
\includegraphics[clip,trim=2.5cm 0.5cm 3.5cm 1.0cm,height=0.27\textwidth]{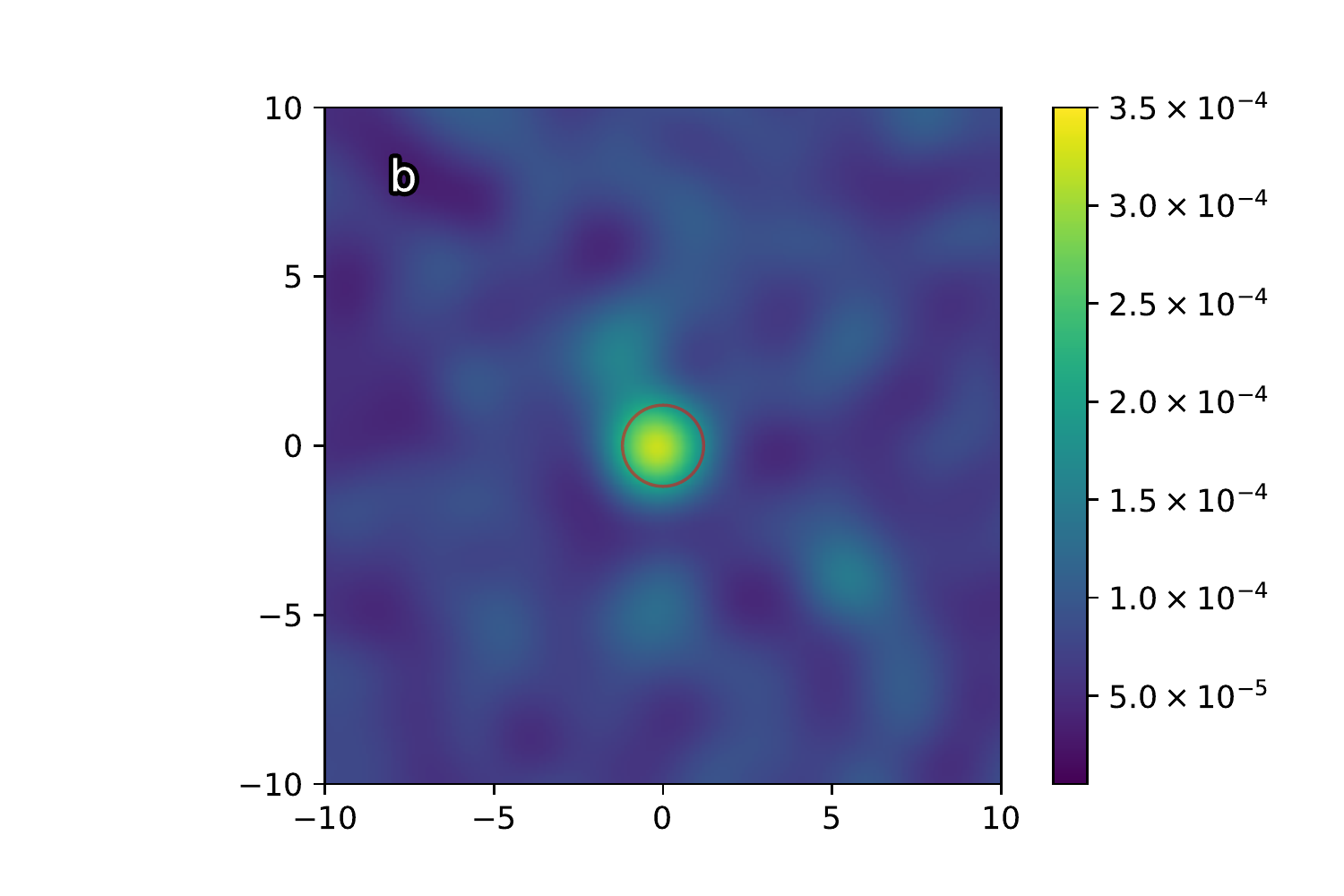}
\includegraphics[clip,trim=2.2cm 0.5cm 0.5cm 1.0cm,height=0.27\textwidth]{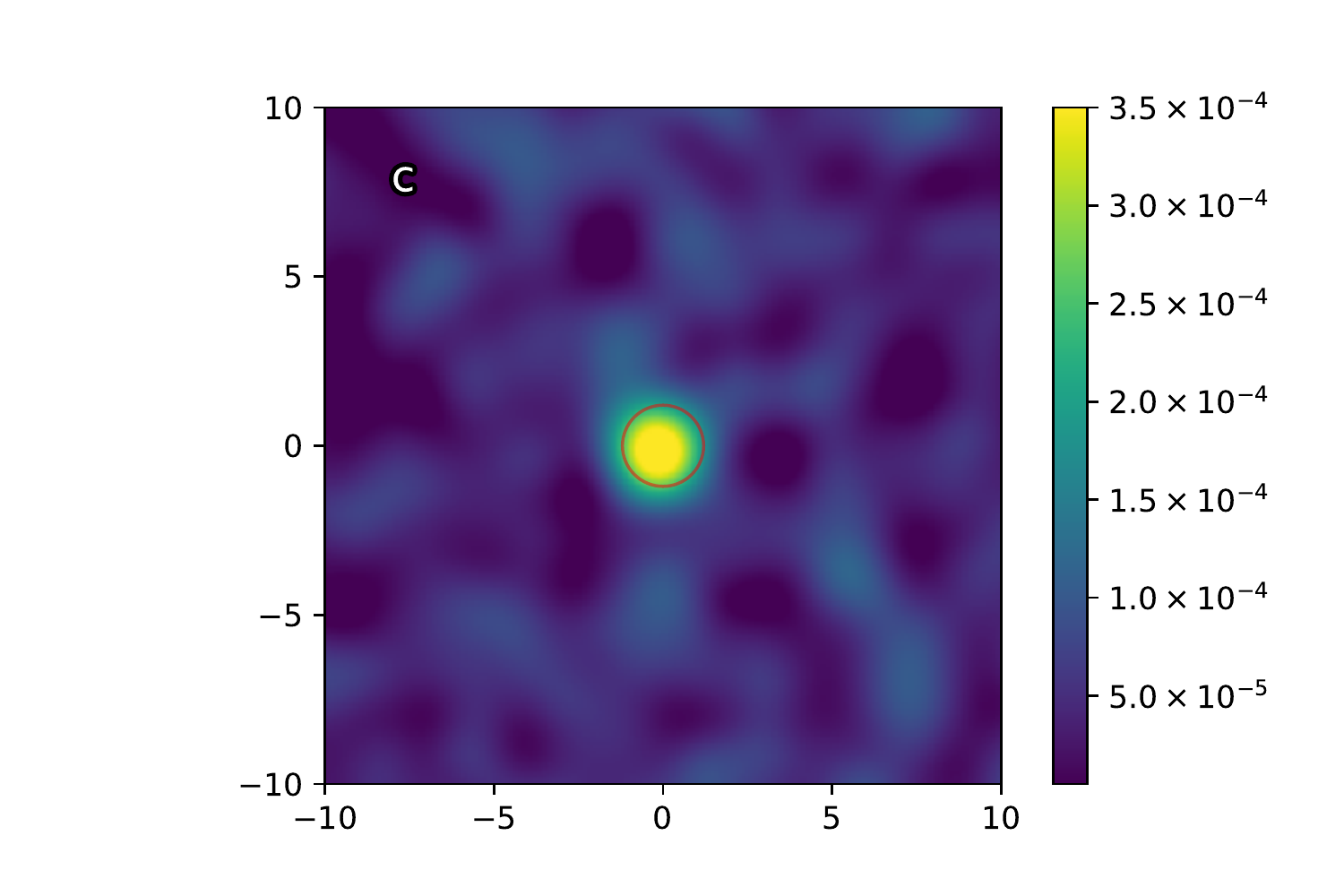}

\includegraphics[clip,trim=2.5cm 0.5cm 3.5cm 1.0cm,height=0.27\textwidth]{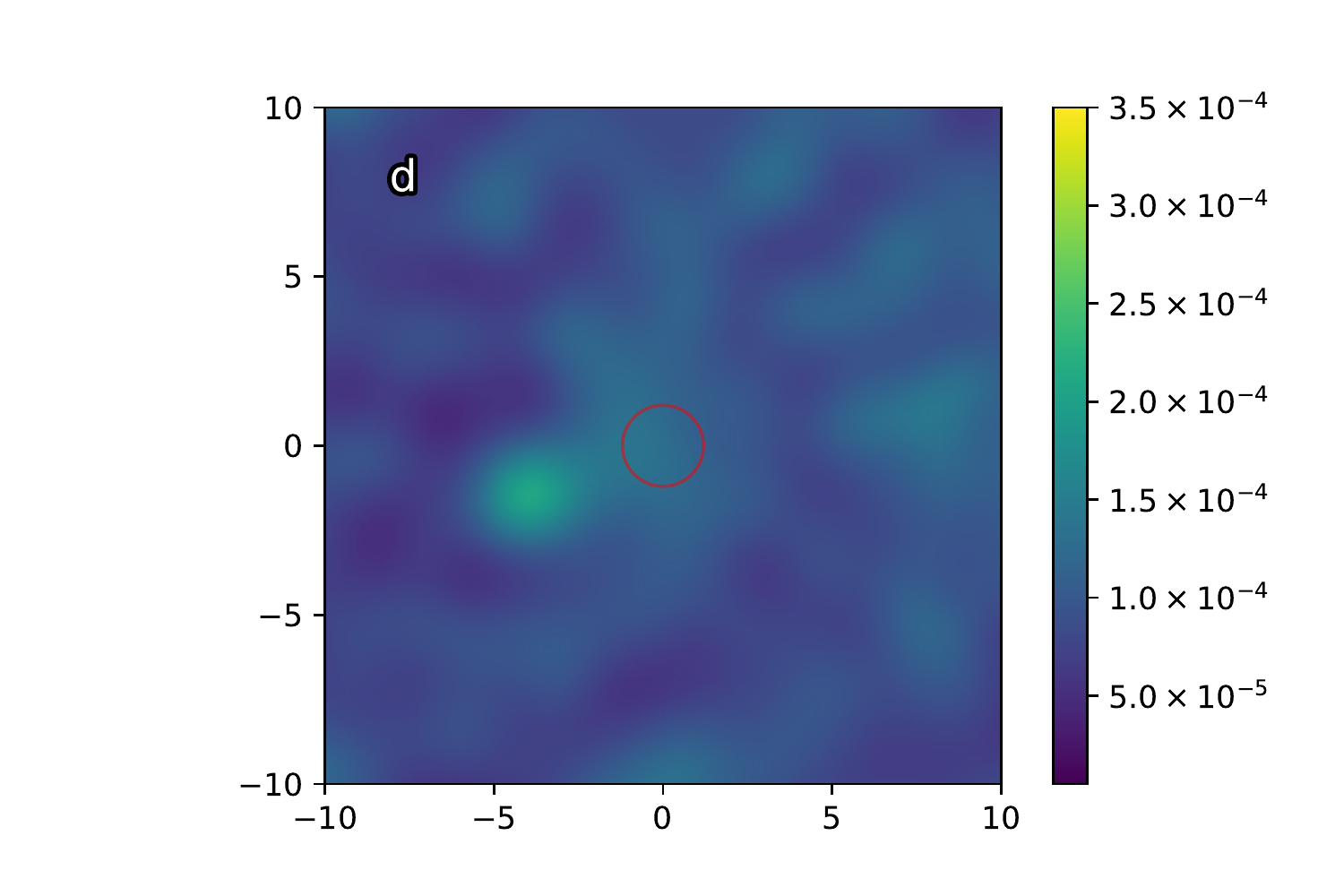}
\includegraphics[clip,trim=2.5cm 0.5cm 3.5cm 1.0cm,height=0.27\textwidth]{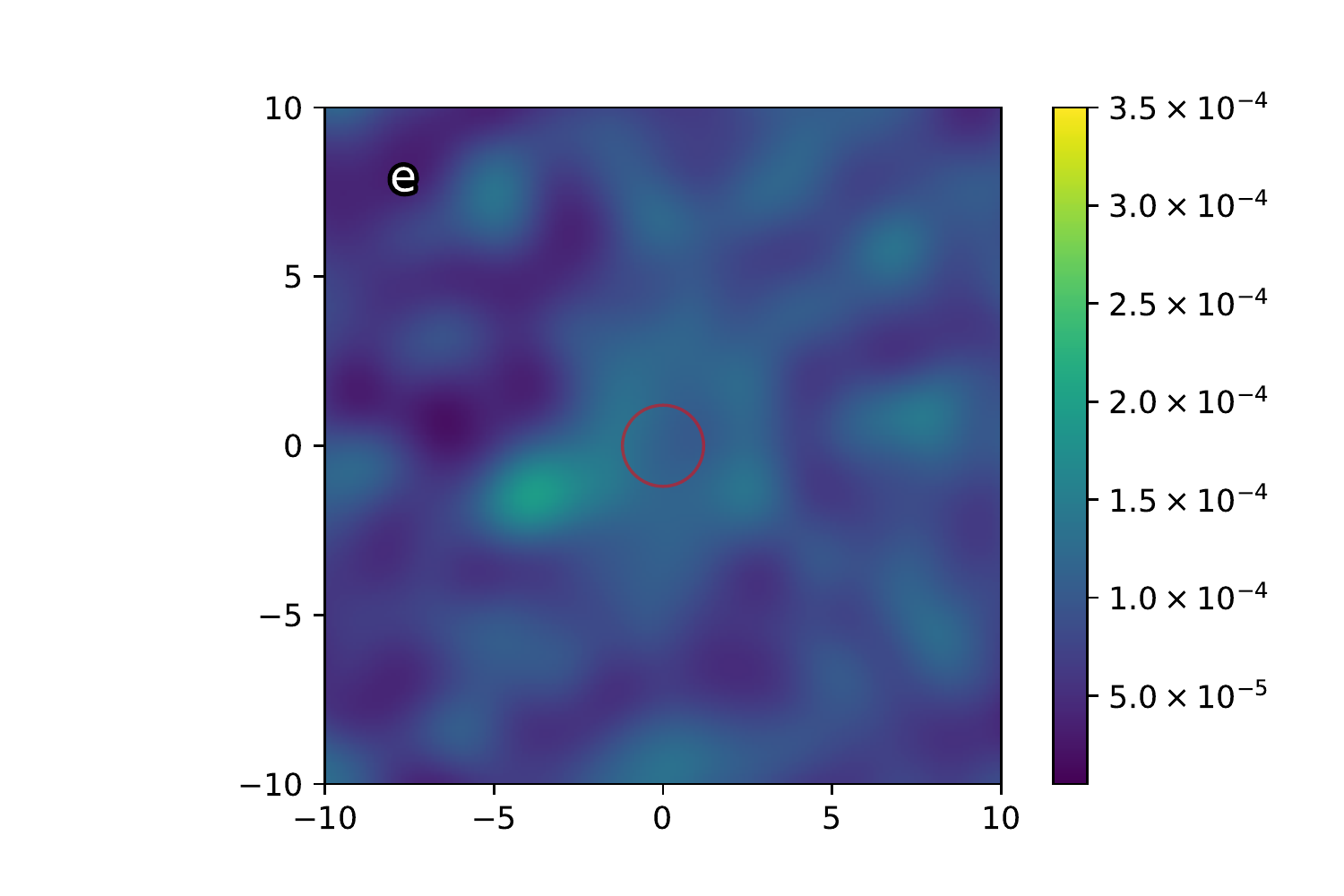}
\includegraphics[clip,trim=2.2cm 0.5cm 0.5cm 1.0cm,height=0.27\textwidth]{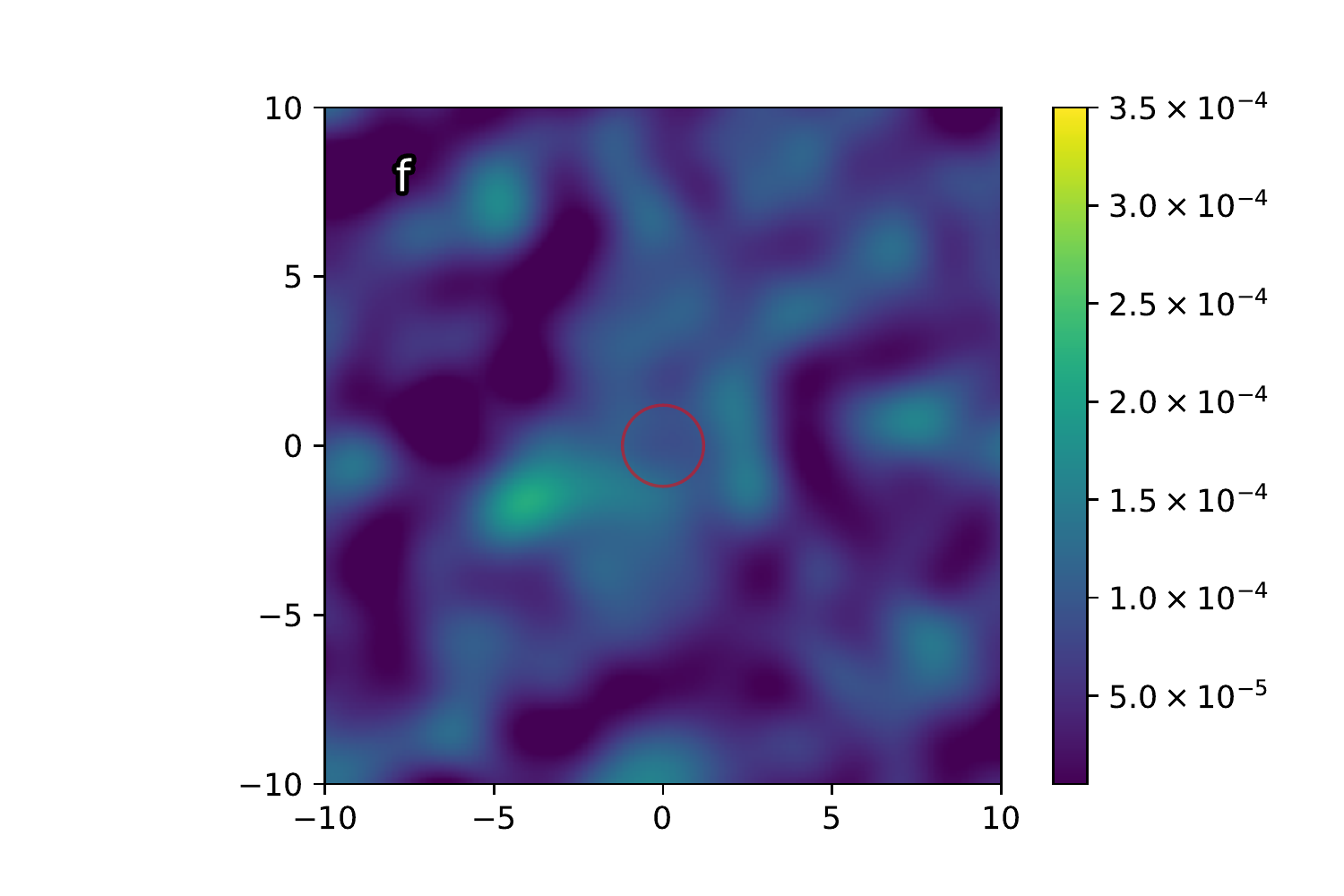}

\includegraphics[clip,trim=2.5cm 0.5cm 3.5cm 1.0cm,height=0.27\textwidth]{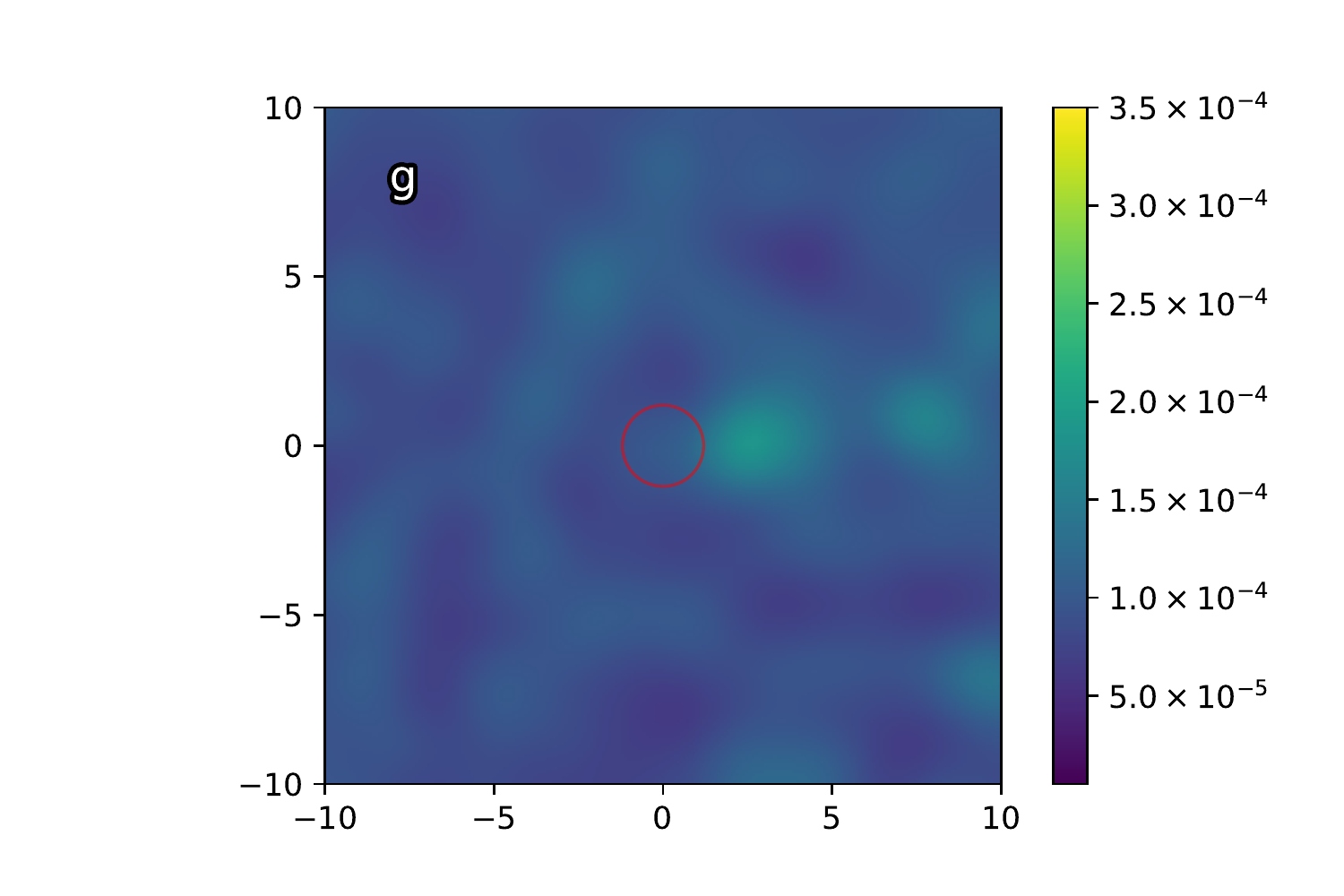}
\includegraphics[clip,trim=2.5cm 0.5cm 3.5cm 1.0cm,height=0.27\textwidth]{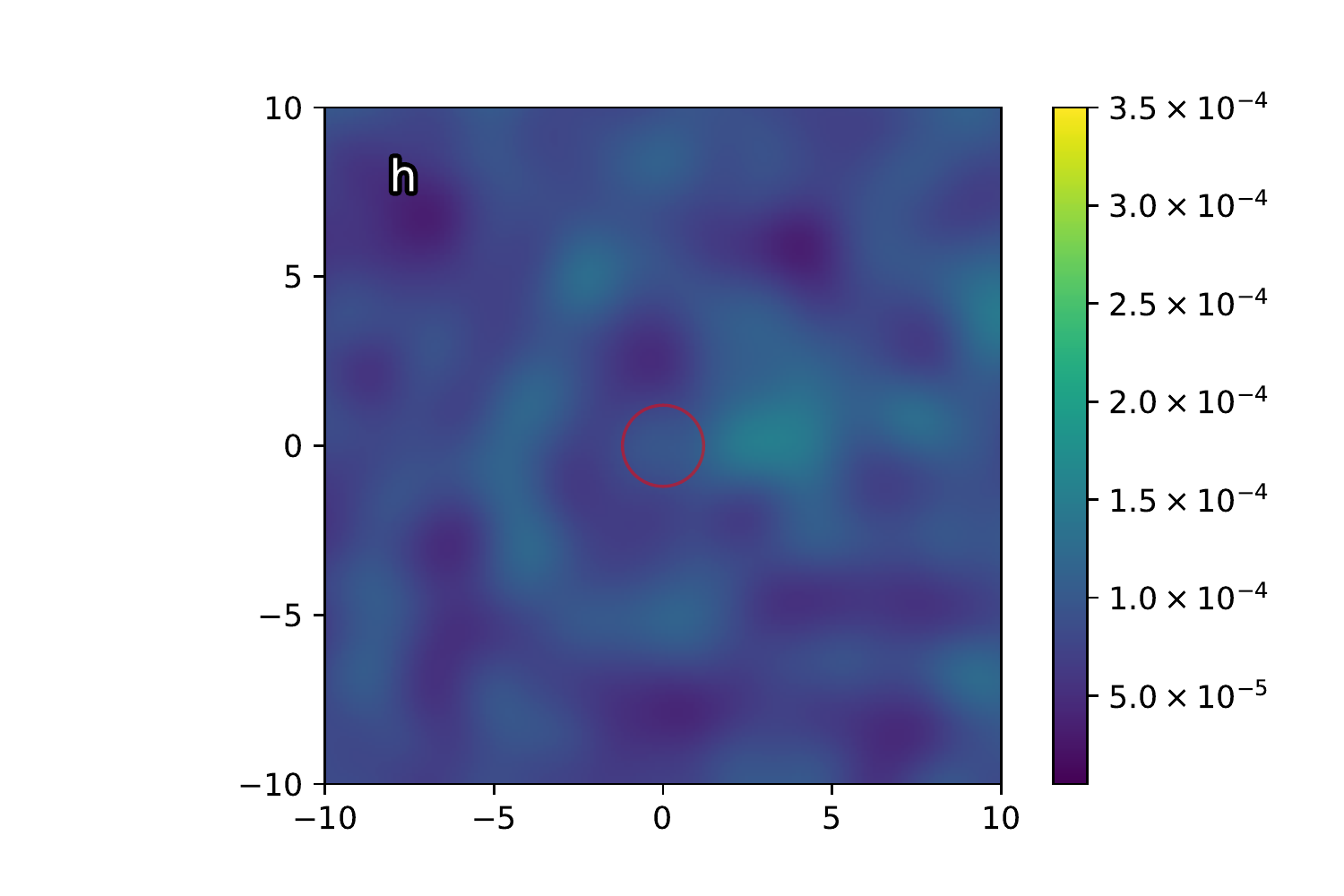}
\includegraphics[clip,trim=2.2cm 0.5cm 0.5cm 1.0cm,height=0.27\textwidth]{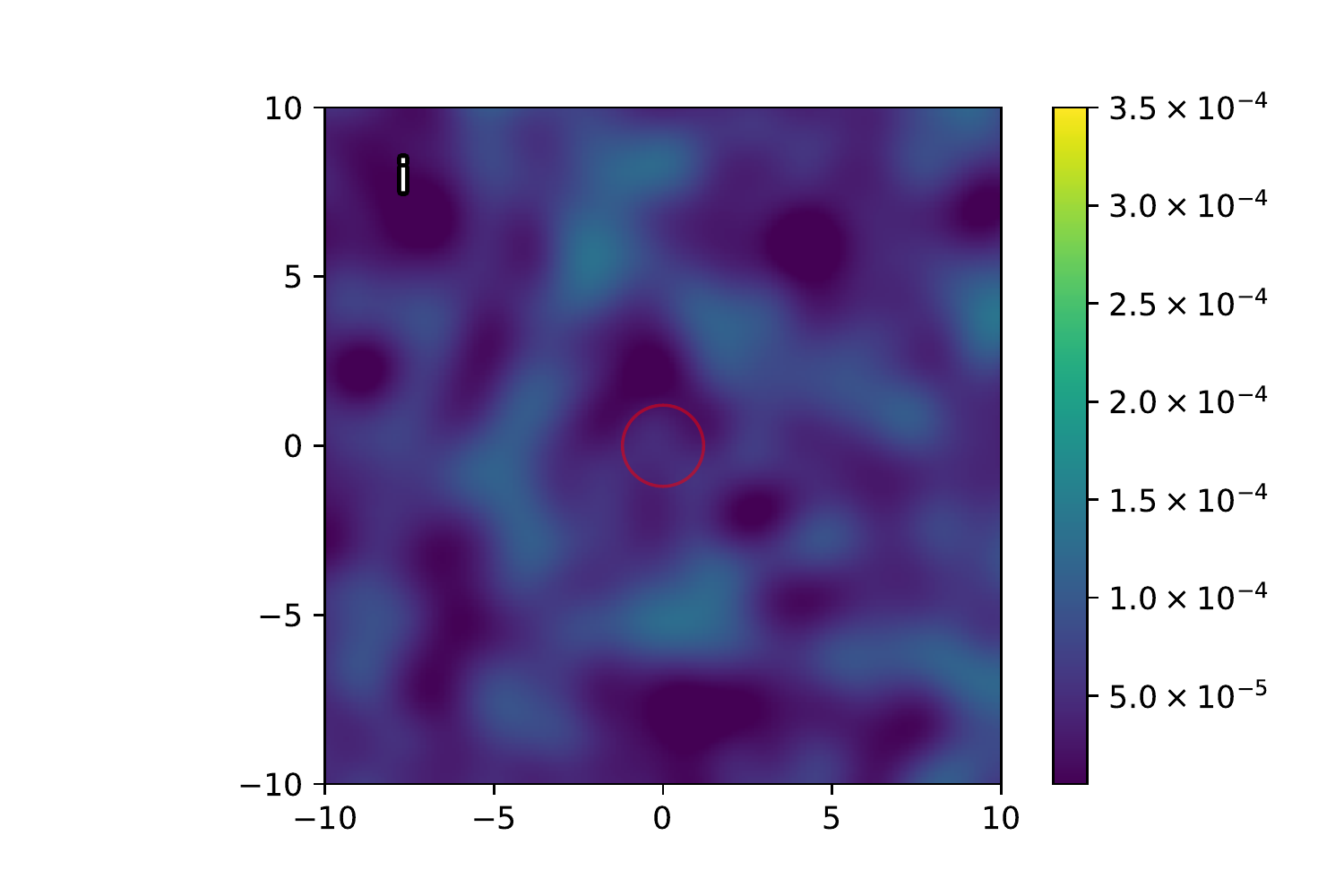}
\caption{Stacks at $250$, $350$, and $500~\micron$ (from left to right) on MaDCoWS (a, b, c) and ACT cluster locations (d, e, f), as well as a set of random locations (g, h, i) in the H-ATLAS data set. See Sect. \ref{sec:submm_emission} for details. The $x$ axis of each plot is aligned with RA, while the $y$ axis is aligned with declination; both are in units of arcmin. The color bars are in units of MJy~sr$^{-1}$. The red circles are $1.2\arcmin$ in radius, the scale inside which we assigned excess flux density as being due to the cluster stack.}
\label{fig:stackedHerschel}
\end{figure*}

\subsubsection{IR and dusty emission}\label{sec:ir_disc}

The {\it Herschel} data, in combination with the ACT $224$~GHz data, suggest that the ACT and MaDCoWS cluster catalogs comprise clusters with different properties and are potentially drawn from different populations; one, dustier population is preferentially sampled by MaDCoWS, and the other, more virialized, is preferentially sampled by ACT as described below. As was discussed in Sect. \ref{sec:submm_emission}, the stacked H-ATLAS data on MaDCoWS clusters show a clear signal in each frequency band, whereas there is no obvious signal when stacking on the ACT clusters (see Fig. \ref{fig:stackedHerschel}). In addition to the stacked emission, we also considered the emission for individual clusters. For each cluster in the H-ATLAS footprint in each of the ACT ($34$ clusters in the H-ATLAS footprint), MaDCoWS ($66$ clusters), and randoms ($66$ clusters) catalog (see Sect. \ref{sec:submm_emission}), we calculated emission within $1.2\arcmin$ radius of the cluster location at each of $250$, $350$, and $500~\mu$m. A histogram of those values is shown in Fig.~\ref{fig:brighthist}.

A two-sample Kolmogorov–Smirnov (KS) test was applied to determine whether the central emission of ACT and MaDCoWS data sets was consistent with the central emission at random points and with each other. At all {\it Herschel} frequencies the distribution of MaDCoWS cluster central emissions is statistically inconsistent with that of the randomly offset central emissions and with that of the ACT clusters. The distribution of ACT cluster central emissions is statistically consistent with the randomly offset central emissions at $500~\micron$, while it is inconsistent at a $p$-value of $0.018$ and $0.016$ at $250$ and $350~\micron$, respectively. The average central emission of the MaDCoWS cluster samples is more than one standard deviation higher than the average central emission of the randomly offset clusters at all frequencies. For the ACT clusters, at all wavelengths the central emission is statistically consistent with the randomly offset clusters. As evidenced by the goodness-of-fit of the gray-body model as shown in Sect.~\ref{sec:submm_emission}, we attribute this emission to the MaDCoWS clusters being dustier on average than the ACT clusters, and identify three causes for this effect. Firstly, in general clusters at high redshift have a greater proportion of blue galaxies in their cores than those at low redshift \citep{Butcher1978,Dressler1980, Brodwin2013}. As blue galaxies tend to be dustier \citep{Casey2014}, and the mean redshift of the ACT clusters is lower than that of the MaDCoWS ($\sim0.5$ vs. $\sim1.01$), the MaDCoWS clusters should be dustier on average. Secondly, at a given redshift, ACT preferentially finds clusters that contain more virialized gas, as virialized gas contributes most strongly to the integrated SZ effect signal \citep{Motl2005, Poole2006, Poole2007, Wik2008, Krause2012}. Finally, since dusty contamination contributes to the SZ infill of clusters, and hence biases SZ surveys like ACT against detecting them, ACT will preferentially select against cluster that contain significant dusty IR emission.


\subsubsection{Radio emission}\label{sec:radio_disc}
The radio data provides additional evidence that the ACT and MaDCoWS cluster catalogs preferentially sample from two different populations. As discussed in Sect.~\ref{sec:radio}, the \madcows clusters have higher intrinsic radio luminosity at $1.4$~GHz as compared to the ACT clusters, even when the ACT clusters are restricted to the same redshift range as the \madcows candidates. Moreover, performing a two-sample KS test on the ACT and MaDCoWS radio fluxes at $1.4$~GHz confirms that they are drawn from different populations. To compare the populations, we first cut the ACT catalog to the same redshift range as the \madcows catalog: $0.7<z<1.5$. We then restricted both cluster catalogs to an intrinsic luminosity corresponding to the standard deviation of the background aperture fluxes, $0.3$~mJy using Eq.~\ref{eq:8}, which roughly represents the noise floor for our background subtracted fluxes. Since the fluxes for both catalogs are taken from the same survey, they should have on average the same noise properties, so that if the non detections are not removed, they will bias the KS test toward high $p$-values, that is, toward determining that the two samples were drawn from the same underlying population. When we perform the two-sample KS test on the ACT and \madcows catalogs with the flux density cut, we obtain $p$-value of $0.025$, which suggests that the \madcows and ACT clusters with radio sources represent different underlying populations. The above analysis ignores what proportion of clusters host radio sources. We also consider the intrinsic luminosities of the ACT and \madcows clusters: We simply consider the percentage of clusters (still restricted to $0.7<z<1.5$) that have intrinsic luminosities greater than the reference aperture flux density $0.3$~mJy converted to luminosity as described above. We used bootstrapping to estimate the uncertainties on this number, which is $67.2\pm 1.7\%$ for the ACT catalog and $58.6\pm 1.0\%$ for the \madcows catalog. 

In summary, this analysis suggests that, restricting the ACT catalog optical photo-redshifts to match those of the \madcows clusters, the ACT clusters host radio sources slightly more frequently than the \madcows clusters, while the radio sources hosted by the \madcows clusters are stronger by a factor of $\approx 2$. We again note that the mean redshift of the ACT clusters is still somewhat lower than that of the \madcows ($\overline{z} = 0.89$ vs. $\overline{z} = 1.01$). 

Further, any contamination of the \madcows catalog (Sect.~\ref{sec:weights}) would in principle bias both the number density of radio sources and, accordingly, also the average inferred intrinsic luminosity, to lower values as radio sources are preferentially found in clusters as opposed to the field \citep{Coble2007}. Assuming an unrealistically high contamination of $10\%$, to set an upper bound on the percentage of clusters with contamination, and that contaminating candidates have radio sources at the rate of the field, which we take to be $10\%$ the rate of the true candidates, we calculate the corrected percentage of \madcows clusters with aperture flux density $>0.3$~mJy as

\begin{equation*}
    \frac{N_{\textsc{cluster, radio}}}{N_{\textsc{cluster}}} = \frac{N_{\textsc{total, radio}}-N_{\textsc{contaminant, radio}}}{N_{\textsc{total}}-N_{\textsc{contaminant}}} \approx 65\%,
\end{equation*}
which is not significantly different from the percentage found for ACT.

\begin{figure}
    \centering
    \includegraphics[clip,trim=1mm 2mm 12mm 0mm,width=\columnwidth]{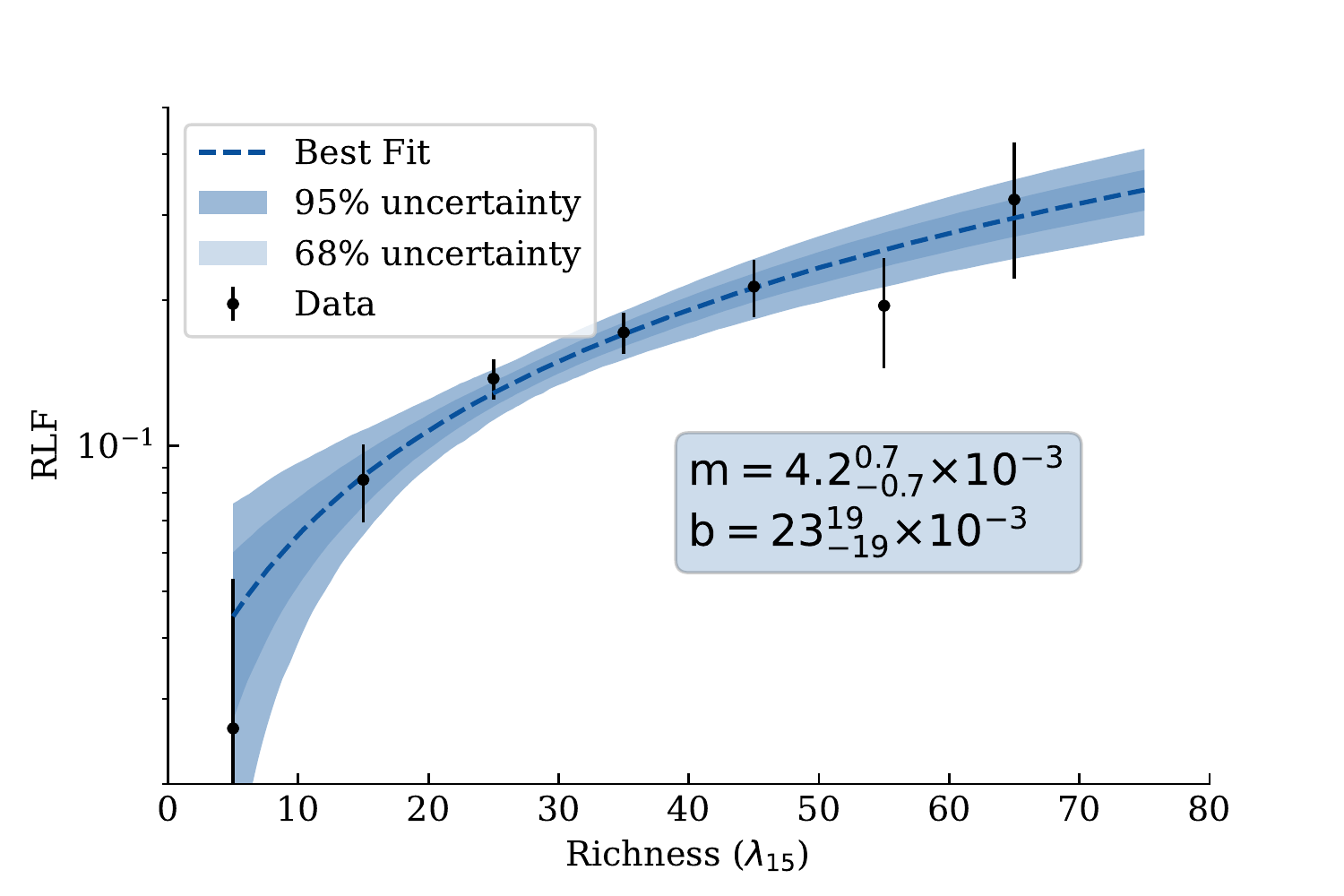}
    \caption{Best fit of a linear model to the RLF of the \madcows clusters determined using NVSS 1.4~GHz fluxes (Sect.~\ref{sec:radio}) vs. richness. The best fit is inconsistent with the null hypothesis of no evolution of the RLF with richness at $5.9\sigma$. In comparison to \citet{Mo2020}, we favor a  higher slope for the relation at $1.2\sigma$, although see the discussion in Sect.~\ref{sec:radio_disc}}
    \label{fig:RLF}
\end{figure}

Finally, we briefly consider the richness evolution of radio luminosity fraction. Following \citet{Mo2020}, we define a cluster to be radio loud if the intrinsic luminosity associated with the cluster is $L_{\textsc{1.4~GHz}} \geq 10^{25} \textsc{W Hz}^{-1}$, and the radio loud fraction (RLF) as the fraction of total clusters that are radio loud. For simplicity we continue to use our $1.2\arcmin$ radial aperture, which corresponds to about $530$ to $625$~kpc, compared to $500$kpc used in \citet{Mo2020}. The effect of this will generally be to raise the RLF as field radio sources are associated with clusters when they lie along the line-of-sight of that cluster, which will in turn flatten the richness-RLF relation. We binned the cluster radio luminosities into 8 bins of richness, and calculated the RLF in those bins, with the uncertainties calculated via bootstrapping. We fit the resulting data to a linear model using a maximum likelihood method, with the uncertainties in the parameters estimated via an MCMC method implemented in \texttt{emcee} \citep{MacKey2013}. The results of that fit are shown in Fig.~\ref{fig:RLF}. We find the best-fit slope to be $m  = 4.1^{+0.7}_{-0.7}\times 10^{-3}$ and the best-fit intercept to be $b = 26^{+18}_{-19}\times 10^{-3}$, and the reduced chi-squared of the fit $\chi_r = 0.45$. The best-fit slope is somewhat higher than \citet{Mo2020}, although only inconsistent at $1.2\sigma$. We are inconsistent with the null hypothesis of no slope at $5.9\sigma$.

\subsection{Infill}

While the average bias from radio emission detected in the ACT ($\textsc{s}_{\textsc{meas}} = 0.8\pm 0.03\%$ and $\textsc{s}_{\textsc{typ}}= 2.0\pm0.1\%$) and \madcows cluster candidates ($\textsc{s}_{\textsc{meas}} =0.45 \pm 0.05 \%$ and $\textsc{s}_{\textsc{typ}}= 0.83\pm0.05\%$) is at a low level, in both cases the samples include high infill tails. For example, 94 ACT clusters show radio emission consistent with a greater than 5\% bias. Interestingly, that number of clusters ($\sim$2.8\% of the $3,335$ ACT clusters with NVSS fluxes) is consistent with the best-fit differential source counts near clusters found in \citet{Coble2007} ($\sim$2.5\%). Of course, the level of infill of the ACT-identified clusters is a biased measure of the true infill, as clusters with less infill are more likely to meet the ACT detection threshold for a given intrinsic mass (\yc). For example, \citet{Gupta2017} estimated that $0.5\%$ and $1.4\%$ of clusters with mass $3\times 10^{14}~\rm M_\odot$ at redshift of $0.25$ had their SZ signal completely suppressed by infill at $150$ and $98$~GHz, respectively. Moreover, accurately computing the infill correction is difficult; propagating the flux density  from lower frequencies means accepting the uncertainty of extrapolating the spectral index across wide frequency ranges, while CMB measurements at $90$ and $150$~GHz frequently lack the spacial resolution to distinguish point sources in clusters from the SZ signal of that cluster. High resolution measurements of the point source flux densities at or near $90$ and $150$~GHz offer the best path forward. Dicker et al. (in prep.) undertook such a measurement using the MUSTANG2 instrument \citep{Dicker2014} on the 100-meter Green Bank Telescope to perform $9\arcsec$ resolution imaging of the SZ effect for a sample of galaxy clusters. This high resolution also allowed them to identify compact sources in the clusters, which would not be resolved by ACT, and to measure their flux density contributions at 90~GHz in order to assess the impact on SZ measurements. They found a similar distribution of infill, with 85\% of clusters showing a change in the  integrated SZ flux less than 5\%, but 10\% of the sample had very high ($>10\%$) infill.
Further, the level of dust infill of the \madcows cluster candidates may be biasing their measured SZ signal. From the model we developed in Sect.~\ref{sec:submm_emission}, we predict that the average dust infill of the SZ signal is $1.5\pm 0.5\%$. This result is limited by the limited IR follow-up available for the \madcows cluster candidates, with {\it Herschel} data for only 66 candidates. While the distribution of the 66 candidates appears normal, we do not have enough data to rule out deviations from that distribution (e.g., a high flux density tail).
While the effect is subdominant to other sources of uncertainty in this work, future large optical and IR surveys (e.g., the Vera C. Rubin observatory \citep{Ivezi2019} and the Euclid telescope \citep{Euclid2020}), as well as next-generation CMB experiments, which aim to place sub-percent constraints on cosmology from cluster abundance counts, will need to use methods that either address or are insensitive to this infill. 
In the near term, higher resolution measurements, both of the point source flux densities and of their spectral indices in the frequency ranges of interest, using, for example, MUSTANG-2, NIKA2, TolTEC, and Atacama Large Millimeter Array (ALMA), would enable improved constraints on the infill of the SZ signal by submillimeter and radio sources.

\subsection{Mass-richness scaling}\label{sec:scaling_disc}

As noted in Sect.~\ref{sec:scaling_results}, our mass-richness scaling relation differs significantly from self-similarity, for which one would expect $p_1 \approx 1$ \citep{Capasso2019, McClintock2019, Bleem2020, Grandis2021}, although it is consistent with the previous measurement from \cite{Gonzalez2019}.  We identify several issues that may bias the mass-richness scaling relation to higher slopes. Firstly, at low richness a portion of the MaDCoWS sample is not well described by a mass-richness scaling relation. If some number of low-richness candidates are either unvirialized systems with low SZ signal or line-of-sight chance superpositions with no intrinsic SZ signal, then the effect would be to bias the mass-richness relation to higher slopes, as these candidates would tilt the lower end of the relation downward (see Sect.~\ref{sec:scaling_results}). In a similar vein, if the richness measure is systematically biased to higher values due to line-of-sight interlopers, then the effect would also be to steepen the mass-richness relation. This effect would be especially apparent as a constant bias in richness is a larger relative effect at lower values for the richness, again having the effect of tilting the scaling relation slope to steeper values. Line-of-sight contaminants are known to bias optically selected clusters \citep{Costanzi2019, Grandis2021}, and correspondingly their mass-richness scaling relation. We would require a larger contamination fraction than either of those two studies to explain the low average weight we observe; however, the \madcows cluster sample is very high redshift, and projection effects are stronger at high redshift \citep{Yee2002, Costanzi2019}. Moreover, it is unclear how often line-of-sight contaminated clusters pass the \madcows selection function as compared to uncontaminated clusters. and it is further unclear how sensitive the \madcows richness measurement is to line-of-sight interlopers. These effects together could cause projection effects to be more severe for the \madcows cluster sample than other samples.

We also assessed the impact of infill from submillimeter and radio sources on the mass-richness scaling relation. As discussed in Sect.~\ref{sec:submm_emission}, we did not have enough {\it Herschel} submillimeter data to infer if infill scales with richness or mass. Therefore, it is possible that the proportionate infill of the SZ signal is higher at low richness than at high richness. We do, however, have $224$~GHz data of many \madcows clusters, which was used to quantify the submillimeter and millimeter source infill. We binned the \madcows clusters in richness and stacked the f220 maps in these bins. There was no trend with richness in the $224$~GHz emission, which suggests that the level of submillimeter and millimeter source infill is not strongly correlated with richness. 

Finally, the higher IR emission of the \madcows cluster candidates is consistent with a gray-body, dusty emission profile, and in general correlates with higher star formation rates (SFRs) \citep[see e.g.,][]{Devlin2009}.
These high SFR clusters may deviate from self-similarity, and as such we would not a priori expect the slope of the mass-richness scaling relation to be unity. 

\subsection{Weights}\label{sec:weights_disc}

As discussed in Sect.~\ref{sec:weights}, the weights for the \madcows candidates are split into two populations with a few high weight clusters (73 with weight $>0.7$) and a large number (544) of candidates with low weights ($<0.5$). The high weight clusters are well characterized by a mass-richness scaling relation; the low weight clusters are not. At very low richness ($\lambda_{15} < 20)$ some of these candidates may be structures smaller than clusters, such as groups or line-of-sight superpositions. 

Additionally, at higher richness, low weight candidates may be mergers or pre-mergers, or otherwise unvirialized but massive systems. Unvirialized gas in these mergers has been shown to suppress the SZ signal from these clusters both in simulations \citep[e.g.,][]{Wik2008,Battaglia2012,Nelson2014} and observations \citep[e.g.,][]{Hilton2018}. While the merger can briefly enhance the SZ signal, the period during which it suppresses it is in general longer than that in which it enhances it. During the merger process, the SZ signal in a merging system can be suppressed $\approx 40\%$ \citep{Dicker2020} compared to a virialized cluster of the same total mass. This has the effect of making the observed SZ mass of the candidate less consistent with the mass-richness scaling relation and more consistent with noise (i.e., the effect is to reduce the weight of the cluster). In \citet{Dicker2020}, which reports on SZ follow-up of high-richness MaDCoWS candidates, $\approx 25\%$ of the MaDCoWS clusters observed were determined to be mergers. This ratio is significantly affected by Malmquist bias, however, as non-mergers have higher SZ brightness than merging systems. Further, \citet{Dicker2020} clusters are generally higher-richness systems ($\lambda_{15} \approx 50$) and it is not clear how the fraction of systems that are undergoing a merger or are pre-merger changes with richness. 
Targeted follow-up is warranted to investigate further. Additionally, stacked measurements of the weak lensing of the CMB, in the vein of \citet{Madhavacheril2020}, are indifferent to clusters that are undergoing mergers. 
As such, comparing the lensing inferred masses of low and high weight clusters would allow one to put constraints on the fraction of low weight clusters that are undergoing mergers. Similarly, the agreement between the average mass of the \madcows cluster sample above richness of $20$ as determined using the mass-richness scaling relation derived in this paper and in \citet{Madhavacheril2020} suggests that mergers and pre-mergers do not dominate the \madcows cluster sample.

Finally, at low richness chance superpositions of galaxies may be causing spurious detections. Such spurious objects are hard to differentiate from low-mass groups in this analysis. In any case, caution must be exercised when using the low weight clusters for cosmology. Inaccurate cluster masses from both spurious clusters and clusters with low SZ signal for their mass (due to merger history, low gas fraction, or unvirialized components along the line-of-sight) will bias cosmological parameters.


\section{Conclusions}\label{sec:conclusions}

In this work we identified co-detections of the ACT and \madcows cluster catalogs. We note the very low rate of co-detections with respect to the size of both the ACT and the \madcows cluster catalogs.  We  used forced photometry to evaluate the mass-richness scaling relation of the \madcows cluster catalog. We quantified the infill of the sample by radio and submillimeter emission. We find that the best-fit scaling relation has a slope of $1.84^{+0.15}_{-0.14}$, higher than the unitary value one would expect for self-similar clusters \citep{Andreon2010} but comparable to the preliminary work of \citet{Gonzalez2019}. We offer some possible reasons for this deviation in Sect.~\ref{sec:scaling_disc}. 
These include, potentially, a bias in the richness measure and some amount of sample impurity with respect to mass.  Instead of mass, the \madcows selection function may preferentially include star-forming systems, leading to deviations from self-similarity. As part of the mass-richness scaling relation, we fit a weight to each cluster, which describes the relative probability that the SZ measurement associated with that cluster was drawn from a normal distribution centered on the mass-richness scaling relation versus a normal distribution centered on zero. We found that a large fraction of the \madcows cluster candidates had weights lower than 50\%. We ascribe the low weight of these clusters, and correspondingly their non-detection in the ACT cluster catalog, to a variety of factors, including suppression of the SZ signal by merger history and other unvirialized components along the line-of-sight, as well as to the \madcows catalog containing very small clusters with SZ signals too low to detect. As for the \madcows non-detections of ACT clusters, we conclude that the \madcows selection function is not mass limited, and thus the overlap of the \madcows catalog with a catalog such as ACT can be very incomplete.

Additionally, by investigating their submillimeter and radio properties, we find evidence that the MaDCoWS candidates have a higher average submillimeter flux density and a higher average intrinsic radio luminosity than their ACT counterparts, even when the ACT clusters are restricted to the same redshift range as the \madcows cluster candidates; however, the \madcows clusters do have a lower average radio flux density in the observed frame of reference, such that the ACT clusters have high radio infill on average. We interpret this as the systems selected by MaDCoWS being on average dustier, while ACT clusters are less dusty and radio loud. We find no evidence that the submillimeter flux density of \madcows clusters evolves with richness, although due to the paucity of {\it Herschel} data this is only based on $220$~GHz stacks. We find that the RLF of the \madcows clusters does increase with increasing richness, and we find that the relation agrees with prior studies \citep{Mo2020}.

Looking to the future, the next generation of MaDCoWS (MaDCoWS2) will  use the deeper CatWISE2020 \citep{Eisenhardt2020} WISE photometry in combination with deeper optical imaging. MaDCoWS2 is designed to extend from $z\sim0.5-2$ with an improved selection function at all redshifts. More in-depth studies of the radio and submillimeter properties of the MaDCoWS clusters in the short term will allow us to better understand the population differences between IR- and SZ-selected clusters and will allow us to better correct the mass-richness scaling relations.   Continuing ACT operations and further data releases will increase the depth of SZ observations, allowing the survey to probe to lower masses and over larger regions. 

Looking further, in the first half of the decade, data from Simons Observatory \citep[SO;][]{zhu2021, thesimonsobservatorycollaboration2019simons,Ade2019}, the CMB-S4 experiment \citep{abitbol2017cmbs4, Abazajian2016, Abazajian2019}, and the Cerro Chajnantor Atacama Telescope-prime \citep[CCAT-prime;][]{ Vavagiakis2018, aravena2019ccatprime, Choi2020} will provide deeper and higher resolution SZ and submillimeter maps over a larger portion of the sky. Additionally, the Vera C. Rubin Observatory \citep{Ivezi2019}, Spectro-Photometer for the History of the Universe, Epoch of Reionization, and Ices Explorer (SPHEREx) \citep{Bock2018, Cooray2018, Crill2020}, and Euclid telescopes \citep{Euclid2019} will provide larger and more accurate IR and optically selected cluster catalogs. 

In the 2030s, the next-generation CMB-S4 experiment \citep{Abazajian2016, abitbol2017cmbs4}, CMB-HD experiment \citep{Sehgal2019,Sehgal2020}, and the Atacama Large Aperture Submillimeter Telescope \citep[AtLAST;][]{Klaassen2020} will potentially provide orders of magnitude increases in SZ map depth over significant fractions of the extragalactic sky, with the latter two providing a transformative leap in subarcminute resolution SZ and submillimeter studies at a significantly lower confusion limit. 


\begin{acknowledgements}
      We thank the referee for the useful and constructive comments that helped improve this work.
      This work was supported by the U.S. National Science Foundation through awards AST-1440226, AST0965625 and AST-0408698 for the ACT project, as well as awards PHY-1214379 and PHY-0855887. Funding was also provided by Princeton University, the University of Pennsylvania, and a Canada Foundation for Innovation (CFI) award to UBC. ACT operates in the Parque Astron\'{o}mico Atacama in northern Chile under the auspices of the Comisi\'{o}n Nacional de Investigaci\'{o}n Cient\'{i}fica y Tecnol\'{o}gica de Chile (CONICYT), now La Agencia Nacional de Investigaci\'{o}n y Desarrollo (ANID). Computations were performed on the GPC and \emph{Niagara} supercomputers at the SciNet HPC Consortium. SciNet is funded by the CFI under the auspices of Compute Canada, the Government of Ontario, the Ontario Research Fund -- Research Excellence; and the University of Toronto. The development of multichroic detectors and lenses was supported by NASA grants NNX13AE56G and NNX14AB58G.   
      {\it Herschel} is an ESA space observatory with science instruments provided by European-led Principal Investigator consortia and with important participation from NASA.
      
      Colleagues at AstroNorte and RadioSky provide logistical support and keep operations in Chile running smoothly. We also thank the Mishrahi Fund and the Wilkinson Fund for their generous support of the project. Zhilei Xu is supported by the Gordon and Betty Moore Foundation. Kavilan Moodley acknowledges support from the National Research Foundation of South Africa. John P. Hughes acknowledges funding for SZ cluster studies from NSF grant number AST-1615657. Dongwon Han, Amanda MacInnis, and Neelima Sehgal acknowledge support from NSF grant number AST-1907657. Crist\'obal Sif\'on acknowledges support from the ANID under FONDECYT grant no.\ 11191125. Luca Di Mascolo is supported by the ERC-StG `ClustersXCosmo' grant agreement 716762.
\end{acknowledgements}

\bibliographystyle{aa}
\bibliography{madcows}

\appendix

\section{Bootstrapping}\label{app:boots}
Throughout the paper, we are presented with a situation in which we have a data set, $\vec{d}$, which we are interested in computing some statistic on, and for which we would like to estimate the uncertainty in that statistic. We use the bootstrapping method to do so: given $\vec{d}$, we resample from that data set with replacement k times, resulting in a superset of data sets, $\{\vec{d}_1, ... \vec{d}_i,... \vec{d}_k\}$ where the $\vec{d}_i$ are the resampled data sets. For each $\vec{d}_i$, we compute a statistic on $\vec{d}_i$, which we call $\mu_i$, generally the mean, or in the case where the data are maps, the central emission. We then have a set of statistics, $\vec{\mu} = \{\mu_1, ... \mu_i,... \mu_k\}$. We then take the average and standard deviation of $\vec{\mu}$ as the average and standard deviation of that statistic on $\vec{d}$.

\section{ACT SZ masses versus CARMA, ACA, MUSTANG2, and NIKA2}\label{app:szmasscomp}
\begin{table*}
  \centering
  \caption{Comparison of the masses inferred using ACT data to those from SZ observations with CARMA \citep{Gonzalez2019}, ACA \citep{DiMascolo2020}, MUSTANG2 \citep{Dicker2020}, and NIKA2 \citep{Ruppin2020}.  
  }
  \begin{tabular}{lccccc}
    \hline\hline
    \noalign{\smallskip}
    Cluster ID & $M_{500,\textsc{ACT}}$ & $M_{500,\textsc{ACA}}$\tablefootmark{a} & $M_{500,\textsc{CARMA}}$\tablefootmark{b} &   $M_{500,\textsc{MUSTANG2}}$\tablefootmark{c}    & $M_{500,\textsc{NIKA2}}$\tablefootmark{d}    \\\noalign{\vspace{1pt}}
               & $[10^{14}~\rm M_\odot]$ & $[10^{14}~\rm M_\odot]$ & $[10^{14}~\rm M_\odot]$ & $[10^{14}~\rm M_\odot]$ & $[10^{14}~\rm M_\odot]$  \\
    \noalign{\smallskip}
    \hline
    \noalign{\smallskip}
    MOO~J0105$+$1324 & $3.53\substack{+0.65\\-0.55}$ &      & $4.03\substack{+0.48\\-0.45}$  & $3.83^{+0.38}_{-0.37}$    &       \\\noalign{\vspace{1pt}}
    MOO~J0129$-$1640 & $3.14\substack{+0.70\\-0.57}$ & $2.57\substack{+0.30\\-0.30}$ &       &         &       \\\noalign{\vspace{1pt}}
    MOO~J0319$-$0025 & $2.38\substack{+0.56\\-0.45}$ &      & $3.11\substack{+0.53\\-0.47}$  &         &       \\\noalign{\vspace{1pt}}
    MOO~J1014$+$0038 & $3.53\substack{+0.57\\-0.50}$ &      & $3.26\substack{+0.32\\-0.30}$  & $3.12^{+0.30}_{-0.30}$    &       \\\noalign{\vspace{1pt}}
    MOO~J1142$+$1527 & $5.00\substack{+0.78\\-0.67}$ &      & $5.45\substack{+0.58\\-0.51}$  & $3.52^{+0.34}_{-0.33}$     & $6.06 \pm 3.47$     \\\noalign{\vspace{1pt}}
    MOO~J1322$-$0228 & $3.30\substack{+0.66\\-0.55}$ &      &       & $3.07^{+0.48}_{-0.58}$    &       \\\noalign{\vspace{1pt}}
    MOO~J1354$+$1329 & $2.05\substack{+0.35\\-0.30}$ &      &       & $2.46^{+0.32}_{-0.35}$    &       \\\noalign{\vspace{1pt}}
    MOO~J1414$+$0227 & $3.04\substack{+0.56\\-0.40}$ & $2.75\substack{+0.32\\-0.32}$ &       &         &       \\\noalign{\vspace{1pt}}
    MOO~J1514$+$1346 & $2.54\substack{+0.43\\-0.37}$ &      & $1.89\substack{+0.68\\-0.79}$  &         &       \\\noalign{\vspace{1pt}}
    MOO~J1521$+$0452 & $3.68\substack{+0.57\\-0.50}$ &      & $3.65\substack{+1.03\\-0.94}$  &         &       \\\noalign{\vspace{1pt}}
    MOO~J2146$-$0320 & $3.16\substack{+0.60\\-0.50}$ & $3.90\substack{+0.54\\-0.76}$ &       &         &       \\\noalign{\vspace{1pt}}
    MOO~J2206$+$0906 & $4.34\substack{+0.83\\-0.70}$ &      & $2.66\substack{+0.93\\-0.74}$   &         &       \\\noalign{\vspace{1pt}}
    MOO~J2231$+$1130 & $3.54\substack{+0.75\\-0.62}$&      & $4.38\substack{+1.51\\-1.37}$   &         &      \\
    \noalign{\smallskip}
    \hline
    \noalign{\medskip}
  \end{tabular}
    \tablefoot{
        \tablefoottext{a}{\citealt{DiMascolo2020}} 
        \tablefoottext{b}{\citealt{Gonzalez2019}}
        \tablefoottext{c}{\citealt{Dicker2020}.  A full treatment of the uncertainty is available in the publication.}
        \tablefoottext{d}{The NIKA2 measurement, as reported in \citealt{Ruppin2020}, relied on a strong, informative prior on the integrated SZ signal from CARMA \citep{Gonzalez2019} and hence should likely not be regarded as an independent constraint. We include it for completeness, and note they include a large, 56\% systematic uncertainty. All three surveys assumed the \citet{Arnaud2010} scaling relation.}
    }  \label{tab:szmasscomp}
\end{table*}

\section{MaDCoWS/ACT co-detections}

\clearpage
\input{catalog}

\end{document}

%% file: catalog.tex
\onecolumn

\begin{longtable}[p!]{clccccc}
\caption{ACT/MaDCoWS Co-detections}\\
\hline\hline\noalign{\smallskip}
ACT Name & ACT $z$ & ACT $z$ Type & $M_{\rm 500,SZ} [10^{14}M_\odot]$ & MaDCoWS Name & MaDCoWS $z$ & $\lambda_{15}$\\\noalign{\smallskip}
\hline\noalign{\smallskip}
ACT-CL J0019.0-0000 & 0.862& spec & $1.68\substack{+0.37\\-0.30}$ & MOO J0018+0000 & 0.88 & 19$\pm$ 5 \\
ACT-CL J0019.8+0210 & 0.856& spec & $2.33\substack{+0.42\\-0.35}$ & MOO J0019+0209 & 0.83 & 29$\pm$ 6 \\
ACT-CL J0023.9-0945 & 1.04$\pm$ 0.03 & phot & $2.34\substack{+0.49\\-0.41}$ & MOO J0024-0944 & --- & ---\\
ACT-CL J0028.1-1005 & 0.99$\pm$ 0.03 & phot & $3.77\substack{+0.67\\-0.57}$ & MOO J0028-1005 & 0.97 & 57$\pm$ 8 \\
ACT-CL J0048.4+1757 & 0.737& spec & $2.03\substack{+0.46\\-0.37}$ & MOO J0048+1757 & 0.82 & 58$\pm$ 8 \\
ACT-CL J0101.7+0030 & 0.94$\pm$ 0.02 & phot & $2.29\substack{+0.42\\-0.35}$ & MOO J0101+0030 & 0.97 & 42$\pm$ 7 \\
ACT-CL J0102.6+0201 & 1.02$\pm$ 0.08 & phot & $2.11\substack{+0.39\\-0.33}$ & MOO J0102+0201 & 1.17 & 20$\pm$ 5 \\
ACT-CL J0105.5+1323 & 1.143& spec & $3.09\substack{+0.54\\-0.46}$ & MOO J0105+1324 & 1.13 & 87$\pm$ 9 \\
ACT-CL J0105.8-1839 & 0.930$\pm$ 0.019 & phot & $4.40\substack{+0.75\\-0.64}$ & MOO J0105-1839 & 0.91 & 65$\pm$ 8 \\
ACT-CL J0120.7-0305 & 1.12$\pm$ 0.03 & phot & $1.74\substack{+0.35\\-0.29}$ & MOO J0120-0304 & 1.24 & 40$\pm$ 6 \\
ACT-CL J0125.3-0802 & 1.06$\pm$ 0.03 & phot & $3.56\substack{+0.55\\-0.48}$ & MOO J0125-0802 & 1.03 & 65$\pm$ 8 \\
ACT-CL J0129.2-1641 & 1.05$\pm$ 0.03 & phot & $2.65\substack{+0.56\\-0.46}$ & MOO J0129-1640 & 1.05 & 49$\pm$ 7 \\
ACT-CL J0131.9+0329 & 0.99$\pm$ 0.03 & phot & $1.70\substack{+0.35\\-0.29}$ & MOO J0132+0329 & 1.14 & 35$\pm$ 6 \\
ACT-CL J0208.1-0935 & 1.07$\pm$ 0.03 & phot & $2.02\substack{+0.46\\-0.37}$ & MOO J0208-0935 & 1.1 & 24$\pm$ 5 \\
ACT-CL J0212.2+0746 & 0.47$\pm$ 0.02 & phot & $5.63\substack{+1.02\\-0.87}$ & MOO J0212+0746 & --- & ---\\
ACT-CL J0221.0+1755 & 1.01$\pm$ 0.03 & phot & $1.89\substack{+0.42\\-0.35}$ & MOO J0221+1755 & --- & ---\\
ACT-CL J0234.5-0107 & 0.535& spec & $2.04\substack{+0.42\\-0.35}$ & MOO J0234-0107 & 1.19 & 27$\pm$ 5 \\
ACT-CL J0239.6-1036 & 0.889$\pm$ 0.017 & phot & $2.58\substack{+0.57\\-0.47}$ & MOO J0239-1035 & 0.94 & 22$\pm$ 5 \\
ACT-CL J0248.4-0925 & 1.04$\pm$ 0.03 & phot & $2.26\substack{+0.51\\-0.41}$ & MOO J0248-0925 & --- & ---\\
ACT-CL J0256.5+0006 & 0.362& spec & $3.94\substack{+0.75\\-0.63}$ & MOO J0256+0006 & --- & ---\\
ACT-CL J0300.2+0125 & 1.27$\pm$ 0.03 & phot & $2.84\substack{+0.47\\-0.40}$ & MOO J0300+0124 & 1.33 & 37$\pm$ 6 \\
ACT-CL J0303.6+1857 & 1.21$\pm$ 0.07 & phot & $3.23\substack{+0.53\\-0.46}$ & MOO J0303+1857 & 1.21 & 45$\pm$ 7 \\
ACT-CL J0308.1-2915 & 1.00$\pm$ 0.03 & phot & $3.01\substack{+0.56\\-0.47}$ & MOO J0308-2915 & 1.01 & 53$\pm$ 7 \\
ACT-CL J0353.3+0832 & 1.16$\pm$ 0.07 & phot & $2.41\substack{+0.50\\-0.41}$ & MOO J0353+0832 & 1.16 & 34$\pm$ 6 \\
ACT-CL J0448.4-1705 & 0.96$\pm$ 0.05 & phot & $3.22\substack{+0.63\\-0.52}$ & MOO J0448-1705 & 0.96 & 79$\pm$ 9 \\
ACT-CL J0934.4+1751 & 0.87$\pm$ 0.03 & phot & $2.86\substack{+0.50\\-0.42}$ & MOO J0934+1751 & 0.96 & 53$\pm$ 7 \\
ACT-CL J1008.7+1147 & 0.259& spec & $4.52\substack{+0.92\\-0.77}$ & MOO J1008+1148 & --- & ---\\
ACT-CL J1014.1+0038 & 1.23& spec & $3.13\substack{+0.49\\-0.43}$ & MOO J1014+0038 & 1.21 & 43$\pm$ 6 \\
ACT-CL J1029.9+0016 & 1.35$\pm$ 0.05 & phot & $2.22\substack{+0.41\\-0.34}$ & MOO J1029+0017 & 1.46 & 31$\pm$ 5 \\
ACT-CL J1048.7+0743 & 0.87$\pm$ 0.03 & phot & $2.36\substack{+0.48\\-0.39}$ & MOO J1048+0743 & 0.94 & 37$\pm$ 6 \\
ACT-CL J1053.2+1052 & 0.89$\pm$ 0.03 & phot & $4.41\substack{+0.71\\-0.61}$ & MOO J1053+1052 & 0.99 & 43$\pm$ 7 \\
ACT-CL J1110.2-0030 & 0.995$\pm$ 0.016 & phot & $2.30\substack{+0.46\\-0.38}$ & MOO J1110-0030 & --- & ---\\
ACT-CL J1139.3+0154 & 1.052$\pm$ 0.017 & phot & $3.13\substack{+0.53\\-0.45}$ & MOO J1139+0154 & --- & ---\\
ACT-CL J1142.1+1345 & 1.14$\pm$ 0.03 & phot & $1.86\substack{+0.36\\-0.30}$ & MOO J1142+1346 & 1.24 & 33$\pm$ 6 \\
ACT-CL J1142.7+1527 & 1.19& spec & $4.41\substack{+0.65\\-0.57}$ & MOO J1142+1527 & 1.12 & 57$\pm$ 7 \\
ACT-CL J1149.4+0921 & 0.91$\pm$ 0.03 & phot & $2.04\substack{+0.41\\-0.34}$ & MOO J1149+0921 & 0.96 & 30$\pm$ 6 \\
ACT-CL J1152.3+1652 & 1.19$\pm$ 0.03 & phot & $2.38\substack{+0.39\\-0.34}$ & MOO J1152+1652 & --- & ---\\
ACT-CL J1205.0+1525 & 1.20$\pm$ 0.03 & phot & $1.59\substack{+0.32\\-0.27}$ & MOO J1204+1525 & 1.08 & 39$\pm$ 6 \\
ACT-CL J1205.3-0245 & 0.98$\pm$ 0.03 & phot & $2.11\substack{+0.46\\-0.38}$ & MOO J1205-0244 & --- & ---\\
ACT-CL J1208.3+0501 & 0.90$\pm$ 0.03 & phot & $1.78\substack{+0.39\\-0.32}$ & MOO J1208+0501 & 0.9 & 43$\pm$ 7 \\
ACT-CL J1241.0+0010 & 0.790$\pm$ 0.015 & phot & $2.19\substack{+0.49\\-0.40}$ & MOO J1241+0011 & 0.84 & 34$\pm$ 6 \\
ACT-CL J1254.9+0947 & 0.80$\pm$ 0.03 & phot & $1.49\substack{+0.34\\-0.27}$ & MOO J1254+0948 & 0.81 & 39$\pm$ 7 \\
ACT-CL J1310.6+1707 & 1.04$\pm$ 0.03 & phot & $1.48\substack{+0.32\\-0.26}$ & MOO J1310+1707 & 1.01 & 49$\pm$ 7 \\
ACT-CL J1322.9-0227 & 0.793& spec & $2.90\substack{+0.55\\-0.46}$ & MOO J1322-0228 & 0.82 & 83$\pm$ 9 \\
ACT-CL J1346.2-0142 & 1.19$\pm$ 0.04 & phot & $2.40\substack{+0.45\\-0.37}$ & MOO J1346-0142 & 1.24 & 74$\pm$ 8 \\
ACT-CL J1354.8+1329 & 1.48$\pm$ 0.07 & phot & $1.83\substack{+0.30\\-0.25}$ & MOO J1354+1329 & 1.48 & 44$\pm$ 6 \\
ACT-CL J1355.8+1607 & 0.97$\pm$ 0.03 & phot & $2.69\substack{+0.43\\-0.37}$ & MOO J1355+1606 & 1.03 & 48$\pm$ 7 \\
ACT-CL J1414.5+0227 & 1.04$\pm$ 0.03 & phot & $2.68\substack{+0.47\\-0.40}$ & MOO J1414+0227 & 1.02 & 41$\pm$ 7 \\
ACT-CL J1418.2+0723 & 1.26$\pm$ 0.03 & phot & $2.06\substack{+0.34\\-0.29}$ & MOO J1418+0723 & 1.31 & 44$\pm$ 6 \\
ACT-CL J1424.9-0141 & 0.948$\pm$ 0.016 & phot & $1.90\substack{+0.45\\-0.36}$ & MOO J1424-0141 & 0.88 & 41$\pm$ 7 \\
ACT-CL J1454.6+0628 & 1.31$\pm$ 0.07 & phot & $1.28\substack{+0.28\\-0.23}$ & MOO J1454+0628 & 1.31 & 27$\pm$ 5 \\
ACT-CL J1455.5+0439 & 0.86$\pm$ 0.03 & phot & $1.91\substack{+0.37\\-0.31}$ & MOO J1455+0439 & 0.88 & 36$\pm$ 6 \\
ACT-CL J1514.7+1346 & 1.059& spec & $2.29\substack{+0.37\\-0.32}$ & MOO J1514+1346 & 1.09 & 73$\pm$ 8 \\
ACT-CL J1521.1+0451 & 1.312& spec & $3.27\substack{+0.49\\-0.42}$ & MOO J1521+0452 & 1.28 & 46$\pm$ 7 \\
ACT-CL J1525.8+1540 & 1.02$\pm$ 0.05 & phot & $3.25\substack{+0.51\\-0.44}$ & MOO J1525+1541 & 1.02 & 52$\pm$ 7 \\
ACT-CL J1536.5+0954 & 0.770& spec & $1.51\substack{+0.33\\-0.27}$ & MOO J1536+0953 & 1.02 & 39$\pm$ 6 \\
ACT-CL J1620.1+1340 & 0.92$\pm$ 0.03 & phot & $1.73\substack{+0.35\\-0.29}$ & MOO J1620+1340 & 0.94 & 33$\pm$ 6 \\
ACT-CL J2121.8+0040 & 0.516& spec & $2.51\substack{+0.63\\-0.50}$ & MOO J2121+0040 & 1.14 & ---\\
ACT-CL J2146.6-0321 & 1.16$\pm$ 0.05 & phot & $3.15\substack{+0.60\\-0.50}$ & MOO J2146-0320 & 1.16 & 50$\pm$ 7 \\
ACT-CL J2204.9-2955 & 1.31$\pm$ 0.06 & phot & $3.72\substack{+0.61\\-0.53}$ & MOO J2205-2955 & 1.31 & 52$\pm$ 7 \\
ACT-CL J2231.9+1131 & 0.81$\pm$ 0.03 & phot & $3.10\substack{+0.62\\-0.52}$ & MOO J2231+1130 & 0.8 & 48$\pm$ 7 \\
ACT-CL J2235.0+1321 & 0.86$\pm$ 0.03 & phot & $5.81\substack{+0.94\\-0.80}$ & MOO J2235+1320 & 0.84 & 58$\pm$ 8 \\
ACT-CL J2316.2+0920 & 0.79$\pm$ 0.03 & phot & $2.26\substack{+0.54\\-0.44}$ & MOO J2316+0920 & 0.86 & 49$\pm$ 7 \\
ACT-CL J2319.8-1856 & 0.93$\pm$ 0.05 & phot & $2.84\substack{+0.59\\-0.49}$ & MOO J2319-1856 & 0.93 & 47$\pm$ 7 \\
ACT-CL J2326.2+0030 & 1.18$\pm$ 0.04 & phot & $1.78\substack{+0.35\\-0.29}$ & MOO J2326+0030 & --- & ---\\
ACT-CL J2332.6-0014 & 0.99$\pm$ 0.03 & phot & $1.85\substack{+0.36\\-0.30}$ & MOO J2332-0014 & --- & ---\\
ACT-CL J2358.8+1836 & 1.40$\pm$ 0.05 & phot & $2.02\substack{+0.39\\-0.32}$ & MOO J2358+1836 & 1.4 & 36$\pm$ 6 \\
ACT-CL J0028.9-4449 & 0.94$\pm$ 0.03 & phot & $1.49\substack{+0.34\\-0.27}$ & MOO J0028-4449 & 0.930 & 29$\pm$ 6 \\
ACT-CL J0151.3-4300 & 1.24$\pm$ 0.03 & phot & $1.88\substack{+0.37\\-0.31}$ & MOO J0151-4300 & --- & ---\\
ACT-CL J0151.4-5954 & 0.95$\pm$ 0.03 & phot & $2.65\substack{+0.50\\-0.42}$ & MOO J0151-5954 & --- & ---\\
ACT-CL J0200.7-3106 & 1.02$\pm$ 0.03 & phot & $3.23\substack{+0.54\\-0.46}$ & MOO J0200-3106 & --- & ---\\
ACT-CL J0244.6-3011 & 1.01$\pm$ 0.09 & phot & $2.18\substack{+0.48\\-0.39}$ & MOO J0244-3011 & 1.370 & 36$\pm$ 6 \\
ACT-CL J0248.3-4931 & 0.96$\pm$ 0.02 & phot & $1.73\substack{+0.37\\-0.31}$ & MOO J0248-4930 & --- & ---\\
ACT-CL J0339.1-3952 & 1.24$\pm$ 0.07 & phot & $2.40\substack{+0.42\\-0.36}$ & MOO J0339-3951 & 1.240 & 43$\pm$ 7 \\
ACT-CL J0434.6-3723 & 0.916$\pm$ 0.017 & phot & $2.77\substack{+0.54\\-0.45}$ & MOO J0434-3723 & 0.850 & 51$\pm$ 8 \\
ACT-CL J2241.3-5339 & 0.94$\pm$ 0.02 & phot & $2.20\substack{+0.48\\-0.39}$ & MOO J2241-5338 & --- & ---\\
ACT-CL J2312.0-4844 & 0.677$\pm$ 0.014 & phot & $2.69\substack{+0.50\\-0.42}$ & MOO J2312-4844 & 0.800 & 41$\pm$ 7 \\
ACT-CL J2332.7-3813 & 1.21$\pm$ 0.03 & phot & $2.27\substack{+0.40\\-0.34}$ & MOO J2332-3813 & --- & ---\\
ACT-CL J2334.2-4324 & 0.574$\pm$ 0.008 & phot & $2.30\substack{+0.46\\-0.38}$ & MOO J2334-4324 & --- & ---\\
ACT-CL J2335.3-3256 & 0.51$\pm$ 0.04 & phot & $3.32\substack{+0.61\\-0.51}$ & MOO J2335-3256 & --- & ---\\
ACT-CL J0204.3-1918 & 1.01$\pm$ 0.03 & phot & $3.42\substack{+0.61\\-0.52}$ & MOO J0204-1918 & 1.100 & 47$\pm$ 7 \\
ACT-CL J0319.4-0025 & 1.194& spec & $1.99\substack{+0.44\\-0.36}$ & MOO J0319-0025 & 1.210 & 33$\pm$ 6 \\
ACT-CL J0842.3+0033 & 1.028$\pm$ 0.016 & phot & $1.94\substack{+0.43\\-0.35}$ & MOO J0842+0033 & 0.980 & 47$\pm$ 7 \\
ACT-CL J0856.4+1736 & 0.81$\pm$ 0.03 & phot & $3.37\substack{+0.57\\-0.49}$ & MOO J0856+1736 & 0.800 & 50$\pm$ 7 \\
ACT-CL J1004.5+1203 & 0.93$\pm$ 0.03 & phot & $2.37\substack{+0.47\\-0.39}$ & MOO J1004+1203 & 0.980 & 35$\pm$ 6 \\
ACT-CL J1106.8+0737 & 1.09$\pm$ 0.08 & phot & $2.30\substack{+0.45\\-0.37}$ & MOO J1106+0737 & 1.090 & 52$\pm$ 7 \\
ACT-CL J1221.5+1604 & 0.82$\pm$ 0.03 & phot & $1.80\substack{+0.38\\-0.31}$ & MOO J1221+1603 & 0.910 & 70$\pm$ 9 \\
ACT-CL J1303.2+1733 & 0.81$\pm$ 0.03 & phot & $2.87\substack{+0.48\\-0.41}$ & MOO J1303+1733 & 0.860 & 49$\pm$ 7 \\
ACT-CL J1426.7+1740 & 0.91$\pm$ 0.03 & phot & $2.90\substack{+0.47\\-0.40}$ & MOO J1426+1741 & 0.930 & 65$\pm$ 8 \\
ACT-CL J2140.5+0248 & 1.02$\pm$ 0.03 & phot & $2.95\substack{+0.61\\-0.50}$ & MOO J2140+0248 & 1.049 & 59$\pm$ 8 \\
ACT-CL J2206.5+0906 & 0.84$\pm$ 0.03 & phot & $3.82\substack{+0.69\\-0.58}$ & MOO J2206+0906 & 0.930 & 53$\pm$ 7 \\
ACT-CL J2320.2-0621 & 0.923& spec & $1.86\substack{+0.42\\-0.34}$ & MOO J2320-0620 & 0.980 & 37$\pm$ 6 \\
ACT-CL J0103.7+0119 & 0.213$\pm$ 0.017 & phot & --- & MOO J0103+0117 & --- & ---\\
ACT-CL J1254.9+0947 & 0.80$\pm$ 0.03 & phot & --- & MOO J1254+0948 & 0.81 & 39$\pm$ 7 \\
ACT-CL J2332.7-3813 & 1.21$\pm$ 0.03 & phot & ---& MOO J2332-3813 & --- & ---\\
ACT-CL J2140.5+0248 & 1.02$\pm$ 0.03 & phot & --- & MOO J2140+0248 & 1.049 & 59$\pm$ 8 \\
\hline
\caption*{These cluster co-detections were made following the analysis described in Section~\ref{sec:crossmatches}. Richness ($\lambda_{15}$) values are those reported by \cite{Gonzalez2019}. Note that some clusters do not have a $\lambda_{15}$ value associated with them, as \cite{Gonzalez2019} did not report richness values for clusters with for clusters with low partial IRAC data.}
\label{Catalog:ACTCOWS}
\end{longtable}

\twocolumn